\documentclass[a4paper,fleqn,usenatbib]{mnras}

\usepackage{mathptmx}

\usepackage[T1]{fontenc}
\usepackage{ae,aecompl}

\usepackage{graphicx}	
\usepackage{amsmath}	
\usepackage{amssymb}	
\usepackage[usenames, dvipsnames]{color}
\usepackage{url}

\defcitealias{Husser2016}{Paper~I}
\defcitealias{Kamann2016}{Paper~II}

\title[Globular cluster kinematics with MUSE]{A stellar census in globular clusters with MUSE \\ {\Large The contribution of rotation to cluster dynamics studied with 200\,000 stars}}

\author[S. Kamann et al.]{%
S. Kamann,$^{1,2}$\thanks{E-mail: s.kamann@ljmu.ac.uk}
T.-O. Husser,$^{1}$
S. Dreizler,$^{1}$
E. Emsellem,$^{3,4}$
P.~M. Weilbacher,$^{5}$
\newauthor
S. Martens,$^{1}$
R. Bacon,$^{4}$
M. den Brok,$^{6}$
B. Giesers,$^{1}$
D. Krajnovi\'{c},$^{5}$
\newauthor
M.~M. Roth,$^{5}$
M. Wendt,$^{5,\,7}$
L. Wisotzki$^{5}$
\\
$^{1}$Institute for Astrophysics, Georg-August-University, Friedrich-Hund-Platz 1, 37077 G\"ottingen, Germany\\
$^{2}$Astrophysics Research Institute, Liverpool John Moores University, 146 Brownlow Hill, Liverpool L3 5RF, United Kingdom\\
$^{3}$ESO, European Southern Observatory, Karl-Schwarzschild Str. 2, 85748 Garching bei M\"unchen, Germany\\
$^{4}$CRAL, Observatoire de Lyon, CNRS, Universit\'{e} Lyon 1, 9 avenue Ch. Andr\'{e}, 69561 Saint Genis-Laval Cedex, France \\
$^{5}$Leibniz-Institute for Astrophysics (AIP), An der Sternwarte 16, 14482 Potsdam, Germany \\
$^{6}$Institute for Astronomy, Department of Physics, ETH Z\"urich, CH-8093 Z\"urich, Switzerland \\
$^{7}$Institut f\"ur Physik und Astronomie, Universit\"at Potsdam, 14476 Potsdam, Germany \\
}

\date{Accepted XXX. Received YYY; in original form ZZZ}

\pubyear{2017}

\begin{document}
\label{firstpage}
\pagerange{\pageref{firstpage}--\pageref{lastpage}}
\maketitle

\begin{abstract}
This is the first of a series of papers presenting the results from our survey of 25 Galactic globular clusters with the MUSE integral-field spectrograph. In combination with our dedicated algorithm for source deblending, MUSE provides unique multiplex capabilities in crowded stellar fields and allows us to acquire samples of up to 20\,000 stars within the half-light radius of each cluster. The present paper focuses on the analysis of the internal dynamics of 22 out of the 25 clusters, using about 500\,000 spectra of 200\,000 individual stars. Thanks to the large stellar samples per cluster, we are able to perform a detailed analysis of the central rotation and dispersion fields using both radial profiles and two-dimensional maps. The velocity dispersion profiles we derive show a good general agreement with existing radial velocity studies but typically reach closer to the cluster centres. By comparison with proper motion data we derive or update the dynamical distance estimates to 14 clusters. Compared to previous dynamical distance estimates for 47~Tuc, our value is in much better agreement with other methods. We further find significant ($>3\sigma$) rotation in the majority (13/22) of our clusters. Our analysis seems to confirm earlier findings of a link between rotation and the ellipticities of globular clusters. In addition, we find a correlation between the strengths of internal rotation and the relaxation times of the clusters, suggesting that the central rotation fields are relics of the cluster formation that are gradually dissipated via two-body relaxation.
\end{abstract}




\section{Introduction}

Over the last decades, dynamical studies have been a cornerstone to improve our understanding of Galactic globular clusters. Radial velocity measurements gave access to their velocity dispersions \citep{Gunn1979,Pryor1993}, revealed rotation \citep{Meylan1986}, and provided the first constraints on the binary fractions of the clusters \citep{Pryor1988}. Furthermore, the comparisons to theoretical models such as those of \citet{Michie1963} or \citet{King1966} allowed for the investigation of the internal structure of the clusters, for example by constraining mass segregation or inferring dynamical mass-to-light ratios \citep[e.g.][]{DaCosta1976,McLaughlin2005a}. However, the high stellar densities inside globular clusters pose a problem to both observers and simulators. To observers, because of the challenges involved when performing spectroscopy in such dense environments, and to simulators, because of the high number of gravitational encounters that have to be considered during the lifetime of a cluster \citep[e.g.][]{Aarseth1974}. Therefore, data-model comparisons were hampered by the simplifications of the models on the one hand and the sparsity of the data on the other hand for a long time. 

Recent years saw a constant change as models became more and more realistic and observations produced steadily increasing sample sizes of stars per cluster. On the theoretical side, it became possible to directly follow the evolution of clusters with realistic particle numbers in N-body models e.g.][]{Baumgardt2003,Wang2016}. Furthermore, Monte Carlo simulations allowed to investigate the evolution of globular clusters for large parameter spaces and proved to yield results consistent with the more accurate N-body models \citep[e.g.][]{Giersz2013,Rodriguez2016a}. Finally, the ongoing work on analytical models has led to a more realistic handling of aspects such as rotation, anisotropy, or mass segregation \citep[][]{Varri2012,Gieles2015,Sollima2017}.

On the observational side, the advent of integral-field spectrographs has enabled us to perform spectroscopy even in the crowded cluster centres \citep{Noyola2008,Luetzgendorf2011, Lanzoni2013,Kamann2014}, while multi-object spectrographs allow for an efficient coverage of the outskirts of the clusters \citep{Lane2011,Bellazzini2012,Lapenna2015a}. Thanks to high-precision astrometry, especially with the \textit{Hubble} space telescope (HST), it became possible to measure proper motions around the cluster centres \citep{Bellini2014,Watkins2015}, while the Gaia satellite promises to provide such information also for the outskirts of the clusters \citep{Pancino2017}.

Consequently, it is now possible to perform very detailed studies of the internal dynamics of Galactic globular clusters. One example in this respect is the quest for the origin of multiple populations that seem to be ubiquitous in globular clusters \citep[e.g.][]{Carretta2009,Milone2017}. As shown by \citet{Renzini2015} or \citet{Bastian2015b}, all the models used to explain the observed spreads in light element abundances have shortcomings. Studying the motions of the stars as a function of their chemical compositions can provide further insight. For example, \citet{Henault-Brunet2015} argued that differences in the rotational patterns of the populations can be used to favour one model over the other. Though challenging, first attempts in this direction have been performed. Anisotropy variations across the various populations were detected by \citet{Richer2013} and \citet{Bellini2015} in 47~Tuc and NGC~2808, respectively. Recently, an enhanced rotational component of the extreme population in M13 was reported by \citet{Cordero2017}.

In general, rotation is found to be quite common in globular clusters. Detailed studies revealing rotation were performed for a number of clusters, such as NGC~3201 \citep{Cote1995a}, $\omega$~Centauri \citep{Ven2006}, M15 \citep{Bosch2006}, NGC~4372 \citep{Kacharov2014}, or recently for M53 \citep{Boberg2017} and 47~Tuc \citep{Bellini2017e}. In the surveys performed by \citet{Lane2010} or \citet{Bellazzini2012}, rotation with amplitudes of a few ${\rm km\,s^{-1}}$ were uncovered in many clusters. The latter study also suggested a link between the rotation amplitude and the metallicities and horizontal branch morphologies of the clusters \citep[but see][]{Kimmig2015}. Remarkably, all clusters studied by \citet{Fabricius2014} displayed rotation, the strengths of which correlated with the global ellipticities of the clusters. Such a correlation is not necessarily expected, given that the ellipticities of globular clusters are low compared to those of galaxies and that other mechanisms such as Galactic tidal forces \citep[e.g.][]{Bergh2008} are likely to affect the appearance of globular clusters. Note that in contrast to previously mentioned studies, \citet{Fabricius2014} looked at the cluster centres, where relaxation times are shortest and rotation thus are most easily diminished by two-body interactions and the outwards transport of angular momentum. Accordingly, models of rotating globular clusters are characterised by an increase of rotation amplitude with radius, up to a maximum that is located beyond the half-light radius \citep[][]{Lagoute1996,Fiestas2006,Varri2012,Jeffreson2017}. 

Despite the evidence for ordered motions, it is general consensus that all globular clusters are mainly supported by the random motions of their constituent stars. Analyses of the velocity dispersion profiles of the clusters have gained a lot of attention over the last years with regard to the question whether globular clusters harbour massive black holes. The results are still controversial \citep[see the discussion in][]{Lanzoni2013}, however, the recent study of \citet{Baumgardt2017} based on a large grid of N-body models and an extensive compilation of literature data suggests that intermediate-mass black holes are the exception rather than the rule in Galactic globular clusters. We emphasize that so far no intermediate-mass black hole has been unambigiously detected in a globular cluster. In clusters such as $\omega$~Centauri that appear as probable hosts because of their central kinematics, other effects such as radial anisotropies \citep{Zocchi2017a} or mass segregation of stellar remnants \citep{Luetzgendorf2013,Peuten2016} may mimic the signature of a massive black hole.

In this paper, we present the first results from a survey of 25 Galactic globular clusters observed with the integral-field spectrograph MUSE \citep{Bacon2010}. The instrument allows us to completely cover the cluster centres, out to approximately the half-light radii. Owing to the large stellar densities, the central regions are the most challenging ones for spectroscopic observations. In \citet{Kamann2013}, we presented a method which allows us to deblend the spectra of overlapping stars and therefore to tap the full potential of integral-field spectrographs in crowded stellar fields. A pilot study of the globular cluster NGC~6397 was presented recently in \citep[][hereafter \citetalias{Husser2016}]{Husser2016} and \citep[][hereafter \citetalias{Kamann2016}]{Kamann2016}. It was based on 18\,000 stellar spectra of 12\,000 stars observed in less than one night of telescope time and illustrated the unique multiplexing capabilities of MUSE in crowded stellar fields and the prospects for studying the internal dynamics of globular clusters with this kind of data. The promising results from this study lead us to conduct the survey that will be presented in the current paper.

This paper is organised as follows. In Sect.~\ref{sec:survey} we outline the design and the scientific aims of our globular cluster survey. The data reduction and analysis are presented in Sects.~\ref{sec:reduction} an \ref{sec:analysis}. The internal dynamics of the sample clusters are presented in Sect~\ref{sec:kinematics} and set into the scientific context in Sect.~\ref{sec:discussion} before we conclude in Sect.~\ref{sec:conclusions}.

\section{The MUSE survey of Galactic globular clusters}
\label{sec:survey}

As part of the MUSE guaranteed time observations, we are carrying out a survey of $25$ globular clusters (PI: S. Dreizler). The clusters were selected to be nearby ($d<15\,{\rm kpc}$), massive (central velocity dispersions $\sigma_{\rm c}\ga5\,{\rm km\,s^{-1}}$) and well visible from Paranal. These criteria ensure short observing times and well resolvable kinematics at the expected velocity accuracy of MUSE of $\gtrsim1\,{\rm km\,s^{-1}}$ \citepalias[c.f.][]{Kamann2016}. For most clusters, we aim to completely cover the central regions, out to approximately their half-light radii. In a few clusters with large half-light radii our coverage will remain incomplete given the large number of pointings that would be required to achieve a complete coverage. An overview of the clusters that are included in our survey and the number of pointings per cluster that were already observed is given in Table~\ref{tab:overview}.

There is a variety of science cases that we plan to pursue with these data, ranging from the analysis of the cluster dynamics over binary studies and chemical analyses to the search for emission-line sources. To enable a comprehensive search for binary stars, each cluster is observed in multiple epochs. The numbers of epochs per cluster that are currently available are included in Table~\ref{tab:overview}. Note that the observations are still ongoing. For some clusters we plan to obtain 15 epochs which will enable us to characterise the orbits of detected binaries. For the remaining clusters, we plan to complete three epochs which will be sufficient for the detection of binary stars. Analyses of the binary populations of the clusters will be presented in follow-up studies (Giesers et al, subm.). The same is true for the analysis of stellar parameters or emission line stars. In addition, there is a huge legacy value of our survey as it represents the first blind spectroscopic survey of the central regions of globular clusters. Interesting objects such as optical counterparts to X-ray sources \citep[e.g.][]{Chomiuk2013}, stars in unexpected locations of the colour-magnitude diagrams \citep[like sub-subgiants, e.g.][]{Mathieu2003,Geller2017}, extremely low mass white dwarfs, or gas clouds in direction of the clusters \citep{Wendt2017c} will be the subject of dedicated publications.

\begin{table}
 \caption{Overview of the globular clusters observed in the MUSE survey.}
 \label{tab:overview}
 \begin{center}\begin{tabular}{ r | c | r | r | c | r | r }
\hline
  NGC & Name & $N_{\rm pointings}$ & $N_{\rm epochs}$ & ToT & $N_{\rm spectra}$ & $N_{\rm stars}$ \\
  &  &  &  & [h] &  &  \\
 (1) & (2) & (3) & (4) & (5) & (6) & (7) \\ \hline
 104 & 47\,Tuc & 10 &    7 &  6.4 & 84558 & 19181 \\
 362 &  & 4 &    2 &  0.7 & 9530 & 6112 \\
 1851 &  & 4 &    4 &  2.2 & 24600 & 9102 \\
 1904 & M\,79 & 4 &  4.5 &  1.9 & * & * \\
 2808 &  & 4 &    2 &  1.2 & 13762 & 6640 \\
 3201 &  & 4 &    8 &  5.8 & 27190 & 4186 \\
 5139 & $\omega$\,Cen & 10 &    4 &  4.9 & 63747 & 27390 \\
 5286 &  & 1 &    1 &  0.2 & * & * \\
 5904 & M\,5 & 4 &    1 &  0.4 & 11252 & 9639 \\
 6093 & M\,80 & 3 &    1 &  0.7 & 5901 & 3929 \\
 6121 & M\,4 & 2 &    1 &  0.1 & 1030 & 926 \\
 6254 & M\,10 & 4 &  2.5 &  1.7 & 15925 & 9071 \\
 6266 & M\,62 & 4 &    2 &  1.1 & 14555 & 11129 \\
 6293 &  & 1 &    2 &  0.1 & 829 & 598 \\
 6388 &  & 4 &    4 &  2.2 & 32942 & 12334 \\
 6441 &  & 4 &    4 &  2.8 & 29192 & 11654 \\
 6522 &  & 1 &    2 &  0.2 & 2889 & 2205 \\
 6541 &  & 4 &    1 &  0.6 & 8757 & 7190 \\
 6624 &  & 1 &    1 &  0.2 & * & * \\
 6656 & M\,22 & 4 &    1 &  1.2 & 14234 & 10272 \\
 6681 & M\,70 & 1 &    2 &  0.8 & 5984 & 4036 \\
 6752 &  & 4 &    2 &  0.8 & 14582 & 7362 \\
 7078 & M\,15 & 4 &    2 &  1.1 & 16068 & 10420 \\
 7089 & M\,2 & 4 &    4 &  2.4 & 25757 & 10881 \\
 7099 & M\,30 & 4 &  3.5 &  2.6 & 21566 & 7547 \\
  &  &  &  &  &  \\
 {\bf total:} &  & 94 &  68.5 &  42.2 & 444850 & 191804 \\
\hline\end{tabular}\end{center}
\medskip
Notes. (1) NGC number. (2) Alternative identifier (if any). (3) Number of pointings. This number roughly corresponds to the covered field of view in arcminutes. (4) Average number of epochs available for each pointing. (5) Total integration time in hours. (6) Number of extracted spectra. For clusters marked with an asterisk the analysis is still pending (see text). (7) Number of stars with at least one extracted spectrum.

\end{table}

\section{Observations and data reduction}
\label{sec:reduction}

\begin{figure*}
	\includegraphics[width=\textwidth]{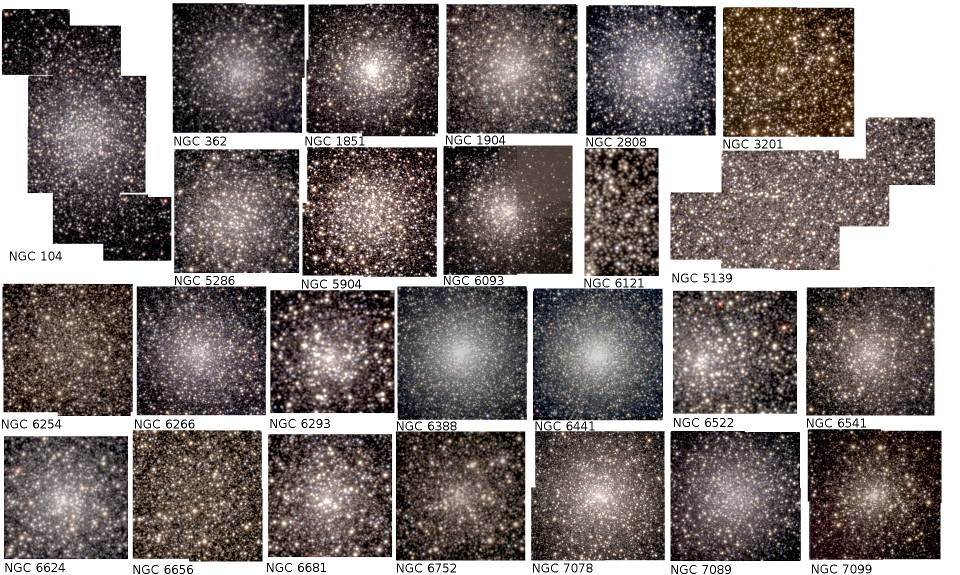}
    \caption{GRI mosaics of all observed clusters, created from the reduced MUSE data and ordered by increasing NGC number. Note that for NGC~104 (47~Tuc) and NGC~5139 ($\omega$~Centauri), two and three pointings, respectively, with larger distances to the cluster centres are not shown. In each mosaic, north is up and east is left.}
    \label{fig:whitelight}
\end{figure*}

For the current paper, we used all data that were observed before October 2016. The data were observed during 14 observing runs (in visitor mode), starting in September 2014. During each epoch, each targeted pointing was observed with three different instrument derotator angles (0, 90, and 180 degrees) in order to level out possible resolution differences between the individual spectrographs. All data were observed in the nominal mode without AO, providing a continuous wavelength coverage from $4750\,{\text \AA}$ to $9350\,{\text \AA}$. The FWHM of the MUSE line spread function is $\sim2.5\,{\text \AA}$ and is approximately constant across the wavelength range, corresponding to a spectral resolution of $R\sim1\,700$ at the blue and $R\sim3\,500$ at the red end. The seeing (measured on the reduced data cubes) was typically good during the observations, with a median value of $0.74\arcsec$ and 95\% of the values in the range ($0.48\arcsec$, $1.10\arcsec$).

The reduction of all cubes was carried out using the standard MUSE pipeline \citep{Weilbacher2012,Weilbacher2014}. For each exposure, the basic steps of the reduction cascade (bias subtraction, tracing of the slices, wavelength calibration, and basic science reduction) were performed in the {\it MuseWise} system developed within the MUSE GTO team \citep{Vriend2015}. The outcome of these steps --- 24 pixel tables, one per spectrograph --- was checked for consistency. We found that a non-negligible fraction of the data ($\sim20\%$) were compromised by inaccuracies in the wavelength calibration. These were detected by determining radial velocities for the telluric absorption bands across various regions of the field of view, which showed systematic shifts of several ${\rm km\,s^{-1}}$ between individual spectrographs. The origin of this issue could be traced back to the refinement of the wavelength solution derived from the calibration data using bright skylines in the science data. For short exposures $\la 60\,{\text s}$ and dense clusters, the skylines were sometimes not visible because of the intensity of the starlight, hence their centroids were measured with large inaccuracies. In version 1.6 of the pipeline, this problem was solved by removing the pixels with the strongest (stellar) signal via asymmetric kappa-sigma-clipping and measuring the centroids of the skylines in the remaining pixels only. We reprocessed all exposures that were affected by this problem with the latest version of the pipeline and adapted the parameters of the kappa-sigma-clipping such that the wavelength accuracy was sufficient across the whole field of view. Again, the radial velocities of the telluric absorption bands were used to verify that the offsets between the spectrographs had vanished.

In the next step of the data reduction process, the 24 pixel tables were flux calibrated and combined into a single pixel table. Finally, the three pixel tables created per epoch and pointing were combined and resampled into the final data cube on which we performed the further analyses. In total, we obtained 351 cubes. After a careful visual selection, 35 cubes were excluded from further analysis, either because of very bad seeing, clouds, or artefacts of currently unknown origin. We did not perform any sky subtraction during the data reduction process because our fields do not contain any patches of empty sky and the observation of dedicated sky exposures would have been too expensive observationally given the number of exposures. We deal with the night sky when extracting the spectra (cf. Sect.~\ref{sec:analysis}).

To illustrate the richness of the MUSE data, we show in Fig.~\ref{fig:whitelight} the collection of whitelight mosaics for all clusters of our sample, created by collapsing the cubes in spectral direction and combining the individual pointings using the \texttt{SWarp} software package \citep{Bertin2002}. Note that for NGC~104 and NGC~5139 we only show the central mosaics while omitting two and three individual pointings at larger distances to the cluster centre, respectively. In NGC~6121, which we only included as a backup target for bad seeing conditions, only two out of the four planned pointings have been observed so far. Apart from this, all planned pointings have been observed and reduced at least once.

\section{Data analysis}
\label{sec:analysis}

In large parts, the analysis of the data cubes was done in a similar way to our pilot study on NGC~6397. Details on the individual steps of the spectrum extraction and the spectral analysis can be found in \citetalias{Husser2016} and \citetalias{Kamann2016}. In the following, we restrict ourselves to a brief summary of both steps and put emphasis on new aspects that have been introduced since the publication of these papers.

\subsection{Spectrum extraction}

\begin{table*}
 \centering
 \caption{Summary of archival \textit{Hubble} data that was analysed in order to obtain source positions and magnitudes for pointings/clusters without coverage in the ACS survey of \citet{Sarajedini2007}.}
 \label{tab:hstdata}
 \begin{tabular}{rcccl}
  \hline
  NGC & Camera & Filters & Proposal ID(s) & Comment \\ \hline
  104 & WFC3 & F225W, F336W & 12971 & Only available for one of the two outer fields. \\
  1904 & WFPC2 & F336W, F439W, F555W & 6095, 6607 & Used catalogue from \citet{Piotto2002}. \\
  5139 & ACS & F435W, F625W & 9442 &  \\
  6266 & WFC3 & F336W & 11609 &  \\
       & WFPC2 & F336W, F439W, F555W, F814W & 8118, 8709 & Used catalogue from \citet{Piotto2002}. \\
  6293 & WFC3 & F390W, F555W, F814W & 12516 &  \\
  6522 & ACS & F435W, F625W & 9690 &  \\
 \end{tabular}
\end{table*}

The extraction of the individual stellar spectra from the data cubes was performed with our dedicated software presented in \citet{Kamann2013}. It determines the positions of the sources as well as the shape of the PSF as a function of wavelength and uses this information to optimally extract the spectrum of each resolved star. For this method to work, an input catalogue of sources is needed which provides astrometry and broad-band magnitudes across the MUSE field of view. Where possible, we used HST data from the ACS survey of Galactic globular clusters \citep{Sarajedini2007,Anderson2008a} as input. However, some of our clusters (NGC~1904, NGC~6266, NGC~6293, NGC~6522) were not included in the survey. In addition, our outer pointings in NGC~104 and NGC~5139 are located outside of the footprint of their ACS observations. In those cases, we obtained archival HST images and analysed them with the \texttt{Dolphot} software package \citep{Dolphin2000}. An overview of the additional HST data that was used can be found in Table~\ref{tab:hstdata}.

The spectra were extracted from the cubes in a multi-step process. First, the subset of sources from the input catalogue that is resolved at the lower spatial resolution of the MUSE data is identified. Using those, in a second step a common PSF and a coordinate transformation from the input catalogue are fitted to the MUSE data. The wavelength dependencies of those quantities are modelled as smooth functions of the wavelength afterwards. Finally, this information is used to extract all spectra. The number of spectra that could be extracted in this way varied with the seeing and the densities of the clusters and was typically between $1\,000$ and $5\,000$ stars per pointing.

\begin{figure}
 \includegraphics[width=\columnwidth]{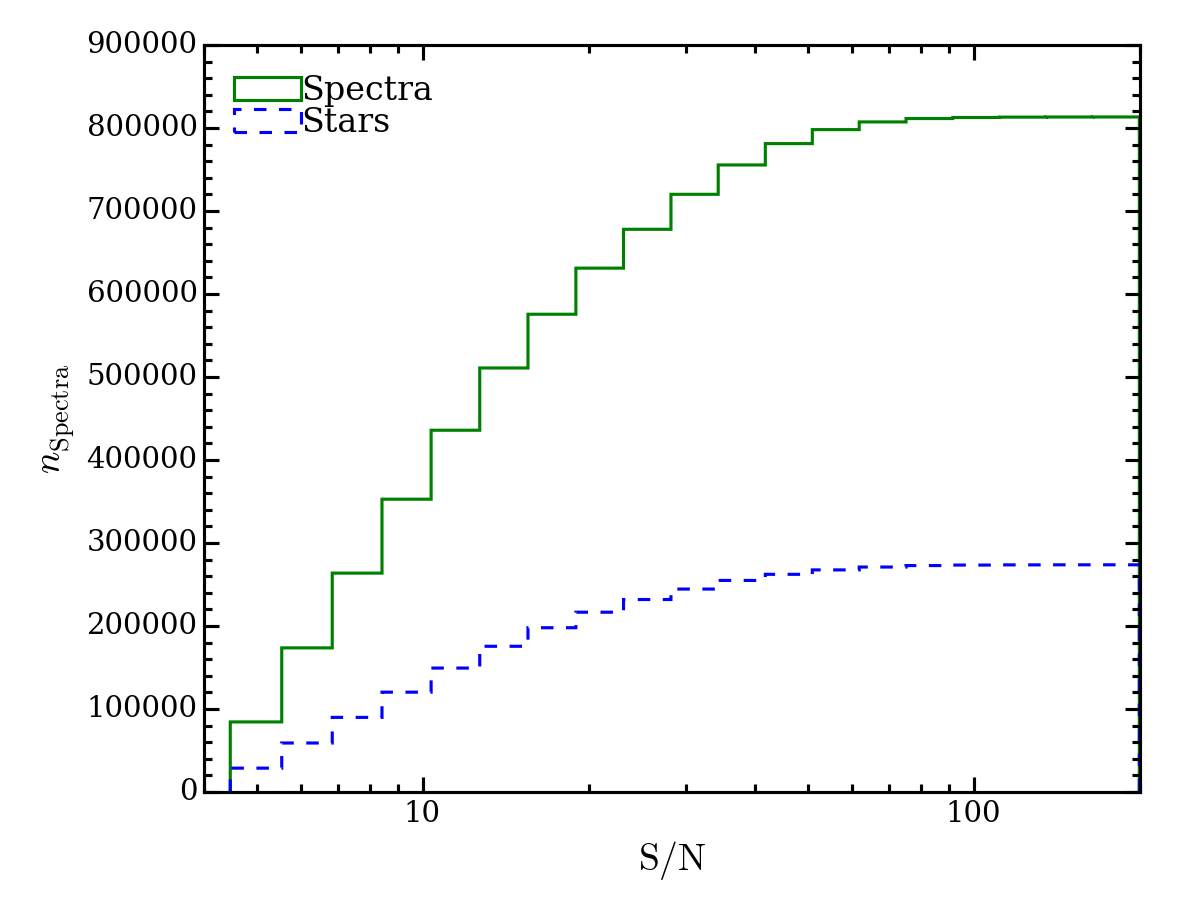}
 \caption{Cumulative S/N histogram of all spectra that were extracted from a data cube with ${\text S/N}>5$ ({\it solid green line}) and of the highest S/N spectra that were extracted for individual stars ({\it dashed blue line}).}
 \label{fig:snr_histogram}
\end{figure}

In contrast to most spectroscopic surveys, we do not perform any pre-selection of the observed stars, but instead aim to obtain a spectrum of every star in the field of view. Consequently, spectra are extracted over a wide range of signal-to-noise ratios (S/N)\footnote{As in \citetalias{Husser2016}, the S/N we provide is the average value for each extracted spectrum determined with the method of \citet{Stoehr2008a}.}, including many spectra for which the S/N is too low to perform any meaningful analysis. In a first cut, we exclude all spectra with ${\text S/N} < 5$ from any further analyses. This left us with about 813\,000 spectra of about 273\,000 stars. The cumulative histograms of the remaining spectra and stars are shown in Fig.~\ref{fig:snr_histogram}. Note that the shape of the histograms, a steep rise at low S/N which gets shallower when moving to higher S/N, is a direct consequence of the luminosity function of the cluster stars. The number of stars per magnitude bin increases when moving to fainter magnitudes.

\subsection{Spectral analysis}
\label{sec:spectral_analysis}

The first step in the analysis of the extracted spectra was to obtain initial values for the stellar parameters of the corresponding stars using the photometric data. For each cluster, an isochrone from the PARSEC database \citep{Bressan2012} was fitted to the colour-magnitude diagram created from the HST data. Via a nearest-neighbour approach in colour-magnitude space, each star was assigned an initial guess for the effective temperature $T_{\rm eff}$ and the surface gravity $\log g$. The initial guesses for the metallicity $[{\rm M/H}]$ were chosen according to published values of the metallicities of the host clusters. Based on these values, a matching template from the G\"ottingen spectral library \citep{Husser2013} was selected and an initial radial velocity was obtained via cross-correlation with the synthetic template. In the last step, a full spectrum fit against the spectral library was performed to obtain final values for the radial velocity, the scaled solar metallicity, and the effective temperature. The surface gravity was held constant at the initial value determined from the isochrone comparison because the spectroscopic determination of $\log g$ at the spectral resolution of MUSE is still an open issue. However, for the present study this does not cause any problems.

For the analysis envisaged in this paper, it is fundamental that the radial velocities are reliable and their uncertainties are known precisely. We performed several checks and corrections to ensure this. As outlined in \citetalias{Husser2016}, the full spectrum fit also includes a component that accounts for telluric absorption. In \citetalias{Kamann2016} we illustrated how this telluric component can be used to check the velocity accuracy across the field of view. For each cube, we selected all spectra with a ${\text S/N} > 30$ and obtained the radial velocities of the fitted telluric components. We then determined the mean telluric velocity and considered it as a measure for the absolute velocity offset of the cube. Hence, all stellar velocities determined from that cube where shifted by the mean telluric velocity. Similarly, we determined the standard deviation of the telluric velocities across the fields of view and considered it as a measure for the accuracy of the wavelength solution. This value was added quadratically to the uncertainties of the stellar radial velocities determined by the full spectrum fit. In the last step, we compared the results obtained during different epochs for the same star. Under the assumption that the uncertainties are correct, the normalised velocity offsets should follow a normal distribution with a standard deviation of one. As in \citetalias{Kamann2016}, we used this information to apply an additional correction as a function of the S/N of the spectra. Spectra of stars that showed obvious radial velocity variations were excluded from this comparison.

\begin{figure}
 \includegraphics[width=\columnwidth]{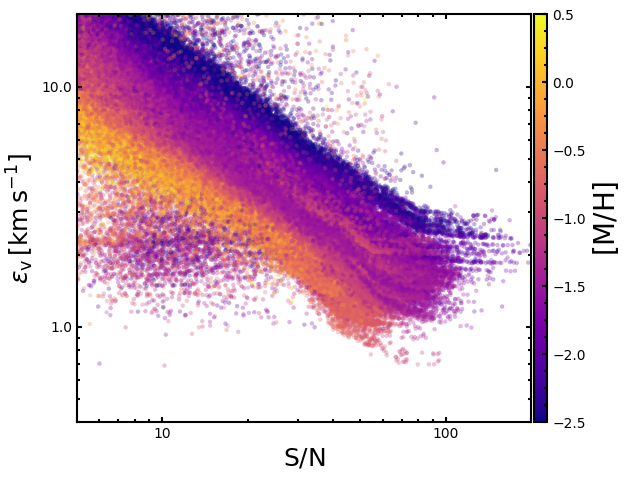}
 \caption{Calibrated uncertainties of the radial velocity measurements as a function of S/N of the spectra, colour-coded by and fitted metallicities.}
 \label{fig:velocity_accuracy}
\end{figure}

To give an idea of the accuracy of the radial velocities, we show the calibrated uncertainties $\epsilon_{\rm v}$ as a function of S/N for all our clusters in Fig.~\ref{fig:velocity_accuracy}. The data have been colour-coded by the metallicities derived from the spectra. As expected, the uncertainties strongly depend on the S/N of the spectra, with an average of $\epsilon_{\rm v}\approx10\,{\rm km\,s^{-1}}$ at ${\rm S/N}=10$ and values accurate to $1$--$3\,{\rm km\,s^{-1}}$ for spectra with ${\rm S/N} \ga 50$. The dependence of $\epsilon_{\rm v}$ on metallicity can be easily explained by the weaker lines in spectra of low-metallicity stars. Note that these uncertainties are measured on individual \emph{spectra} and that for most \emph{stars}, more than one spectrum is already available. The uncertainties of the averaged stellar velocities will decrease by $\sqrt{n}$ on average, with $n$ the number of averaged spectra.

\subsection{Cluster memberships}
\label{sec:membership}

Given the large area on the sky covered by our MUSE data, our stellar samples will be contaminated by stars that are not cluster members. As in \citetalias{Kamann2016}, we used a modified version of the expectation maximisation method presented by \citet{Walker2009} to separate cluster members and non-members in a homogeneous way. For the line of sight towards each cluster, we generated a realisation of the contaminant population using the Besan\c{c}on model of the Milky Way \citep{Robin2003}. The available photometry was used to constrain the list of simulated stars to the apparent magnitude range of our data. The simulation provided us with an expected distribution of the non-member stars in the radial velocity-metallicity plane towards each cluster. The expectation maximisation method was then used to compare the measurements of $[{\rm M/H}]$ and $v_{\rm r}$ of every star to the probability densities expected for the cluster and the non-member stars. Each star was assigned a probability of cluster membership such that the overall likelihood was maximised under the boundary condition that the fraction of cluster to non-member stars decreases monotonically with increasing distance to the cluster centre. For more details on the method, including the formulae that have been used, we refer to \citet{Walker2009} and \citetalias{Kamann2016}.

\begin{figure}
 \includegraphics[width=\columnwidth]{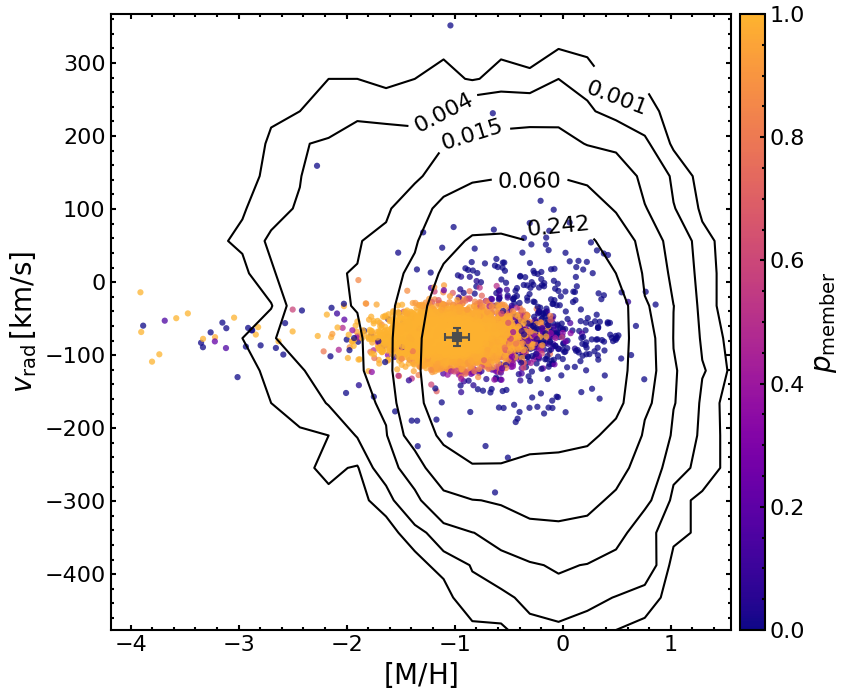}
 \caption{Membership probabilities of all stars observed towards NGC~6266 as a function of the measured metallicities and radial velocities. The contours show the density of expected Milky Way stars (in ${\rm stars}/{\rm arcmin}^2/{\rm km\,s^{-1}}/{\rm dex}$). The estimated mean velocity and metallicity of NGC~6266 and their estimated intrinsic spreads are indicated as a light-grey square.}
 \label{fig:ngc6266_membership}
\end{figure}

The efficiency of the membership determination is illustrated in Fig.~\ref{fig:ngc6266_membership} for NGC~6266, a cluster which is prone to contamination by Milky Way stars because of the high surface density of Milky Way stars in its line of sight. The Milky Way model returns a broad metallicity distribution (indicated by the black contours in Fig.~\ref{fig:ngc6266_membership}), with the metal-poorer part overlapping significantly with the cluster population in the $v_{\rm r}$-${\rm [M/H]}$ plane. Still the algorithm is able to identify the cluster population. Also visible in Fig.~\ref{fig:ngc6266_membership} is that the distribution of observed foreground stars nicely matches the distribution predicted by the Milky Way model, indicating the predictive power of the latter.

We considered all stars with $p_{\rm member}<0.5$ as non-members. In total, this affected $4.1\%$ of our sample stars. As our sample includes clusters with different densities and in the bulge as well as in the halo of the Galaxy, the non-member fraction varied strongly, with the highest fraction of $32.0\%$ obtained for NGC~6522 and the lowest fraction of $0.4\%$ obtained for NGC~104. For the latter, contamination from the Small Magellanic Cloud (SMC) is also not an issue. The few apparent SMC stars we find in the sample have negligible membership probabilities because of the difference in radial velocities compared to NGC~104 ($-18.7\,{\rm km\,s^{-1}}$, SMC: $145.6\,{\rm km\,s^{-1}}$).

\subsection{Selection of the final sample}

Before performing any kind of analysis with data like ours, one has to define which subsample of stars is included in the analysis. This is always a trade-off between trying to exclude as many of the useless spectra (e.g. low S/N, strong cosmic-ray hits, residuals from saturated neighbours) and avoiding to clip the sample by too much. In our case, we chose a S/N cut of 5 because it is the lower limit for trustworthy radial velocity measurements. However, even among the stars above this threshold there will be some that suffer from one of the above problems. For example, the velocity measured from a spectrum with ${\rm S/N}=20$ can be strongly shifted by an undetected cosmic-ray. As a visual inspection of all spectra is infeasible given our sample size, we introduced additional criteria to ensure that only reliable results enter the analysis of the cluster dynamics. These criteria are listed in the following.

\begin{itemize}
 \item We made use of the circumstance that we have two independent measurements for the radial velocity determined from each spectrum. First, the reliability of each cross-correlation measurement was judged using the $r$-statistics as defined by \citet{Tonry1979} and all spectra with $r<4$ were clipped. Then, we determined the difference in radial velocities measured from the cross-correlation and the full-spectrum fit and only considered spectra were this difference was $<3\sigma$, with $\sigma$ being the combined uncertainty of the two measurements.
 \item In some cases the full-spectrum fit returned unreasonably small uncertainties because of discontinuities in the $\chi^2$ space. These stars are visible in the lower right corner of Fig.~\ref{fig:velocity_accuracy}. Such cases were also excluded.
 \item Finally, we calculated broadband magnitudes from the MUSE spectra and in the same passband that was used in the extraction process and calculated the differences $\Delta m$ between input and recovered magnitudes. A large difference might indicate flux contamination from a nearby star or another problem. We defined a magnitude accuracy for each spectrum as $(1 + \Delta m/2\sigma_{\rm \Delta m})^{-2}$, where $\sigma_{\rm \Delta m}$ is the standard deviation of the magnitude differences of stars with comparable input magnitudes extracted from the same cube. Only spectra with an (empirically determined) magnitude accuracy $>0.8$ were accepted.
\end{itemize}

These selection criteria left us with $444\,850$ spectra of $191\,804$ stars. The number of spectra and stars for each individual cluster are provided in Table~\ref{tab:overview}. Note that for three clusters (NGC~1904, NGC~5286, and NGC~6624) no numbers are provided. For those clusters, the spectral analysis was still pending when writing this paper because of problems with the isochrone fit. Hence they have been omitted from the current analysis.

Finally, we removed stars with low probabilities of cluster membership ($p_{\rm member}<0.5$, cf. Sect.~\ref{sec:membership}) or which showed radial velocity variations. As mentioned earlier, binary studies are also foreseen with the MUSE data (Giesers et~al., in prep.). They are still work in progress, however, we already have a preliminary binary probability for each star that was observed at least twice. We excluded stars with binary probabilities $>0.5$. We also verified that the results of the following analysis are not sensitive on which stars are excluded as binaries, likely because of the low overall binary fraction in globular clusters.

\section{Cluster kinematics}
\label{sec:kinematics}

In order to study the dynamics of each cluster in a spatially resolved way, we applied two different binning schemes. First, we investigated the radial profiles by binning the stars according to their distances to the cluster centres. The radial bins were chosen such that each bin contained at least 100 stars and covered an annulus of $\log(r/{\rm arcsec})\geq0.2$. Second, we used the Voronoi tesselation code of \citet{Cappellari2003} to create two-dimensional maps of the kinematics around the cluster centres. As the Voronoi binning code requires a regular grid as input, we pre-binned the data into quadratic cells of $3\arcsec\times3\arcsec$ size. The size of the Voronoi bins was chosen such that on average 100 stars were included in a bin\footnote{Because of the low number of stars in NGC~6121 and NGC~6293, the number of stars per Voronoi bin were reduced to 50 and 30, respectively}. Note that no weighting with luminosity, mass, or S/N of the stars was applied when determining the Voronoi bins. Instead, each star was considered one measurement.

In each (radial or Voronoi) bin, we followed the maximum-likelihood approach introduced by \citet{Pryor1993} to obtain the dynamical properties. The method is based on the assumption that the probability of finding a star with a velocity of $v_i \pm \epsilon_i$ at projected distance $r_i$ to the cluster centre can be approximated as
\begin{equation}
p(v_i, \epsilon_i, r_i) = \frac{1}{2\pi\sqrt{\sigma_{\rm r}^2 + \epsilon_i^2}} \exp \left\{ \frac{(v_i - v_{\rm 0})^2}{-2(\sigma_{\rm r}^2 + \epsilon_i^2)} \right\}\,,
\label{eq:probability}
\end{equation}
where $v_{\rm 0}$ and $\sigma_{\rm r}(r_i)$ are the heliocentric radial velocity and the intrinsic dispersion profile of the cluster, respectively. A common approach for solving eq.~\ref{eq:probability} that we follow here is to minimise the negative log-likelihood,
\begin{equation}
 -\log\lambda = -\log\prod\limits_{i=1}^N p(v_i, \epsilon_i, r_i) = -\sum\limits_{i=1}^N\log p(v_i, \epsilon_i, r_i)\,.
\end{equation}

Equation~\ref{eq:probability} illustrates that it is crucial that the uncertainties, $\epsilon_i$, of the radial velocity measurements are correct. Systematically underestimating the uncertainties would lead to an overestimate of the intrinsic velocity dispersion and vice versa. As outlined in Sect.~\ref{sec:spectral_analysis}, we used repeated measurements of the same stars to calibrate our uncertainties, so we believe they are accurate. Nevertheless, we will discuss the influence of the uncertainties on the dispersion curves below in Sect.~\ref{sec:dispersions}.

One limitation of this approach is that it only strictly applies to stellar systems where the line of sight velocity distribution (LOSVD) can be described as a Gaussian of width $\sigma_{\rm r}(r_i)$, whereas rotation or higher moments of the LOSVD \citep[e.g. $h_3$ and $h_4$ in the parametrisation of][]{Marel1993} are neglected. As one of our aims in this work is to study cluster rotation, we extended the approach as outlined in the following. A common way to study the rotation of galaxies is \emph{kinemetry} \citep{Copin2001,Krajnovic2006}. It is based on the assumption that the variation of a kinematic quantity $K$ (in our case the mean velocity $v_{\rm 0}$) with position angle $\theta$ can be modelled as a harmonic expansion, i.e. as
\begin{equation}
 K(r, \theta) = c_0(r) + \sum\limits_{n=1}^{N} c_n(r) \cos\left[n(\theta - \phi_{N}(r))\right]\,,
\end{equation}
where the $c_i$ and $\phi_i$ define the amplitudes and orientations of the individual components. The simplest case of a pure $n=1$ cosine law corresponds to a rotating disk, whereas higher order moments quantify deviations from the disk model. As the rotation velocities expected in globular clusters are $1$--$2$ orders of magnitude below those in galaxies, we restrict ourselves to the simple disk model. While there is no physical reason why globular clusters should behave as rotating disks, our assumption is justified by the observation that in galaxies -- which typically have much more complex dynamics than globular clusters -- the $n=1$ term usually dominates \citep[see][]{Krajnovic2005}. This implies that we can study rotation in the radial bins by adding an angular dependence to the mean velocity of eq.~\ref{eq:probability}, i.e. by replacing
\begin{equation}
 v_{\rm 0} \longrightarrow \bar{v}(r_i, \theta_i) = v_{\rm 0} + v_{\rm rot}(r_i)\sin(\theta_i - \theta_{\rm 0}(r_i)).
\end{equation}
Here, $v_{\rm rot}(r)$ and $\theta_{\rm 0}(r)$ denote the projected rotation velocity and axis angle as a function of projected distance $r$ to the cluster centre. The axis angle as well as the position angle $\theta_i$ of a star are measured from north through east. Note that according to eq.~\ref{eq:probability}, $\theta_{\rm 0}(r)$ is oriented such that the maximum (minimum) velocity is measured at $\theta_{\rm 0} + \pi/2$ ($\theta_{\rm 0} - \pi/2$).

Having all formulae at hand, we used the MCMC algorithm of \citet{Foreman-Mackey2013} to minimise $\log \lambda$ in every bin. For the four parameters we aimed to constrain per \emph{radial} bin, we used constant probabilities across the following ranges as uninformative priors.
\begin{equation*}
 |v_{\rm 0}| < 1\,000\,{\rm km\,s^{-1}}\,, \\
 \sigma_{\rm r} > 0\,{\rm km\,s^{-1}}\,, \\
 v_{\rm rot} > 0\,{\rm km\,s^{-1}}\,, \\
 0  \leq\theta_{\rm 0} < 2\pi\,.
\end{equation*}
In cases, where \emph{Voronoi} binning was used, we enforced $v_{\rm rot}\equiv\theta_{\rm 0}\equiv0$ while keeping the same priors for $v_{\rm 0}$ and $\sigma_{\rm r}$ as in the radial binning scheme. Hence, we only obtained a mean velocity and a velocity dispersion in each Voronoi bin. However, rotation can still be studied by simply comparing the mean velocities between adjacent bins (see below). Compared to the radial bins, the Voronoi maps have the advantage that no assumptions about the rotation field are required. For both binning schemes, the uncertainties were estimated by running 100 chains with 500 steps each in which the parameters were varied slightly around their most likely values. The values we provide in the following are the median values and the 16th and 84th percentiles of the resulting parameter distributions.

\begin{figure*}
 \includegraphics[width=\textwidth]{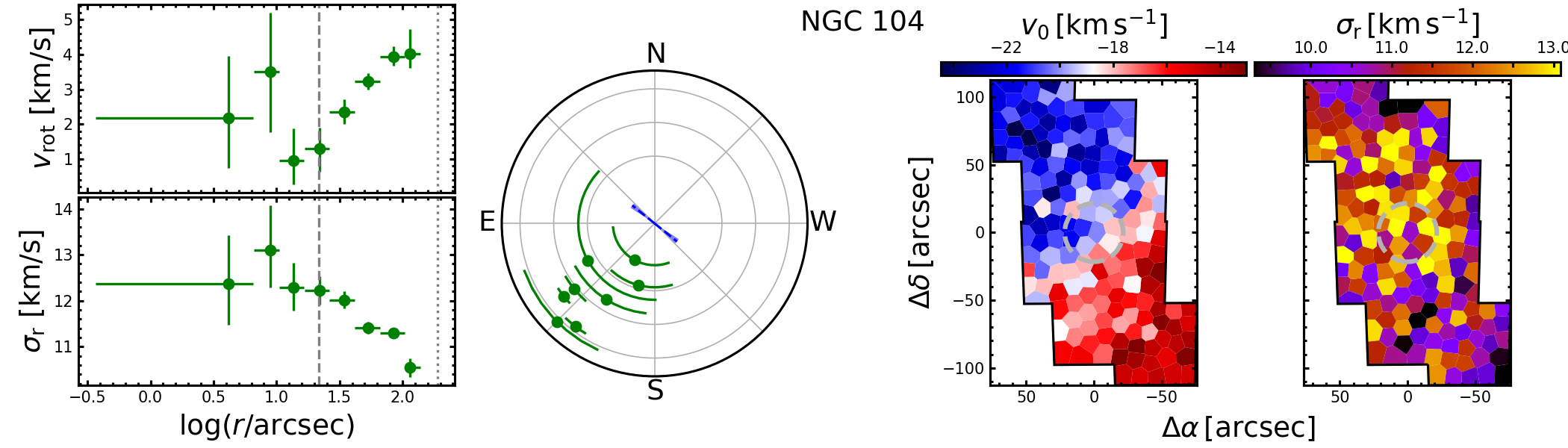}
 \includegraphics[width=\textwidth]{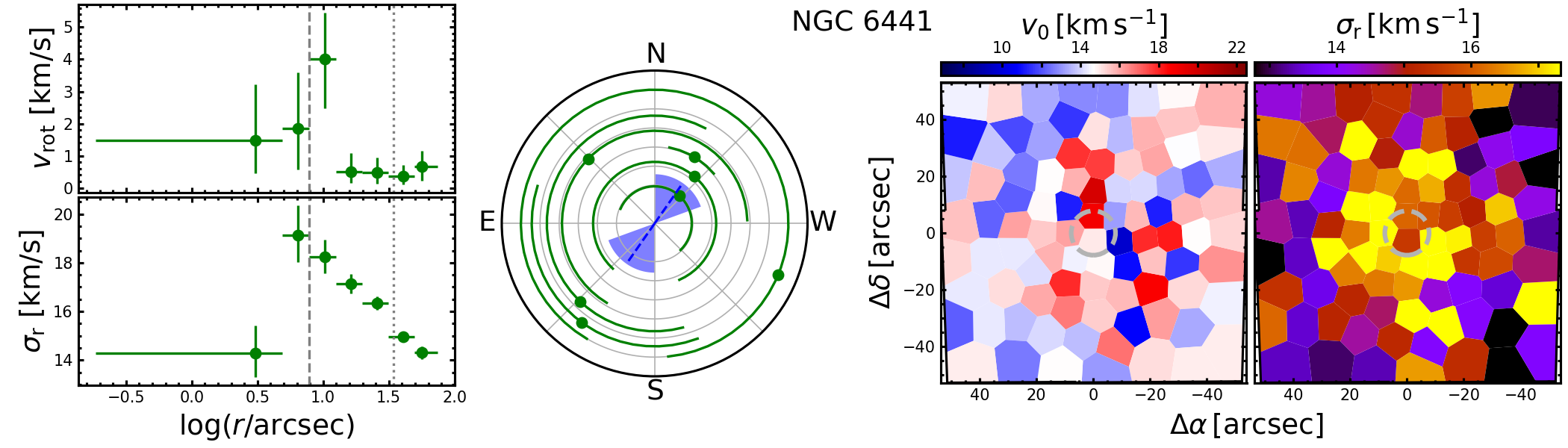}
 \includegraphics[width=\textwidth]{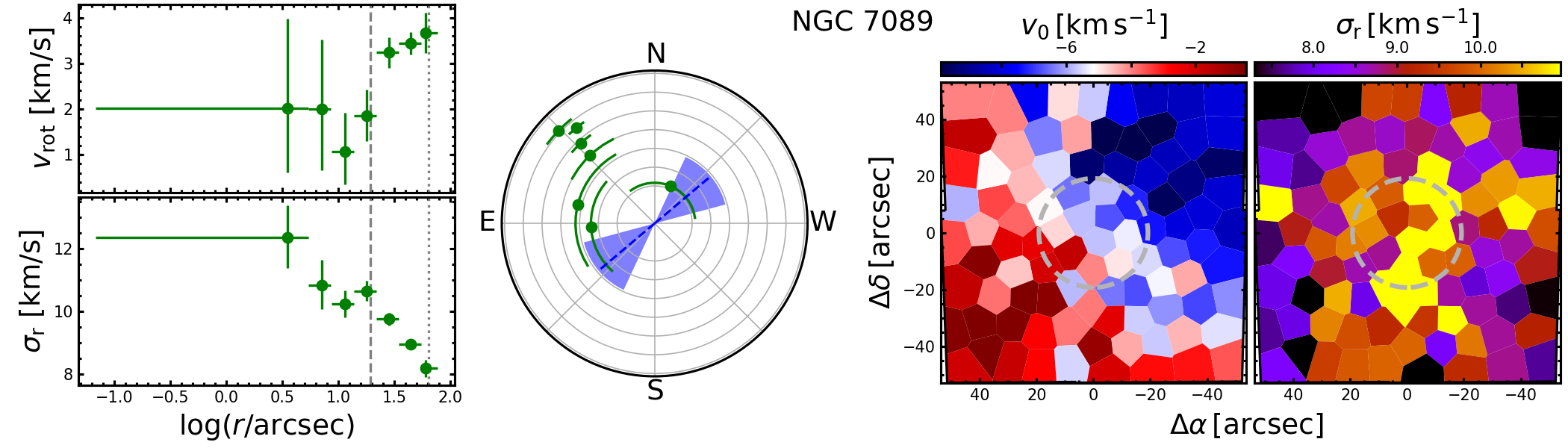}
 \caption{Results of the kinematic analysis for three clusters from our sample, NGC~104 ({\it top}), NGC~6441 ({\it middle}), and NGC~7089 ({\it bottom}). The left panels show the radial rotation and dispersion profiles, respectively. The dashed and dotted vertical lines indicate the core and half-light radii of each cluster, all values were taken from \citet{Harris1996}. The central panel shows the position angle of the rotation curve and its uncertainty for each radial bin. A blue dashed line is used to indicate the cluster's photometric semi-major axis angle as determined in Sect~\ref{sec:ellipticity}, with the blue-shaded area indicating the uncertainty and the length of the line scaling with cluster ellipticity. The right panels show Voronoi binned maps of the mean velocity and the velocity dispersion across the footprint covered by the MUSE data. The dashes circles indicate again the core radii of the clusters. Similar plots for the remaining clusters of our sample are presented in Appendix~\ref{sec:app:kinematics}.}
 \label{fig:example_kinematics}
\end{figure*}

In Fig.~\ref{fig:example_kinematics}, we showcase the results from this analysis for three clusters from our sample. The results for the remaining clusters are displayed in Fig~\ref{fig:app:kinematics} in Appendix~\ref{sec:app:kinematics}. In addition, all values from the analysis of the radial bins are provided in Table~\ref{tab:app:radial_profiles} in Appendix~\ref{sec:app:radial_profiles}.\footnote{The radial profiles are also made available as machine-readable data at \url{https://musegc.uni-goettingen.de/}.}

\subsection{Rotation profiles}
\label{sec:rotation}

\begin{table*}
 \caption{Kinematic properties of the sample clusters as derived from the MUSE data.}
 \label{tab:rotation}
 \begin{center}\begin{tabular}{ l | c | c | c | c | c | c | c | c | c}
\hline
  NGC & $\langle\theta_{\rm 0}\rangle$ & $P.A._{\rm kin}-90$ & $\lVert \bigtriangledown v \rVert$ & $\sigma_{\rm r,\,0}$ & $m_{\rm eff.}$ & $(v/\sigma)_{\rm HL}$ & $\lambda_{\rm R,\,HL}$ & $r_{\rm max}/r_{\rm HL}$ & $d_{\rm Dyn.}$\\
  & (${\rm deg.}$) & (${\rm deg.}$) & (${\rm km\,s^{-1}\,arcmin^{-1}}$) & (${\rm km\,s^{-1}}$) & ($M_{\rm \odot}$) &  &  &  & (${\rm kpc}$) \\
 (1) & (2) & (3) & (4) & (5) & (6) & (7) & (8) & (9) & (10) \\ \hline
 104 & $134.1\pm3.6$ & $126.6\pm7.1$ & $2.9\pm0.1$ & $12.4\pm1.0$ & $0.78$ &$0.17\pm0.01$ &$0.19\pm0.01$ &0.58 & $4.4\pm0.1$\\ 
 362 & $-42.0\pm117.6$ & &  $0.8\pm0.3$ & $7.7\pm0.6$ & $0.79$ &$<0.04$ &$<0.02$ &1.00 & $9.2\pm0.8$\\ 
 1851 & $3.2\pm3.5$ & &  $2.0\pm0.2$ & $9.1\pm0.8$ & $0.81$ &$0.08\pm0.01$ &$0.08\pm0.01$ &1.00 & $10.8\pm0.4$\\ 
 2808 & $-47.0\pm2.4$ & $-50.3\pm8.1$ & $4.8\pm0.3$ & $13.9\pm1.1$ & $0.82$ &$0.13\pm0.01$ &$0.14\pm0.01$ &1.00 & $9.6\pm0.1$\\ 
 3201 & $-115.7\pm188.9$ & &  $0.4\pm0.1$ & $5.4\pm0.5$ & $0.70$ &$<0.01$ &$<0.01$ &0.49 &\\ 
 5139 & $9.9\pm4.3$ & $13.7\pm3.6$ & $1.8\pm0.1$ & $22.8\pm1.9$ & $0.71$ &$0.17\pm0.01$ &$0.19\pm0.01$ &0.54 & $5.2\pm0.4$\\ 
 5904 & $-54.3\pm20.3$ & &  $2.2\pm0.2$ & $7.3\pm0.8$ & $0.79$ &$0.14\pm0.01$ &$0.14\pm0.01$ &0.84 & $6.9\pm0.5$\\ 
 6093 & $-131.9\pm18.0$ & &  $3.0\pm0.4$ & $10.5\pm0.8$ & $0.78$ &$0.06\pm0.02$ &$0.08\pm0.01$ &1.00 &\\ 
 6121 & $-145.5\pm9.4$ & &  $1.2\pm0.5$ & $6.8\pm0.7$ & $0.73$ & &  &  & \\ 
 6254 & $142.8\pm17.9$ & &  $0.4\pm0.1$ & $5.6\pm0.6$ & $0.74$ &$<0.01$ &$<0.01$ &0.84 &\\ 
 6266 & $92.9\pm21.8$ & &  $1.4\pm0.3$ & $15.0\pm1.1$ & $0.80$ &$0.03\pm0.02$ &$0.04\pm0.00$ &1.00 & $6.2\pm0.5$\\ 
 6293 & $-59.9\pm15.0$ & &  $5.9\pm1.4$ & $7.1\pm0.7$ & $0.77$ &$0.19\pm0.03$ &$0.21\pm0.02$ &0.82 &\\ 
 6388 & $-41.8\pm16.8$ & &  $1.5\pm0.3$ & $19.1\pm1.4$ & $0.89$ &$<0.03$ &$0.02\pm0.01$ &1.00 & $11.0\pm0.8$\\ 
 6441 & $-29.6\pm20.8$ & &  $0.7\pm0.3$ & $14.3\pm1.1$ & $0.92$ &$<0.01$ &$<0.01$ &1.00 & $12.5\pm1.2$\\ 
 6522 & $-129.5\pm163.6$ & &  $2.6\pm1.0$ & $9.3\pm0.9$ & $0.79$ &$<0.01$ &$<0.01$ &0.91 &\\ 
 6541 & $-98.1\pm5.1$ & $-93.1\pm12.7$ & $3.1\pm0.3$ & $11.5\pm1.1$ & $0.75$ &$0.14\pm0.01$ &$0.17\pm0.01$ &1.00 &\\ 
 6656 & $-79.1\pm23.9$ & &  $1.1\pm0.3$ & $9.2\pm0.9$ & $0.69$ &$0.03\pm0.01$ &$0.05\pm0.01$ &0.47 & $3.5\pm0.1$\\ 
 6681 & $98.7\pm43.7$ & &  $1.0\pm0.4$ & $8.4\pm0.8$ & $0.74$ &$<0.03$ &$<0.01$ &1.00 & $9.2\pm1.2$\\ 
 6752 & $139.1\pm41.8$ & &  $0.5\pm0.2$ & $8.4\pm0.7$ & $0.73$ &$<0.01$ &$<0.02$ &0.74 & $4.2\pm0.6$\\ 
 7078 & $150.9\pm10.4$ & &  $2.4\pm0.3$ & $14.0\pm1.1$ & $0.75$ &$0.09\pm0.01$ &$0.10\pm0.01$ &1.00 & $10.3\pm1.7$\\ 
 7089 & $41.7\pm2.7$ & $38.1\pm7.1$ & $4.6\pm0.2$ & $12.3\pm1.0$ & $0.79$ &$0.21\pm0.01$ &$0.22\pm0.01$ &1.00 &\\ 
 7099 & $-115.4\pm27.1$ & &  $0.5\pm0.2$ & $8.1\pm0.8$ & $0.72$ &$<0.01$ &$<0.02$ &1.00 & $6.4\pm0.5$\\ 
\hline
\end{tabular}
\end{center}
\medskip
\begin{flushleft}
Notes. (1) NGC number (2) Mean position angle of the rotation axis, measured from north to east. (3) Kinematic position angle (i.e. the direction of the positive rotation amplitude), determined from the Voronoi maps as described in the text. Note that for an easier comparison with $\langle \theta_{\rm 0} \rangle$, the kinematic position angles were offset by $-90\,{\rm deg}$. (4) Gradient of the rotation amplitude with increasing distance to the cluster centre. (5) Central velocity dispersion. (6) Effective stellar mass of the dynamic measurements. (7) Value of $(v/\sigma)$ (cf. eq.~\ref{eq:vsigma}) measured at the halflight radius (or maximum radius covered by MUSE data in cases where the halflight radius was not covered). (8) Value of $\lambda_{\rm R}$ (cf. eq.~\ref{eq:lambdar}) measured at the half-light radius or maximum radius covered by MUSE data. (9) Maximum radius in covered by the MUSE data relative to the halflight radius of each cluster. (10) Dynamical distance, determined by comparison with the data of \citet{Watkins2015}.\end{flushleft}

\end{table*}

\begin{figure*}
 \includegraphics[width=\textwidth]{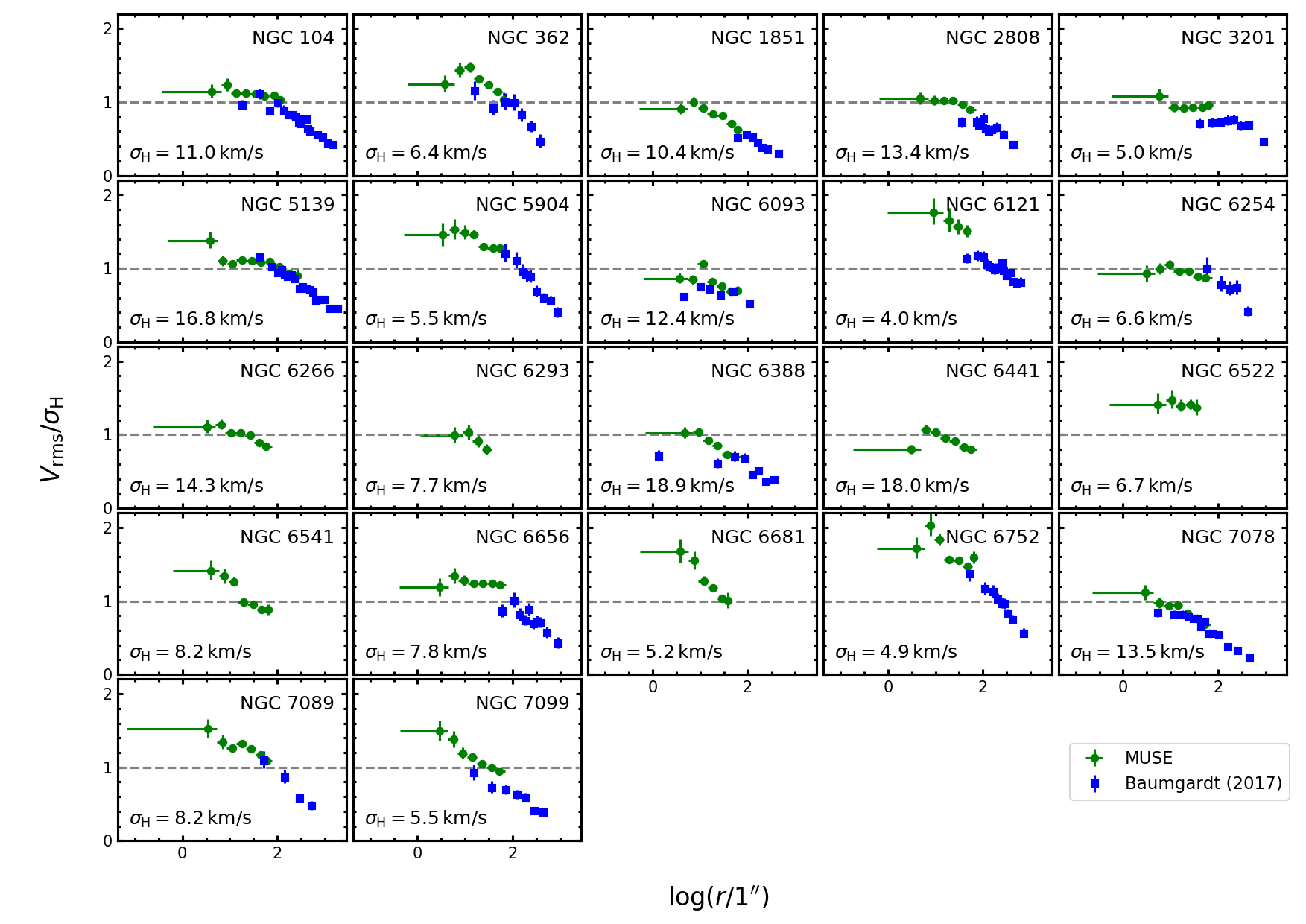}
 \caption{Comparison of the profiles of the second velocity moment determined from the MUSE data {\it (green squares)} and the literature compilation of \citet{Baumgardt2017} {\it (blue circles)}. For clarity, each profile has been normalised to the central velocity dispersion provided by \citet{Harris1996}, the value of which is provided in the lower left of each panel.}
 \label{fig:dispersion_comparison_rv}
\end{figure*}

Thanks to the complete coverage of the central regions, the MUSE data are very well suited to investigate rotation in the clusters. A visual inspection of the radial profiles of $v_{\rm rot}$ or the Voronoi-binned maps of $v_{\rm 0}$ in Figs.~\ref{fig:example_kinematics} and \ref{fig:app:kinematics} reveals that many of the clusters in our sample rotate. If we perform a visual classification of our sample into non-rotating or rotating clusters, we find that about 60\% (13/22) of the clusters show obvious rotation, while the remaining ones appear consistent with no rotation. Further inspection of the radial profiles of the rotating clusters reveals a pronounced similarity in that the rotation signal increases with distance to the cluster centre. It tends to disappear inside the core radius and steadily increases between the core and the half-light radius. This behaviour is in general agreement with the evolutionary globular cluster models of \citet{Fiestas2006} or the equilibrium models of \citet{Lagoute1996} or \citet{Varri2012}. It was also found in detailed studies of individual clusters like NGC~104 or NGC~5139 \citep[e.g.][see discussion below]{Meylan1986,Merritt1997,Ven2006,Sollima2009}. Beyond the half-light radii, our data lack the radial coverage to investigate any further trends. In this respect, it is interesting to note that some of the clusters that we visually classified as non-rotating (NGC~3201, NGC~6121, and NGC~6254) have large core radii so that our data coverage is basically restricted to the areas inside the core radii. Hence we cannot exclude the possibility that the clusters rotate at larger radii. This, and the fact that projection effects may also limit the amount of visible rotation in some clusters (the inclination of most clusters is not known, but see \citet{Ven2006} for NGC~5139 or \citet{Bellini2017e} for NGC~104), leads us to conclude that the fraction of rotating clusters in our sample is probably significantly higher than 60\%.

From the radial profiles depicted in Figs.~\ref{fig:example_kinematics} and \ref{fig:app:kinematics}, we inferred (variance-weighted) mean position angles of the rotation axes which are provided in Table~\ref{tab:rotation}. We also tried to infer the orientations of the rotation fields using the Voronoi maps which we created for the clusters. To this aim, we applied the method presented in appendix C of \citet{Krajnovic2006}.It works by creating a bi(anti)symmtric representation of a Voronoi map for various position angles and determining the position angle for which the difference to the actual Voronoi map is minimized (in a $\chi^2$ sense). However, in cases where the rotation velocity becomes comparable to the random velocity offsets per Voronoi bin ($1$--$2\,{\rm km\,s^{-1}}$ in our case), the latter start to dominate the $\chi^2$ statistics and the inferred position angle is poorly constrained (indicated by a $3\sigma$ uncertainty interval that covers almost the full circle). We discarded such measurements and provide in Table~\ref{tab:rotation} the kinematic position angles only for the strongly rotating clusters that gave useful constraints. Note that the kinematic position angle points in the direction of the highest velocity (measured from north to east), hence by definition it is offset by $+90^\circ$ from the axis angle $\theta_{\rm 0}$. The comparison in Table~\ref{tab:rotation} shows that in the cases where the 2D analysis was possible, the results for the two angles are in good agreement.

Central rotation in globular clusters was also studied by \citet{Fabricius2014} who detected a signal in all of the 10 (northern) targets of their sample. While also using integral-field spectroscopy, their approach to measuring rotation was completely different as they fitted a linear two-dimensional polynomial to the mean velocities measured in integrated light and obtained a velocity gradient and a rotation axis angle from the coefficients of the fit. In order to compare our results with \citet{Fabricius2014}, we used again the radial profiles. By fitting a line to the rotational velocities as a function of projected distance $r$ to the cluster centre and obtaining its slope we obtained a velocity gradient $\lVert \bigtriangledown v \rVert$, which is provided in Table~\ref{tab:rotation} for all clusters of our sample.

There are three clusters in common between our sample and the study of \citet{Fabricius2014}: NGC~5904 ($\lVert \bigtriangledown v \rVert=2.1\pm0.1\pm0.1\,{\rm km\,s^{-1}}$, ${\rm P.A._{kin.}}=58.5\pm2.8\pm5.6\,{\rm deg}$), NGC~6093 ($2.3\pm0.1\pm0.1\,{\rm km\,s^{-1}}$, $139\pm3.7\pm3.5\,{\rm deg}$), and NGC~6254 ($1.0\pm0.1\pm0.1\,{\rm km\,s^{-1}}$, $63.5\pm9.0\pm14.7\,{\rm deg}$), where the brackets indicate their values for the velocity gradient and the kinematic position angle. When comparing position angles, note that \citet{Fabricius2014} again derived kinematic position angles and did not discriminate the senses of rotation, hence their angles are offset by either $+90^\circ$ or $-90^\circ$ from our values. After accounting for these offsets, we find that despite the different approaches used, the results for the common cluster are in good agreement, only the velocity gradients we derive for NGC~6093 and NGC~6254 are slightly higher and lower, respectively. A visual check of their velocity fields (their Fig.~1) confirms that also the senses of rotation agree.

Larger samples of clusters have also been studied for rotation by \citet{Lane2010}, \citet{Bellazzini2012}, \citet{Lardo2015}, and \citet{Kimmig2015}. The first three studies follow on from one another and obtain similar conclusions, hence we restrict ourselves to a comparison with the studies of \citet{Bellazzini2012} and \citet{Kimmig2015}. However, any comparison is complicated because in contrast to out work, those studies focused on deriving global values for the clusters instead of rotation profiles. In addition, as the studies are based on multi-object spectroscopy, the majority of data was taken beyond the half-light radii of the clusters. We do find a general agreement in the sense that the clusters that show stronger rotation in the works of \citet{Bellazzini2012} and \citet{Kimmig2015} do so as well in our data. There is one notable exception though, which is NGC~6441. \citet{Bellazzini2012} and \citet{Kimmig2015} report rather high rotation amplitudes for this cluster (albeit with a large formal uncertainty in the latter study) whereas it is among the least rotating clusters in our sample (cf. Fig.\ref{fig:example_kinematics}). The strong rotational signal in the study of \citet{Bellazzini2012} is mainly caused by four stars at distances $>5\arcmin$ to the cluster centre. As our data does not cover such large radii, it remains open whether this discrepancy is intrinsic to the dynamics of the cluster.

The clusters from our sample that have the most extensive literature data are NGC~104, NGC~5139, and NGC~7078. \citet{Bianchini2013} used these data to derive rotation profiles for all three clusters and found the rotation profiles to peak at distances $\sim1.5\times$ the half-light radii. They obtained global position angles of $136^{\circ}$ (NGC~104), $12^{\circ}$ (NGC~5139), and $106^{\circ}$ (NGC~7078). While the former two are in excellent agreement with our values, a significant offset is observed for the latter one. However, \citet{Bianchini2013} also noted that for NGC~7078 the position angle changed with distance to the cluster centre, and decreased from an initial value of $\sim260^{\circ}$ inside the core radius to $\sim100^{\circ}$ when the entire radial velocity sample was used. This behaviour is in qualitative agreement with the results derived from the MUSE data (cf. Fig.~\ref{fig:app:kinematics}). In addition, our data confirm the increase of the rotation velocity inside the core radius towards the cluster centre of NGC~7078 that was observed by \citet{Bianchini2013}. These two features led \citet{Bosch2006} to conclude that NGC~7078 contains a decoupled core which might be a consequence of the late evolutionary stage of this core-collapse cluster. We note that \citet{Bianchini2013} also report centrally increasing rotation for NGC~104 and NGC~5139 which is not confirmed by the MUSE data (cf. Figs.~\ref{fig:example_kinematics} and \ref{fig:app:kinematics}). However, no uncertainties are given for the central rotation values of \citet{Bianchini2013}, hence the significance of this discrepancy remains unknown.

\subsection{Dispersion \& $V_{\rm rms}$ profiles}
\label{sec:dispersions}

In the globular cluster literature, the velocity dispersion $\sigma_{\rm r}$ is often equated with the central second velocity moment $V_{\rm rms}=\sqrt{\sigma_{\rm r}^2 + v_{\rm rot}^2}$, implying that no rotation exits, i.e. $v_{\rm rot}\equiv0$. While the analysis of the previous sections showed that the velocity dispersion is the dominant contribution to the second velocity moment in globular clusters, rotation should not be neglected. Hence we will in the following distinguish between the velocity dispersion profiles provided in Figs.~\ref{fig:example_kinematics} and \ref{fig:app:kinematics} and the profiles of the central second velocity moment $V_{\rm rms}$, calculated according to the formula provided above.\footnote{Note that our calculation of both quantities is based on the assumption that the velocity dispersion is Gaussian.} The central velocity dispersions, $\sigma_{\rm r,\,0}$, for the sample clusters are provided in Table~\ref{tab:rotation}. They were determined as the weighted average of all dispersion measurements inside the core radius (or the value of the innermost bin in cases where no bin fell into the core radius).

The $V_{\rm rms}$ profiles derived from the MUSE data are shown in Fig.~\ref{fig:dispersion_comparison_rv} and show a range of radial dependencies. In particular, we note that some clusters (NGC~362, NGC~6441, and NGC~6752) show a central dip in their profiles. While such a feature could be artificially caused by crowding effects -- contamination by background light tends to bias the measured radial velocities towards the cluster mean -- we do not think that this is the case. None of the clusters for which this effect is observed has a particularly steep surface brightness profile (see compilations by \citealt{Trager1995} or \citealt{Noyola2006}) so the crowding towards the centre only increases moderately. On the other hand, clusters with very steep surface brightness profiles, such as NGC~7078 or NGC~7099 show the opposite behaviour with a central cusp. Such a cusp is also observed in NGC~3201 and NGC~5139 which instead have large core radii. The latter case deserves particular attention because of the ongoing controversy about an intermediate-mass black hole in its centre. Our central value of $V_{\rm rms}=23.1^{+2.1}_{-1.7}\,{\rm km\,s^{-1}}$ lies above the models by \citet{Zocchi2017a} without a black hole and is closer to the model of \citet{Baumgardt2017} containing a black hole. However, dedicated modelling will be needed before drawing further conclusions.

Also included in Fig.~\ref{fig:dispersion_comparison_rv} are the dispersion profiles from the compilation of literature data provided by \citet{Baumgardt2017}. A comparison to the MUSE data shows that for the vast majority of the clusters in common between the two samples, the profiles are in good agreement in the region where they overlap. Only in NGC~6121, and to a lesser extent also in NGC~3201, our second velocity moments are significantly larger than the literature values. We can think of two issues that could have artificially increased our profiles, undetected binaries or an underestimation of our velocity uncertainties. The impact of the latter increases as the velocity dispersion decreases, hence it is suspicious that the offsets are observed for the two clusters with the lowest central velocity dispersion measurements in the literature. For NGC~6121, our analysis is based on two pointings without repeated observations, therefore our uncertainties calibration is less certain than for the other clusters. However, NGC~3201 is among the clusters with the richest data sets (cf. Table~\ref{tab:overview}), with on average seven repeated observations per star, making it unlikely that the uncertainties were systematically underestimated. In addition, our ongoing analysis of binary properties (Giesers et al., in prep.) suggests that in NGC~3201 all binaries with orbital velocities comparable to the cluster dispersion or higher are detected and removed from any further analysis. This is not the case for NGC~6121. Compared to more dense clusters, NGC~6121 has a relatively high binary fraction of 5--10\% in the core \citep{Sommariva2008c,Milone2012}.

To check whether binaries might be responsible for the offset between our profile and the literature data, we used the code provided by \citet[][described in detail in \citealt{Cottaar2012d}]{Cottaar2014c} to create stellar populations with the tabulated velocity dispersion of NGC~6121 and varying binary fractions. The distributions of mass ratios, log-periods, and eccentricities were modeled as power-laws, using the indices found by \citet{Kiminki2012}. We note that these distributions were obtained in a study of massive stars in OB associations. However, the corresponding values for globular clusters are poorly constrained observationally and therefore are a major source of uncertainty in the inferred binary fractions. We cut the period distribution at $100\,{\rm years}$ to take into account the low survival rates of wide binaries in the dense cluster environments. For simplicity, we further assumed a common mass of $0.8\,{\rm M_\odot}$ for all observed stars. We then drew random samples from the distribution, assigned each star a velocity uncertainty that was randomly drawn from our observed sample, and tried to recover the velocity dispersion using our maximum-likelihood approach. We found that on average, we overestimated the intrinsic velocity dispersion by $0.3\,{\rm km\,s^{-1}}$ and $0.6\,{\rm km\,s^{-1}}$ for binary fractions of 5\% and 10\%, respectively. Hence, while binaries may be responsible for part of the discrepancy, it is unlikely that they are the sole explanation.

\begin{figure}
 \includegraphics[width=\columnwidth]{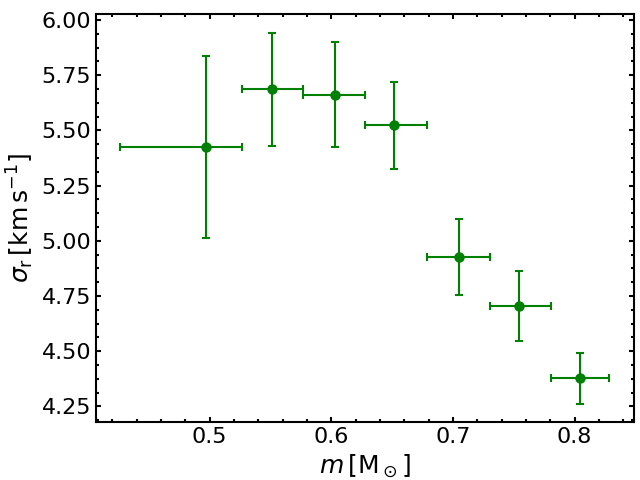}
 \caption{Velocity dispersion of the globular cluster NGC~3201, measured across the full MUSE mosaic and for different bins in stellar mass of the probed stars.}
 \label{fig:mass_segregation_ngc3201}
\end{figure}

\begin{figure*}
 \includegraphics[width=\textwidth]{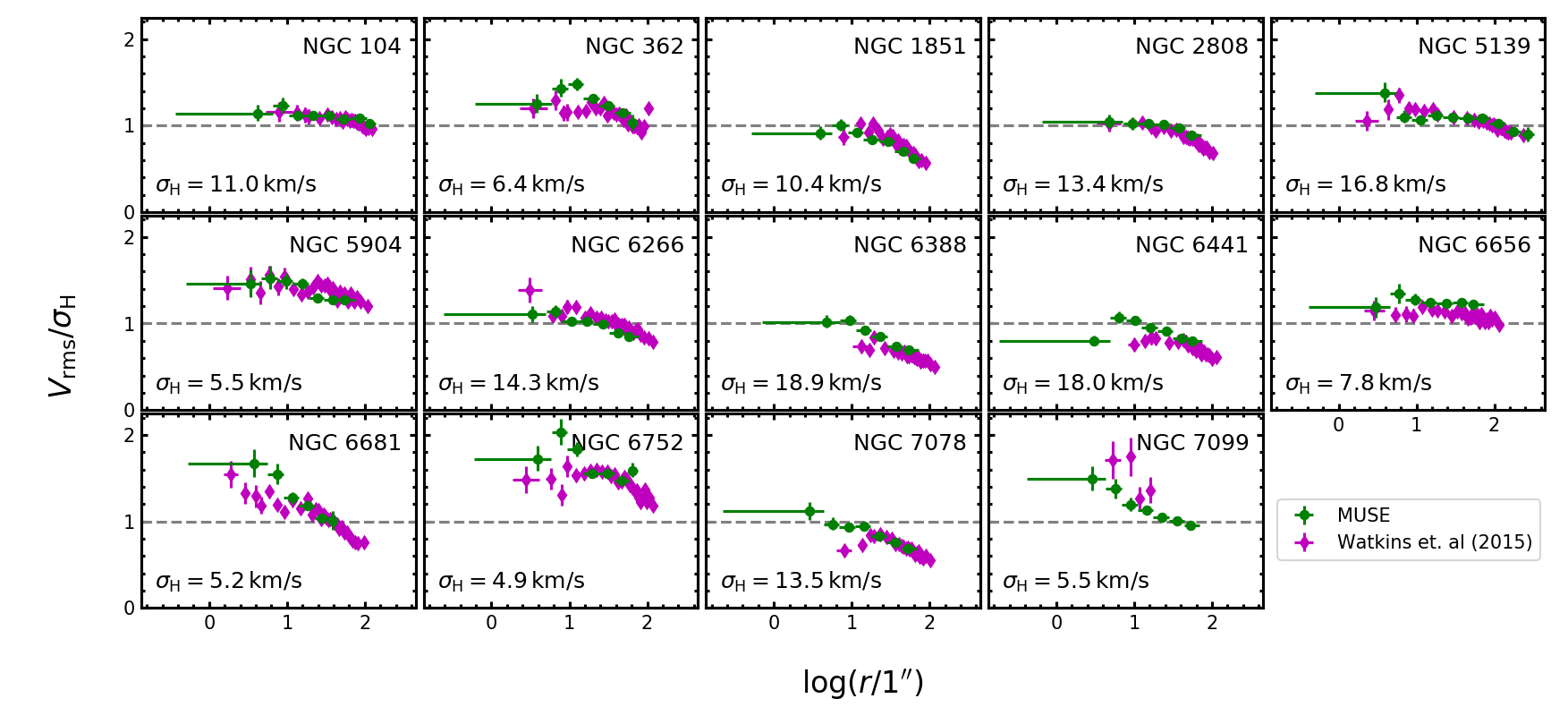}
 \caption{Comparison of the profiles of the second velocity moment determined from the MUSE data ({\it green circles}) and the proper motion data of \citet[][{\it purple diamonds}]{Watkins2015} for clusters present in both samples. As in Fig.~\ref{fig:dispersion_comparison_rv}, the profiles have been normalised to the central dispersion values of \citet{Harris1996} which are provided in each panel. The distances used to convert the proper motion profiles to ${\rm km\,s^{-1}}$ were also taken from \citet{Harris1996}.}
 \label{fig:dispersion_comparison_pm}
\end{figure*}

Mass segregation may provide a physical explanation for the observed offset in NGC~3201 and NGC~6121. The literature profiles are based mainly on giant stars whereas the bulk of our sample in those two clusters consists of less massive main-sequence stars. We used our isochrone fits to obtain giant masses as well as effective masses for our samples. Effective masses were determined by calculating the weighted means of the masses of all stars that entered our analysis. The weight for each star was calculated as $1/(\sigma_{\rm r}^2 + \epsilon_i^2)$, where $\sigma_{\rm r}^2$ is the intrinsic dispersion at the position of the star and $\epsilon_i$ is the uncertainty of its final velocity measurement. As evident from eq.~\ref{eq:probability}, the weights take into account that stars with poorly determined velocity provide less stringent constraints on the measured likelihoods. In Table~\ref{tab:rotation}, the effective weights are listed for all of our clusters. For NGC~3201 and NGC~6121, we find $0.70\,{\rm M_\odot}$ and $0.73\,{\rm M_\odot}$, respectively. The giant masses in both clusters are $\sim 0.85\,{\rm M_\odot}$. According to the model of \citet{Bianchini2016}, the mass differences would correspond to a difference of $\lesssim5\%$ in the velocity dispersion, which is smaller than the observed difference. However, when we split up the sample of NGC~3201 in five mass bins as shown in Fig.~\ref{fig:mass_segregation_ngc3201}, the dispersion increases towards lower masses and the value that we obtain for the highest mass bin is in good agreement with the literature data. We sound a note of caution that such an analysis of the mass dependent kinematics is very challenging because lower masses correspond to lower luminosities and hence higher uncertainties of the measured radial velocities. So even small inaccuracies in the determination of the uncertainties can change the observed trend. For this reason, we postpone a dedicated study on mass segregation to a future publication. However, to facilitate a comparison of models to our data, we provide effective masses also for all of our radial bins in Table~\ref{tab:app:radial_profiles}.

Five clusters from our sample are missing in the compilation of \citet{Baumgardt2017} for which no dedicated radial velocity studies have been carried out so far\footnote{In addition, no comparison data for NGC~6266 is shown in Fig.~\ref{fig:dispersion_comparison_rv}. The cluster was studied by \citet{Luetzgendorf2013} using integrated-light spectroscopy instead of individual radial velocities. A comparison between these two approaches is foreseen at a later stage.}. For those clusters, our work provides the first detailed analysis of the cluster kinematics using radial velocities.

We also compared the profiles of the second velocity moment to the proper-motion data of \citet{Watkins2015} for the 14 clusters present in both samples. To convert the proper motion measurements from ${\rm mas\,yr^{-1}}$ to ${\rm km\,s^{-1}}$, we used the cluster distances from \citet{Harris1996}. As can be seen from Fig.~\ref{fig:dispersion_comparison_pm}, the profiles are generally in good agreement. For some of the denser clusters (especially NGC~6388, NGC~6441, and NGC~7078), there is a trend of higher central values in the MUSE profiles compared to the proper motion profiles. The origin of this discrepancy is currently not clear. It occurs in the regions with the highest crowding, which can potentially affect both types of measurement. Therefore, a detailed comparison of the individual stellar velocities would be very helpful.

\subsection{$v/\sigma$ and $\lambda_{R}$ profiles}
\label{sec:vsigma_lambdar}

The ratio between rotation and dispersion, hereafter referred to as $(v/\sigma)$, is commonly used to evaluate the importance of rotation to the overall cluster dynamics. However, in the globular cluster community, no well-defined method has been established on how to measure $(v/\sigma)$. Commonly, an overall rotation amplitude is measured for the observed radial range and compared to the central velocity dispersion of the cluster. This approach is comparable to the calculation of the ratio between the maximum rotation velocity and the central dispersion that has been used a while ago for galaxy analyses. \citet{Binney2005} mentioned the weaknesses of this approach and proposed an enhanced treatment which has subsequently been used by the SAURON project \citep[e.g.][]{Emsellem2007, Cappellari2007}. It is particularly well suited for integral-field data as it uses the entire footprint of an observation (typically after Voronoi binning the data) instead of only individual values. In this formalism, $(v/\sigma)$ is calculated as \citep[see][eq.~10]{Cappellari2007}
\begin{equation}
 \left(\frac{v}{\sigma}\right)^2 = \frac{\langle v^2 \rangle}{\langle \sigma_{\rm r}^2 \rangle} = \frac{\sum\limits_{n=1}^N F_n v_n^2}{\sum\limits_{n=1}^N F_n \sigma_{{\rm r},\,n}^2}\,,
 \label{eq:vsigma}
\end{equation}
where the sum is over all Voronoi bins and $F_n$ denotes the flux per Voronoi bin. The velocity $v_n$ is measured relative to the systemic velocity of the object. As noted by \citet{Emsellem2007}, a potential shortcoming of the ($v$/$\sigma$) value determined this way is that similar values may be obtained for objects with very different velocity fields. For this reason, they introduced the $\lambda_{\rm R}$ parameter, calculated as
\begin{equation}
 \lambda_{\rm R} = \frac{\langle r |v| \rangle}{\langle r\sqrt{v^2 + \sigma_{\rm r}^2} \rangle} = \frac{\sum\limits_{n=1}^N r_n F_n |v_n|}{\sum\limits_{n=1}^N r_n F_n \sqrt{\sigma_{{\rm r},\,n}^2 + v_n^2}}\,.
 \label{eq:lambdar}
\end{equation}
In this case, $r_n$ is the projected distance of a Voronoi bin to the object centre. \citet{Emsellem2007} showed that $\lambda_{\rm R}$ is a proxy for the spin parameter and that it provides a clear distinction between galaxies that show large scale rotation and those that do not. The simulations of \citet{Jesseit2009} further showed that $\lambda_{\rm R}$ is a robust tracer for the true angular momentum content of galaxies.

Both approaches use the integrated fluxes $F_n$ as a proxy for the stellar masses contained in each Voronoi bin. We obtained integrated fluxes by summing up the luminosities of the stars that fell in each Voronoi bin. To avoid biases because of incompleteness of the catalogues in the central regions, we only considered stars brighter than a given magnitude cut above which we considered the catalogue to complete. This limit was found to be in the range $V=18$ ... $20$, depending on the cluster under consideration. We verified that the results were not sensitive on the chosen cut.

In principle, we could just run the formalism on our data, using the Voronoi maps of $v$ and $\sigma$ created for each of the clusters. However, a complication when working with globular clusters instead of galaxies is that the internal velocities are lower by 1--2 orders of magnitude. As visible from eqs.~\ref{eq:vsigma} and \ref{eq:lambdar}, both values are prone to biases if the random variations in $v_n$ due to the measurement uncertainties are comparable to the variations because of rotation. For this reason, we developed a method to estimate the strength of the bias for our data and correct for it. It is outlined in Appendix~\ref{app:vs_lambdar} \citep[see also Appendix B in][]{Emsellem2007}.

To minimise the impact of uncertainties in the bias correction on our results, we decided to create new maps from the radial profiles instead of using the existing ones. The assumption of a cosine law in each radial bin helps to level off bin-to-bin variations and hence decreases the random velocity offsets per bin. To create two-dimensional maps from the radial profiles, we first calculated Voronoi bins of equal brightness from our photometric catalogues. For each bin, we then determined a mean velocity and a velocity dispersion from the radial profile of the given cluster. The errors from the radial bins were accordingly propagated to the Voronoi bins. In this way, we obtained Voronoi maps with strongly reduced velocity uncertainties per bin. Consequently, the values we obtained for $(v/\sigma)$ and $\lambda_{\rm R}$ were less affected by biases and the corrections we had to apply were smaller (see Fig.~\ref{fig:app:vs_lambdar_bias} in Appendix~\ref{app:vs_lambdar}).

\begin{figure}
 \includegraphics[width=\columnwidth]{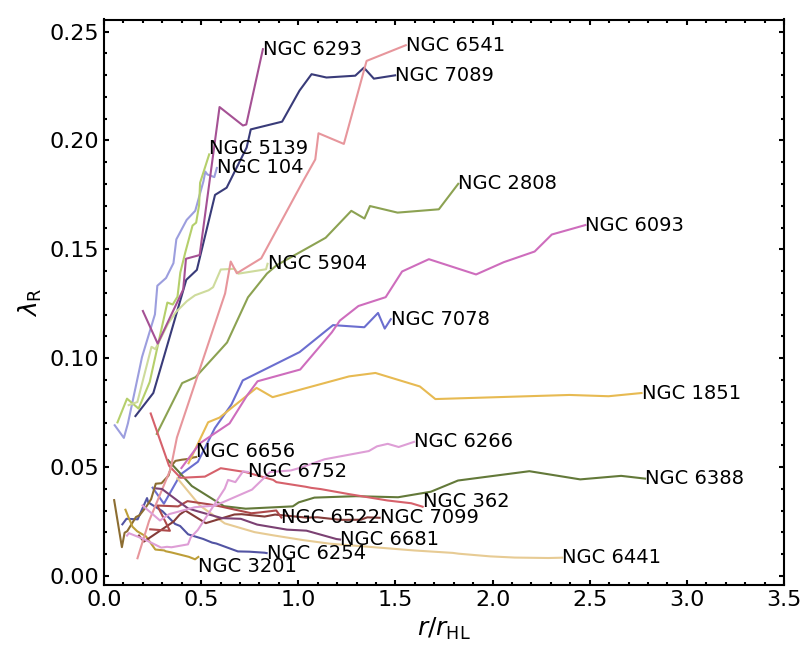}
 \caption{The $\lambda_{\rm R}$ profiles for the clusters from the MUSE sample. The radii have been normalised to the half-light radius of each cluster, taken from \citet{Harris1996}. Due to the limited coverage of the MUSE data of NGC~6121, the cluster was excluded from this comparison (see text).}
 \label{fig:lambdar_profiles}
\end{figure}

As \citet{Emsellem2007} showed, the values for $(v/\sigma)$ and $\lambda_{\rm R}$ should be closely linked and follow the relation
\begin{equation}
 \lambda_{\rm R} = \frac{\kappa\,(v/\sigma)}{\sqrt{1 + \kappa^2\,(v/\sigma)^2}}\,.
\end{equation}
We found that our data was well reproduced by this relation with $\kappa=1.1\pm0.1$, in good agreement with the values determined by \citet{Emsellem2007} for observed galaxies and two-integral Jeans models. Therefore, we will restrict ourselves to a discussion of the $\lambda_{\rm R}$ profiles in the following. We show an overview of the $\lambda_{\rm R}$ profiles obtained for the clusters in our sample in Fig.~\ref{fig:lambdar_profiles}. The values of $\lambda_{\rm R}$ at the halflight radius (or at the maximum radius for clusters where the half-light radius is beyond the radial range covered by the MUSE data) are also provided in Table~\ref{tab:rotation}. We will use these values as measurements for the rotational support of the individual clusters in the discussion below. For completeness, we also included the values of $(v/\sigma)$ at the halflight/maximum radii in Tab.~\ref{tab:rotation}.

The profiles shown in Fig.~\ref{fig:lambdar_profiles} confirm the visual impressions from Figs.~\ref{fig:example_kinematics} and \ref{fig:app:kinematics} in the sense that clusters where the rotation field was already visible by eye also show the steepest $\lambda_{\rm R}$ profiles. Among the clusters with the strongest rotational support in our sample are NGC~104, NGC~5139, NGC~6541, NGC~7089. On the other hand NGC~6441 and NGC~7099 seem to have barely any rotational support.

While the curves depicted in Fig.~\ref{fig:lambdar_profiles} show a variety of shapes, a common behaviour seems to be an almost linear increase of $\lambda_{\rm R}$ out to approximately the halflight radius and a flattening beyond that. This behaviour reflects the increase in rotation velocity that we observe for the majority of our sample. We will revisit the $\lambda_{\rm R}$ analysis below when we discuss rotation in connection with other properties of the clusters.

\subsection{Dynamical distances}

\begin{figure}
 \includegraphics[width=\columnwidth]{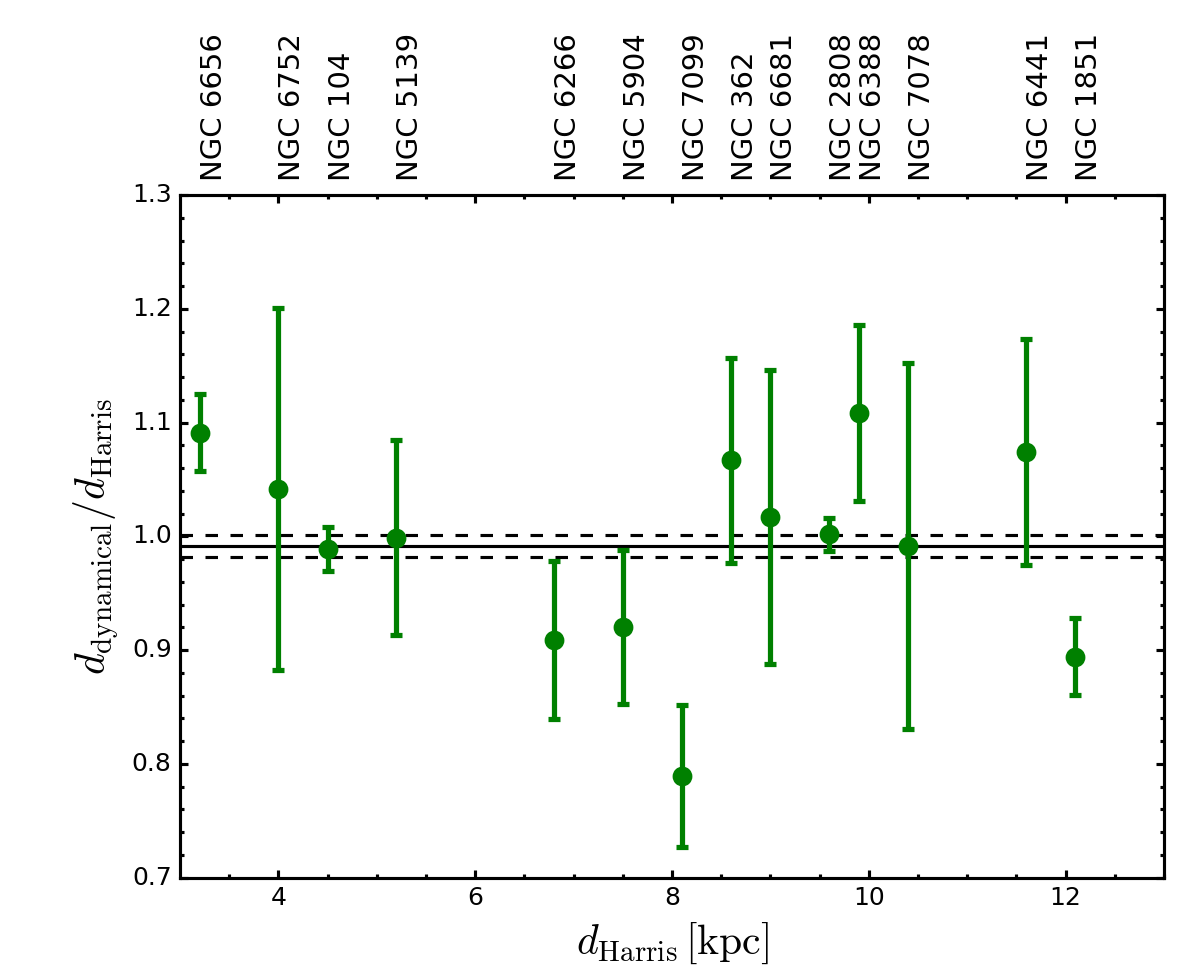}
 \caption{Ratio of the dynamical distances determined by matching the proper motion profiles of \citet{Watkins2015} to the MUSE data and the distances in \citet{Harris1996}, as a function of the latter. The mean ratio and its uncertainty are indicated by the solid and dashed lines, respectively. The name of each cluster is provided at the top of the figure.}
 \label{fig:dynamical_distances}
\end{figure}

The availability of both radial velocity and proper motion profiles also enables the determination of dynamical distances for a sub-sample of our clusters. To this aim, we determined the radial ranges common to both studies and interpolated the proper motion profiles to the same binning as the MUSE profiles. For each bin, the ratio of the two values was determined and converted to a distance. Finally, the weighted mean of the distances determined in the individual bins was used as the distance to the cluster, while the standard deviation between the bins was used as uncertainty of the distance measurement. A comparison of the dynamical distances to the values of \citet{Harris1996} is shown in Fig.~\ref{fig:dynamical_distances}, our values are also listed in Table~\ref{tab:rotation}. We found a mean value of the distance ratios of $0.99\pm0.01$, indicating an excellent agreement with the literature data. Three clusters show significant ($>2\sigma$) deviations from a 1:1 relation, namely NGC~1851, NGC~6656, and NGC~7099. NGC~7099 has the least amount of proper motion data, limiting the reliability of our measurement. The other two clusters will be discussed below.

\vspace*{5mm}

Dynamical distances have also been determined by \citet{Watkins2015b} and \citet{Baumgardt2017} who also found good agreement with the literature data, although the distances determined by \citet{Baumgardt2017} were on average 8\% lower. There are two clusters for which our distances deviate significantly from the values determined by \citet{Watkins2015b} and \citet{Baumgardt2017}, NGC~104 ($4.5\pm0.1\,{\rm kpc}$ compared to $4.15\pm0.08\,{\rm kpc}$ and $3.95\pm0.05\,{\rm kpc}$) and NGC~6656 ($3.5\pm0.1\,{\rm kpc}$ compared to $2.84\pm0.16\,{\rm kpc}$ and $2.66\pm0.10\,{\rm kpc}$). For NGC~104, our measurement is in better agreement with other methods \citep[e.g.][]{Woodley2012}. As the same proper motion data is used in all three studies, this suggests that the existing radial velocity dispersion data lead to an underestimation of the true distance of NGC~104. As argued by \citet{McLaughlin2006a} or \citet{Bogdanov2016}, this is probably caused by a bias of the existing radial velocities towards the cluster mean. This is likely to happen if the stellar spectra are contaminated by nearby stars or the unresolved cluster light. As our approach to extract stellar spectra explicitly takes those contributions into account, it should be robust against such a bias. However, it is worth noting that \citet{McLaughlin2006a} also found a bias in their sample of stars observed with a Fabry-Perot instrument, which in principle allows one to use a similar deblending approach as we did for the MUSE data \citep[see][]{Gebhardt1994a}. Yet another advantage of our approach is the usage of HST imaging which enables us to identify faint stars and blends that would not be resolved at seeing-limited resolution. The good agreement of our dynamical distance estimate with other available distance measurements makes us confident that our analysis is robust against the aforementioned bias.

The situation is less clear in NGC~6656 because the distance to this cluster is much less certain compared to NGC~104. The value in the \citet{Harris1996} catalogue is based on \citet{Cudworth1986} who measured a distance of $3.2\pm0.3\,{\rm kpc}$ based on horizontal branch brightness and internal dynamics. Again, it could be that the existing radial velocity measurements lead to an underestimation of the true distance. In the near future, Gaia will hopefully settle this issue. The same is true for NGC~1851, where all three measurements of the dynamical distance are in agreement ($\sim10.5\,{\rm kpc}$) but are significantly below the value of $12.1\,{\rm kpc}$ determined by \citet{Walker1998} from RR Lyrae stars.

\section{Discussion}
\label{sec:discussion}

\begin{figure}
 \includegraphics[width=\columnwidth]{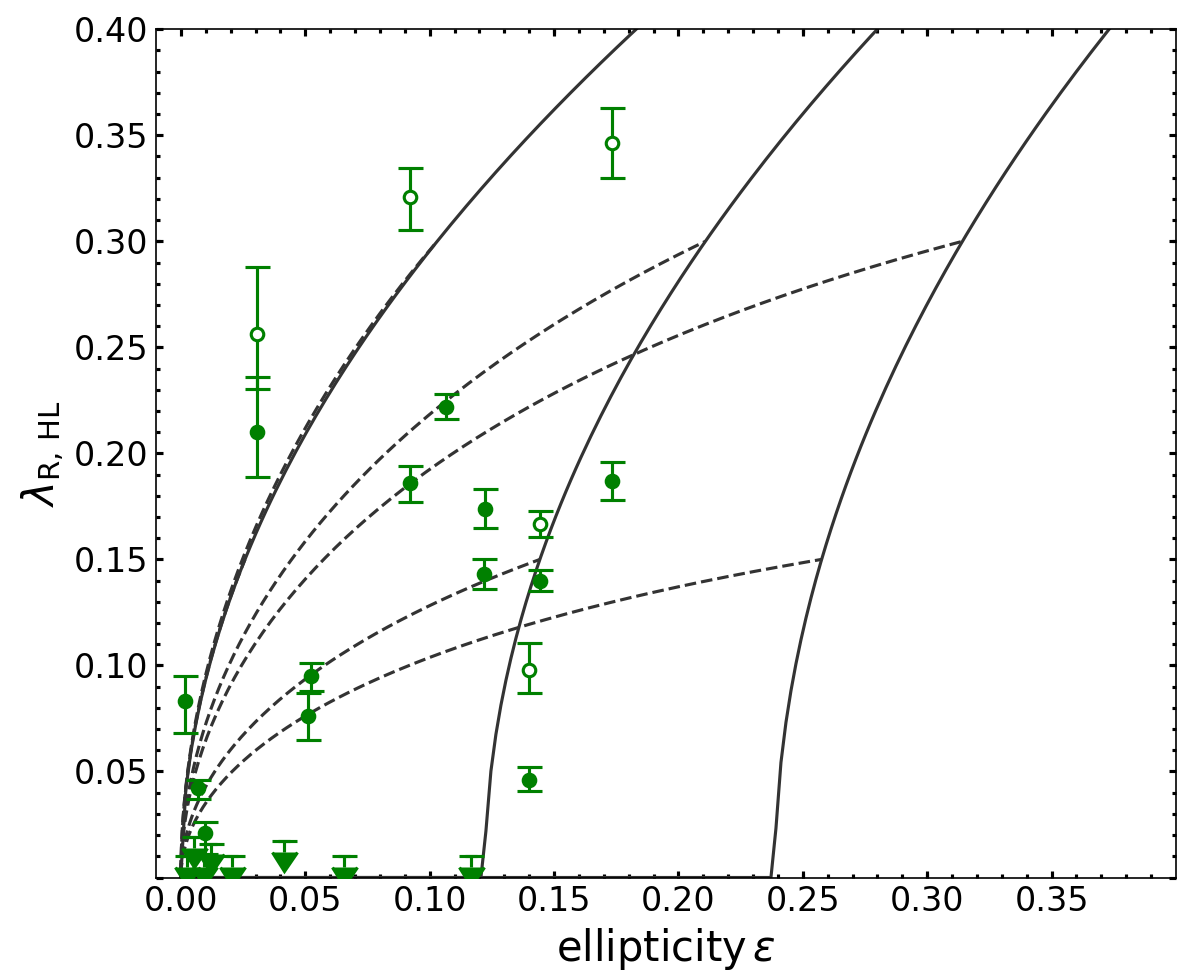}
 \caption{Rotation strengths as a function of the global ellipticities of the clusters, using $\lambda_{\rm R}$ as a measure of the importance of rotation. The $\lambda_{\rm R}$-values have been calculated at the halflight radius $r_{\rm HL}$ where possible, otherwise at the maximum radius covered by the MUSE data. In the latter cases, we also show the values obtained by linear extrapolation to $r_{\rm HL}$ as open symbols. The ellipticities have been taken from \citet{Harris1996}. The solid lines show the prediction for an edge-on oblate rotator in the isotropic case and for anisotropies of $\beta=0.1$ and $0.2$ (from left to right, using the formulae of \citet{Cappellari2007} and \citet{Emsellem2007}). The dashed lines indicate the behaviour expected when varying the inclination.}
 \label{fig:rotation_vs_ellipticity}
\end{figure}

Our analysis of the rotation fields in the previous section showed that the majority of the clusters in our sample show rotation. This finding is in agreement with previous studies that searched for rotation in globular clusters \citep[e.g.][]{Bellazzini2012, Fabricius2014}. We now aim to investigate if the rotation signals show any correlations with other parameters of the sample clusters.

\subsection{Is rotation related to ellipticity?}
\label{sec:ellipticity}

The work of \citet{Fabricius2014} revealed a correlation between the global ellipticity of a cluster and the strength of its central rotation. Interestingly though, neither \citet{Bellazzini2012} nor \citet{Lardo2015} found a correlation between rotation and ellipticity in their analyses. Under the assumption that globular clusters can be described by the same dynamical model, such as an isotropic oblate rotator, stronger rotation would be expected in more flattened systems. However, various effects can dilute a possible correlation, the most important ones being anisotropies, inclination effects, or tidal forces from the Milky Way \citep[see][for an estimate of the impact of the latter]{Bergh2008}.

To verify if our data suggest a link between rotation and ellipticity, we show in Fig.~\ref{fig:rotation_vs_ellipticity} the $\lambda_{\rm R,\,HL}$ values from Table~\ref{tab:rotation} as a function of the ellipticities of the clusters. For those clusters where the radial coverage of the MUSE data ended before reaching the halflight radius $r_{\rm HL}$, we also included the results that would be obtained by scaling the original values by $r_{\rm HL}/r_{\rm max.}$ (open symbols in Fig.~\ref{fig:rotation_vs_ellipticity}). This was done in light of the linear increase in $\lambda_{\rm R}$ that many clusters show for $r<r_{\rm HL}$ in Fig.~\ref{fig:lambdar_profiles}.

The data shown in Fig.~\ref{fig:rotation_vs_ellipticity} suggest a trend of increasing rotation for more elliptical clusters. The statistical significance of a correlation is not high though, a Spearman rank test returns a probability of $3\%$ that the quantities are uncorrelated. If we use the extrapolated values instead for the clusters where we do not reach the halflight radius, this probability decreases to $<1\%$. We find a similarly strong correlation of ellipticity with the velocity gradient $\lVert \bigtriangledown v \rVert$. This can be verified in Table~\ref{tab:correlations}, where the results from the various Spearman correlation tests are summarised. Hence, our results seem to confirm the finding of \citet{Fabricius2014} that more elliptical clusters show stronger rotation, albeit with low significance.

Also shown in Fig.~\ref{fig:rotation_vs_ellipticity} is the predicted behaviour of an isotropic oblate rotator, using the formulae from \citet{Cappellari2007} and \citet{Emsellem2007}. Most of our targets lie significantly below the expectation for an isotropic oblate rotator seen edge-on. As inclination effects play a minor role in the isotropic case (see dotted lines in Fig.~\ref{fig:rotation_vs_ellipticity}), mild anisotropies could provide an explanation for the excess in ellipticity at a given value of $\lambda_{\rm R}$ (see the model curves for anisotropies of $\beta=0.1$ and $0.2$). While \citet{Watkins2015} found that the anisotropies in the centres of most clusters are small, they might be more important in the outskirts of the clusters were the ellipticities were measured \citep[e.g.][]{Giersz1997,Tiongco2016a,Zocchi2016}.

Another result that lead \citet{Fabricius2014} to conclude that there is a link between rotation and flattening of the clusters was that for most of their targets, the kinematic angles and the position angles of the photometric semi-major axes agreed within their relative uncertainties. To check if this is also true for our sample, we performed the same analysis as \citet{Fabricius2014} and determined photometric position angles from a principal component analysis of the available photometry. As in Sect.\ref{sec:vsigma_lambdar}, we applied a magnitude cut to the photometric catalogues to avoid biases due to incompleteness and only considered radii $<80^{\prime\prime}$ where the catalogues covered the full circle. The photometric position angles are compared to our measurements of the rotation-axis angles in Figs.~\ref{fig:example_kinematics} and \ref{fig:app:kinematics}. In the case of isotropic oblate rotators, one would expect both angles to be separated by $90^{\circ}$. This is indeed the case for the vast majority of the rotating clusters in our sample. A notable exception is NGC~5139, where the axes seem to be aligned. This might indicate that in NGC~5139 the central kinematics are more complex than in the remaining clusters of our sample, which is also suggested by the detection of (mild) radial anisotropy around its centre \citep[][]{Ven2006,Marel2010}. Such complex kinematics are observed in early-type galaxies, including dwarf spheroidals \citep[e.g.][]{Ebrova2015,Kacharov2017} and massive ellipticals \citep[e.g.][Krajnovi\'{c} et~al. in prep.\footnote{see \url{http://www.eso.org/sci/meetings/2015/StellarHalos2015/talks_presentation/emsellem_M3G.pdf}}]{Tsatsi2017}, and are commonly explained by invoking mergers --- a possibility that is also often used to explain the peculiar chemistry of NGC~5139 (see Sect.~\ref{sec:populations} below).

\subsection{Rotation and globular cluster formation}

\begin{table}
 \caption{Results of the Spearman correlation tests between the properties of the rotation fields and various cluster parameters.}
 \label{tab:correlations}
 \begin{center}\begin{tabular}{ l | l | c | c | l }
\hline
  Par. 1 & Par. 2 & $r_{\rm s}$ & $p$ & Ref. \\
 (1) & (2) & (3) & (4) & (5) \\ \hline
 $\lVert \bigtriangledown v \rVert$ & $e$ & 0.521 & 0.015 & b \\
  & $\sigma_{\rm 0}$ & 0.322 & 0.15 & a \\
  & $M_{\rm V,t}$ & -0.409 & 0.065 & b \\
  & $[{\rm Fe/H}]$ & -0.064 & 0.78 & b \\
  & HB-index & 0.092 & 0.7 & c \\
  & $\log t_{\rm h}$ & 0.399 & 0.073 & b \\
  & $N_1/N_{\rm tot}$ & -0.549 & 0.022 & d \\
\hline
 $\lambda_{\rm R,\, HL}$ & $e$ & 0.475 (0.606) & 0.03 (0.0036) & b \\
  & $\sigma_{\rm 0}$ & 0.322 (0.257) & 0.16 (0.26) & a \\
  & $M_{\rm V,t}$ & -0.534 (-0.506) & 0.013 (0.019) & b \\
  & $[{\rm Fe/H}]$ & -0.154 (-0.151) & 0.5 (0.51) & b \\
  & HB-index & 0.014 (0.051) & 0.95 (0.83) & c \\
  & $\log t_{\rm h}$ & 0.548 (0.642) & 0.01 (0.0017) & b \\
  & $N_1/N_{\rm tot}$ & -0.531 (-0.488) & 0.028 (0.047) & d \\
\hline
\end{tabular}
\end{center}
\medskip
Notes. (1) Primary parameter, taken from Table~\ref{tab:rotation}. (2) Secondary parameter: $e$ -- global photometric ellipticity, $\sigma_{\rm 0}$ -- central velocity dispersion, $M_{\rm V,\,t}$ -- absolute cluster magnitude, [Fe/H] -- metallicity, HB-index -- horizontal branch morphology, $\log t_{\rm h}$ -- logarithm of relaxation time at half-light radius in years. (3) Spearman correlation coefficient. The values in brackets were obtained  when extrapolating $\lambda_{\rm R,\,HL}$ to the halflight radius for all clusters with $r_{\rm max.}/r_{\rm HL} < 1$ in Table~\ref{tab:rotation}. (4) Two-sided p-value. (5) Reference for secondary parameter: (a) this work, (b): \citet{Harris1996}, (c): \citet{Mackey2005} (d): \citet{Milone2017}.
\end{table}

\begin{figure*}
\includegraphics[width=\textwidth]{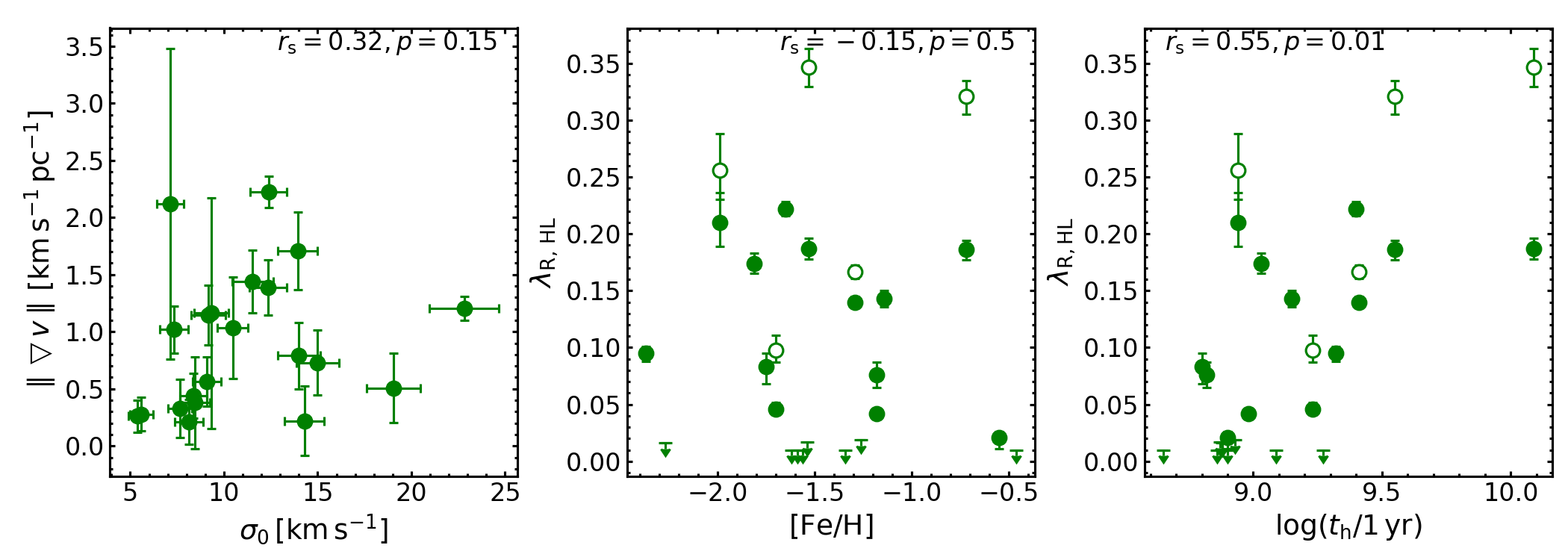}
 \caption{Relations between rotation strengths and various fundamental properties of the clusters discussed in the text, namely the central velocity dispersion $\sigma_{\rm 0}$ ({\it left}), the metallicity $[{\rm Fe/H}]$ ({\it middle}), and the logarithm of the relaxation time at the half-light radius $\log t_{\rm h}$ ({\it right}). As in Fig.\ref{fig:rotation_vs_ellipticity}, we show the extrapolated values of $\lambda_{\rm R,\,HL}$ as open symbols for clusters without coverage of the halflight radii. To the top of each panel, we provide the Spearman correlation coefficient and the two-sided p-value of the correlation. For the origin of the various cluster properties, see Table~\ref{tab:correlations}.}
 \label{fig:correlations}
\end{figure*}

Several correlations between the rotation of a cluster and other fundamental properties are discussed in the literature. \citet{Bellazzini2012} and \citet[][who added a few additional clusters to the sample of \citealt{Bellazzini2012}]{Lardo2015}, report increasing rotation velocities $v_{\rm rot}$ for more metal-rich clusters and clusters with redder horizontal branches, where the colour of the horizontal branch was quantified using the HB-index as given by \citet{Lee1990} or \citet{Mackey2005}. However, no correlation with metallicity was found by \citet{Kimmig2015} and also our data do not confirm these results. As can be verified from the central panel of Fig.~\ref{fig:correlations} and the values in Table~\ref{tab:correlations}, no correlations exist in our sample with the cluster metallicities or the morphologies of their horizontal branches. The latter is surprising as the correlation between rotation velocity and horizontal branch morphology is the most significant one in the data set of \citet{Bellazzini2012}. When comparing our rotation results with those of \citet{Bellazzini2012}, it is striking that the largest discrepancies are observed for NGC~6388 and NGC~6441, the most metal-rich clusters of our sample, for which we obtain a much weaker rotation signal. As the horizontal branch morphology is linked to the metallicity \citep[e.g.][]{Lee1990}, this may partly explain why we do not see a correlation with either quantity in our data. Again, we emphasise the different radial regimes probed by our study and by \citet{Bellazzini2012}.

\citet{Bellazzini2012} also find $v_{\rm rot}$ to correlate with the luminosity and the central velocity dispersion of a cluster. The latter is in agreement with the correlation between the velocity gradient $\lVert \bigtriangledown v \rVert$ and the velocity dispersion reported by \citet{Fabricius2014}, who speculated that undetected rotation had spuriously increased the velocity dispersion measurements in the literature. As the velocity dispersions obtained in this work are robust against such biases, we are able to test this hypothesis. We find only a mild correlation between either $\lambda_{\rm R,\,HL}$ or $\lVert \bigtriangledown v \rVert$ and the updated values of the velocity dispersion, with false-alarm probabilities of around $15\%$ (see Fig.~\ref{fig:correlations} and Table~\ref{tab:correlations}). Yet our study confirms that the rotational support increases with the luminosity of a cluster (expressed as an anti-correlation between $\lambda_{\rm R,\,HL}$ and the absolute magnitude $M_{\rm V, t}$ of a cluster in Table~\ref{tab:correlations}. Under the assumption that globular clusters have comparable mass-to-light ratios \citep[e.g.][]{Baumgardt2017}, this would imply that the rotational support increases in more massive clusters. The correlation with velocity dispersion might be secondary then, because more massive clusters tend to have higher velocity dispersions.

As can be seen in the right panel of Fig.~\ref{fig:correlations} and verified in Table~\ref{tab:correlations}, in our sample the relaxation time at the halflight radius, $t_{\rm h}$ \citep[taken from][]{Harris1996}, shows the strongest correlation with $\lambda_{\rm R,\,HL}$, especially when using the extrapolated values for the clusters without coverage of the halflight radius.\footnote{We sound a note of caution that the calculation of the relaxation times included the same halflight radii $r_{\rm HL}$ \citep[see formulae in][]{Djorgovski1993} we used as a reference to measure $\lambda_{\rm R}$, cf. Sect~\ref{sec:vsigma_lambdar}. However, no correlation exists between $\lambda_{\rm R}$ and $r_{r\rm HL}$ itself in our data.} This is not completely unexpected, as angular momentum is transported outwards by relaxation processes. Hence, our data support a scenario in which the globular cluster inherited angular momentum from the collapsing molecular cloud which would then be slowly dissipated away by two-body relaxation. Indeed, recent theoretical studies \citep[][]{Lee2016,Mapelli2017} suggest that massive star clusters are born with significant amounts of rotation. In addition, rotation is found in both intermediate-age \citep{Davies2011o,Mackey2013b} and young massive \citep{Henault-Brunet2012b} clusters, supporting such a scenario. In the future, systematic studies of the rotation properties as a function of cluster age will be crucial to make further progress in this direction.

The aforementioned scenario might also explain the discrepancies we observe in comparison to the study of \citet{Bellazzini2012}. Whereas our data focuses on the central cluster regions where dissipation dominates because of the short relaxation times, \citet{Bellazzini2012} probe the outskirts of the clusters where rotation patterns imprinted during the formation of the cluster would be longer lived and hence possible differences connected to the metal-content of the molecular cloud are still observable.

In view of the dissipation of angular momentum, it is interesting to look at the core-collapsed clusters of our sample. However, no clear picture emerges. While NCG~6522, NGC~6681, and NGC~7099 are indeed among the least rotating clusters of our sample, rather strong rotation fields are observed in NGC~7078 and possibly also NGC~6293. Therefore, it remains unclear how strongly the central rotation is affected by core-collapse.

In light of the evidence for a decoupled core in NGC~7078 \citep{Bosch2006}, we also checked whether any other cluster from our sample shows a similar rotation behaviour (enhanced rotation in combination with a change in position angle) in its centre. We observe similar features in our data of NGC~362, NGC~5904, NGC~6254, and NGC~6266, suggesting that this feature is not unique to NGC~7078. Two of the four clusters (NGC~362 and NGC~6266) are labelled as possible core-collapse clusters in \citet{Harris1996}, so the feature does not seem to be specific to core-collapse clusters. However, dedicated modelling will be needed to investigate this further.

\subsection{Rotation and multiple populations}
\label{sec:populations}

Finally, we briefly discuss our findings with regard to the open issue of the origin of multiple populations inside the clusters. Several authors discussed the possibility of finding dynamical signatures stemming from one or the other formation scenario. \citet{Gavagnin2016} investigated the possibility that iron-complex globular clusters were created from mergers in dwarfs galaxies and argued that in this case, the clusters should be rotating and flattened. Our sample includes several clusters with reported spreads in heavy elements, namely NGC~1851 \citep{Yong2008}, NGC~5139 \citep[e.g.][]{Norris1995}, NGC~6656 \citep[][but also see \citealt{Mucciarelli2015}]{Marino2009}, and NGC~7089 \citep{Lardo2013}. We note that all of those clusters show a clear rotation signal in our data but no clear distinction can be made with respect to the ``normal'' clusters. \citet{Gavagnin2016} predict that in merger remnants, the rotation velocity should increase out to  about the half-light radius and stay approximately constant beyond it. The increase at small distances to the centre is also visible in our data. Beyond the half-light radius, we lack the data for a detailed comparison. However, in NGC~1851, we see evidence that the rotation velocity decreases again.

For a follow-up study, we plan to investigate if the various populations inside the clusters show different rotation patterns. According to the simulations of \citet{Henault-Brunet2015}, a second population forming from the ejecta of a primordial one should rotate faster than the latter while the opposite should be true in the early disk accretion scenario proposed by \citet{Bastian2013}. While the chances of still finding such differences are higher beyond the half-light radius (because of the longer relaxation times), the detection of clear rotation fields even around the cluster centres might indicate that differences are still observable within the MUSE footprints. A first step into the direction of studying chemically-resolved rotation properties has been done by \citet{Cordero2017} who found the extreme population of NGC~6205 to have a higher rotation rate than the other two populations (although the sample sizes per population were still small).

\citet{Mastrobuono-Battisti2016} also studied the evolution of a rotating disk of second-population stars in a globular clusters. According to their simulations, a positive correlation would be expected between the cluster ellipticities and the relaxation times. Indeed, in our sample such a correlation is observed -- which is not very surprising as both quantities correlate with the strength of the cluster rotation. \citet{Mastrobuono-Battisti2016} further speculate about a correlation between ellipticity and the fraction of second-population stars. When we use the fractions of first population stars reported by of \citet{Milone2017} at face value, we indeed see some evidence for an anti-correlation (cf. Table~\ref{tab:correlations}), which would support such a scenario. But this finding may just be a consequence of the increase of rotation with cluster mass that we observe as \citet{Milone2017} find the fraction of second population stars to also increase with cluster mass. In a forthcoming publication, we plan to investigate possible relations between rotation and multiple populations in more detail.

\section{Conclusions}
\label{sec:conclusions}


We presented the first results from our MUSE survey of 25 Galactic globular clusters and showed that the data is well suited to investigate the central dynamics of the clusters in the sample. In comparison to previous radial velocity samples, a big advantage of our approach is a complete coverage of the cluster centres, out to approximately the half-light radii of our targets. In combination with a data analysis that accounts for stellar blends, this enables us to obtain average sample sizes of currently around 9\,000 stars per cluster. Our data sets represent the largest spectroscopic samples that have been obtained so far for the vast majority of our targets.

Despite the limited spectral resolution of MUSE and the complexity of the instrument (which analyses the light using 24 spectrographs), we are able to achieve a velocity accuracy of $1$--$2{\rm km\,s^{-1}}$, well below the velocity dispersions of the clusters in our sample. The uncertainties for the individual stellar velocities will be higher depending on the S/N of the extracted spectra, of course.  However, reliable velocity measurements are still possible at ${\rm S/N}\sim  5$--$10$.

We constructed two-dimensional maps as well as radial profiles of the average velocities and velocity dispersions for each cluster. They show that the majority of our sample rotates in the central cluster regions. Our finding that the amount of rotation correlates with the ellipticities of a cluster confirms the results by \citet{Fabricius2014} that central rotation affects the overall appearance of a cluster. Further, we find a clear correlation between the rotation strengths and the relaxation times at the half-light radii. This finding supports a scenario in which the clusters are born with a significant amount of angular momentum which is dissipated over their lifetimes as a result of two-body relaxation.

The velocity dispersion profiles show a good agreement with profiles available in the literature (with the exception of NGC~6121 and possibly NGC~3201) but reach closer to the cluster centres in most cases. By means of a comparison with the proper motion data of \citet{Watkins2015}, we derived dynamical distances for 14 clusters. They are usually in good agreement with previous estimates, albeit a few exceptions are found again. For NGC~104, our updated value for the dynamical distance is in better agreement with other distance estimates, suggesting that the existing radial velocity data near the cluster centre underestimated the true velocity dispersion.

\section*{Acknowledgements}
We thank the anonymous referee for a careful reading of our manuscript and for providing us with a report that contained many good suggestions to further improve the paper.
We thank Nate Bastian and the members of the MUSE consortium not listed among the co-authors for inspiring discussions and support during stages of this work.
SK, SD, and PMW acknowledge support from the German Ministry for Education and Science (BMBF Verbundforschung) through project MUSE-AO, grants 05A14BAC, 05A14MGA, and 05A17MGA.
SK and SD also acknowledge support from the German Research Foundation (DFG) through projects KA 4537/2-1 and DR 281/35-1.
Based on observations made with ESO Telescopes at the La Silla Paranal Observatory under programme IDs 094.D-0142, 095.D-0629, 096.D-0175, and 097.D-0295.
Based on observations made with the NASA/ESA Hubble Space Telescope, obtained from the data archive at the Space Telescope Science Institute. STScI is operated by the Association of Universities for Research in Astronomy, Inc. under NASA contract NAS 5-26555.
This research made use of Astropy, a community-developed core Python package for Astronomy (Astropy Collaboration, 2013).



\bibliographystyle{mnras}
\bibliography{muse_gc_kinematics_skamann}



\appendix

\section{Kinematics plots for remaining clusters}
\label{sec:app:kinematics}
\clearpage
\begin{figure*}
 \includegraphics[width=\textwidth]{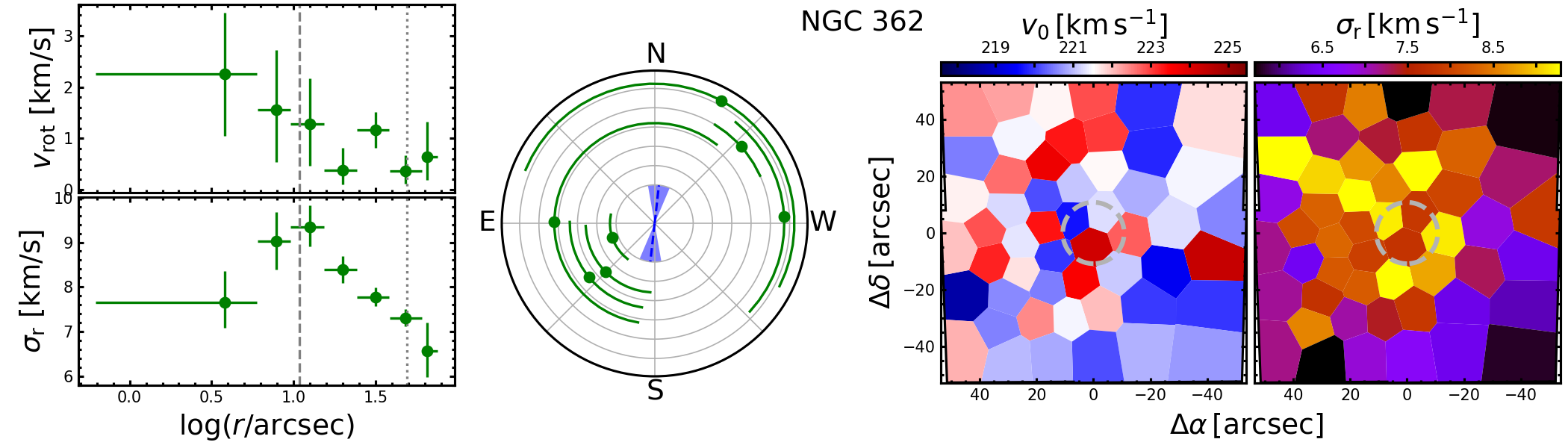}
 \includegraphics[width=\textwidth]{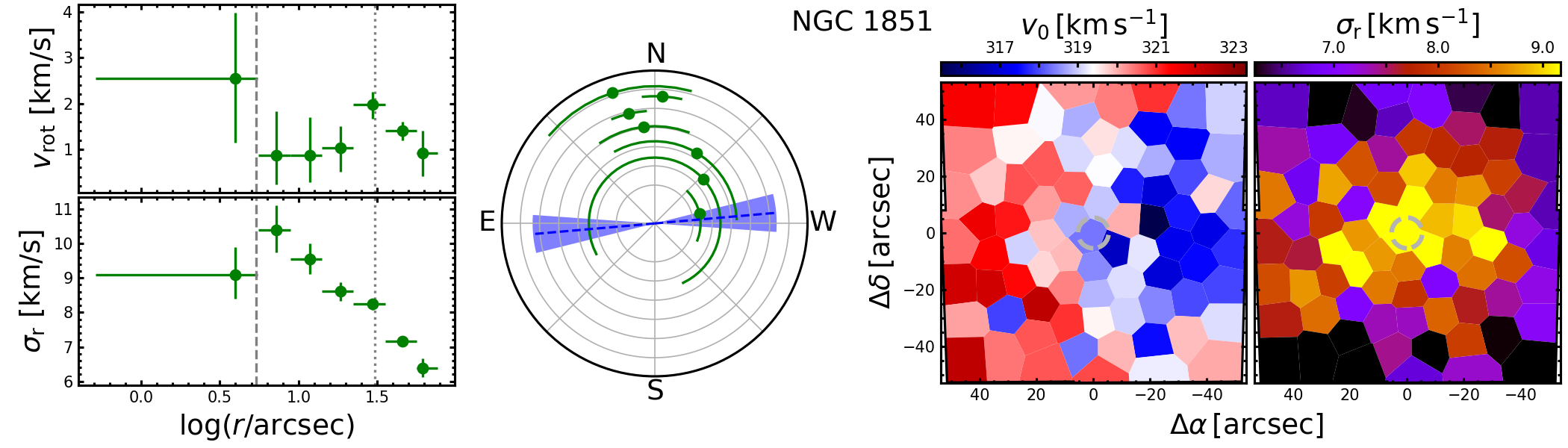}
 \includegraphics[width=\textwidth]{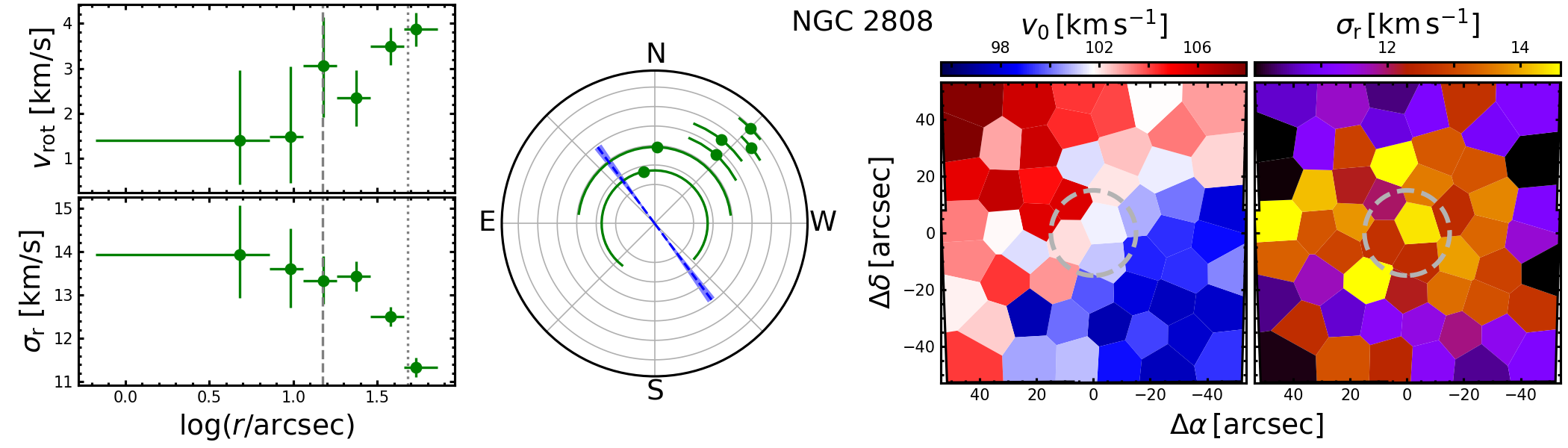}
 \includegraphics[width=\textwidth]{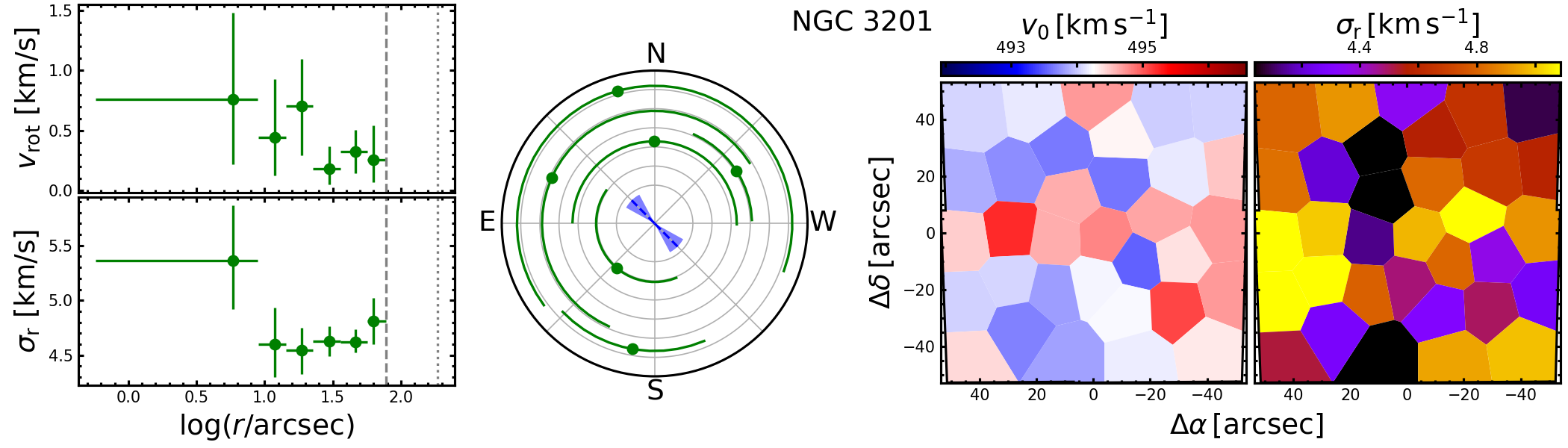}
 \caption{As Fig.~\ref{fig:example_kinematics} but for NGC~362, NGC~1851, NGC~2808, and NGC~3201.}
 \label{fig:app:kinematics}
\end{figure*}
\clearpage

\begin{figure*}
 \includegraphics[width=\textwidth]{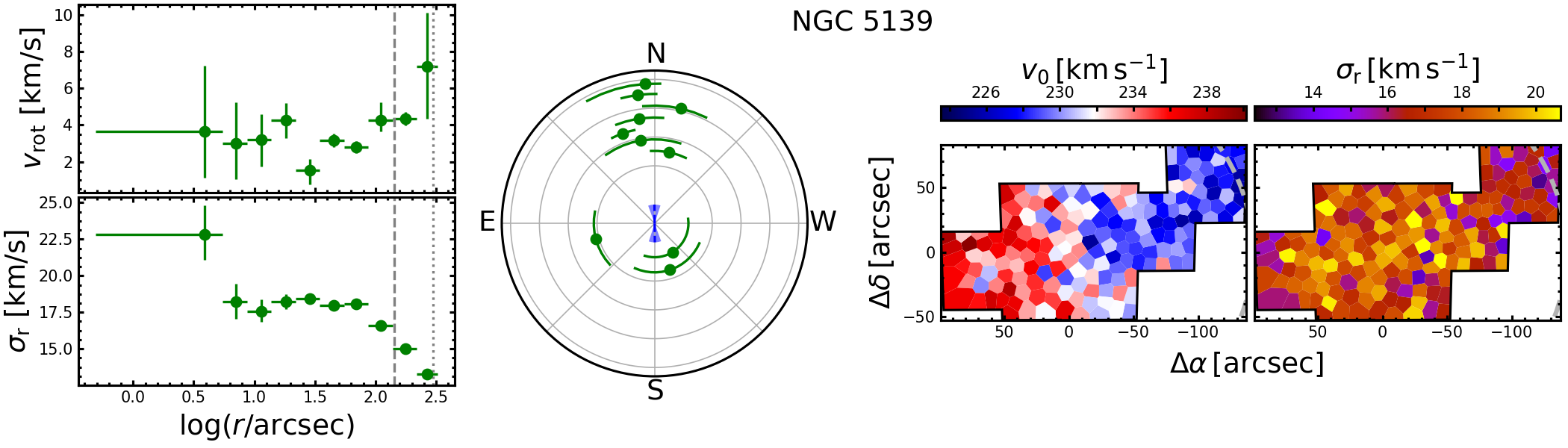}
 \includegraphics[width=\textwidth]{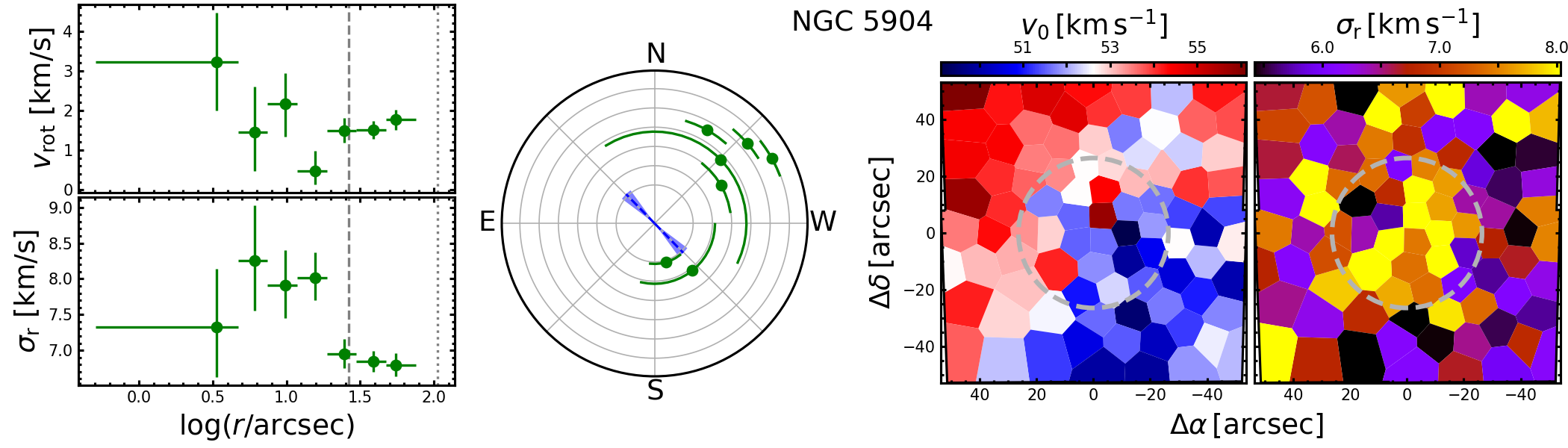}
 \includegraphics[width=\textwidth]{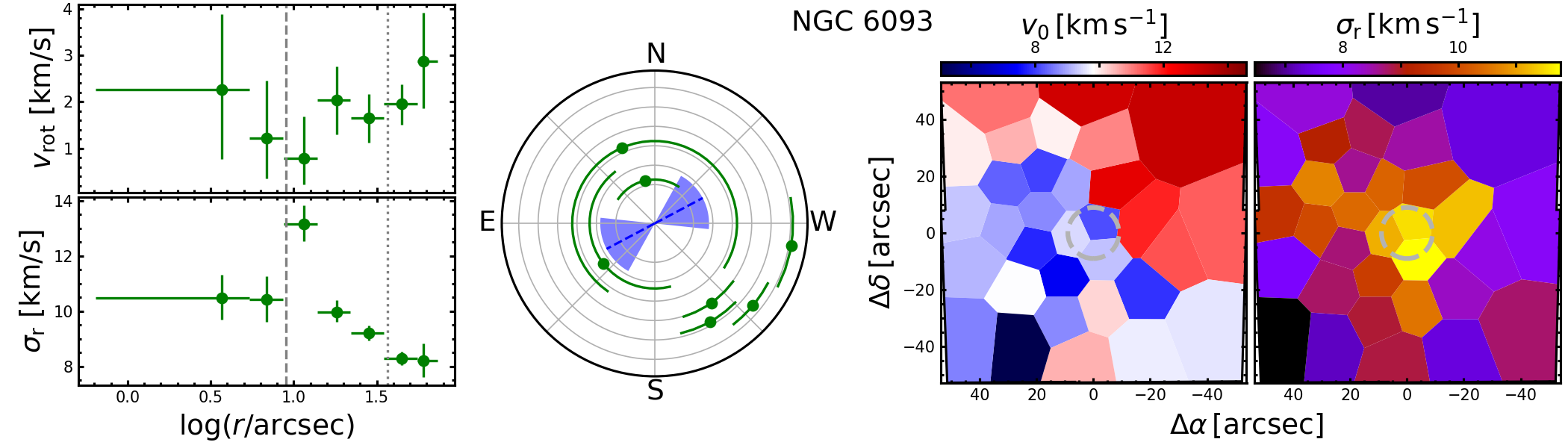}
 \includegraphics[width=\textwidth]{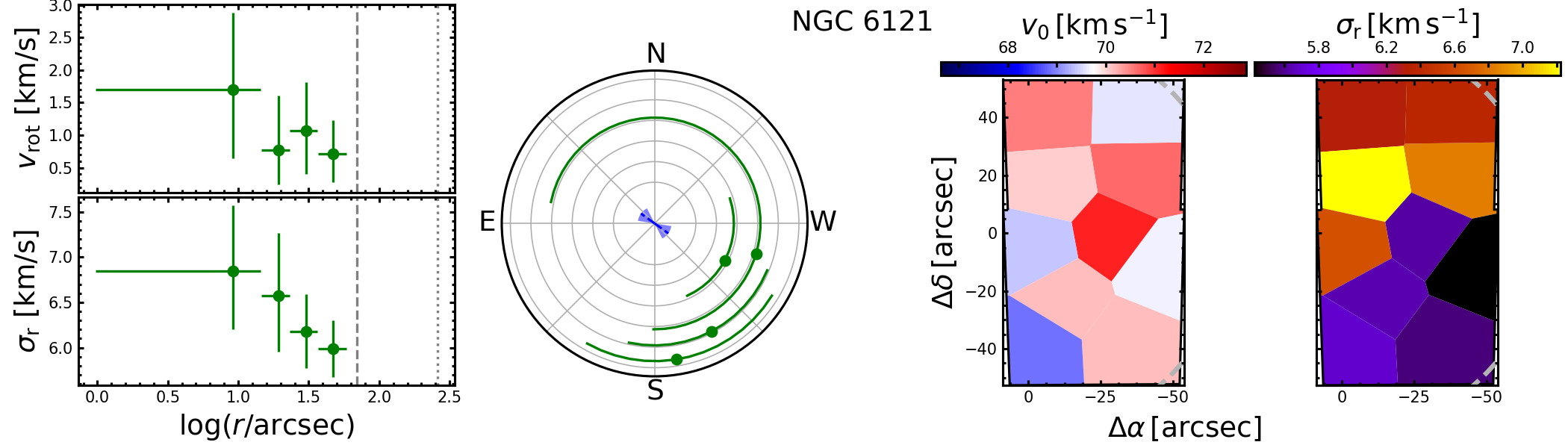}
 \contcaption{As Fig.~\ref{fig:example_kinematics} but for NGC~5139, NGC~5904, NGC~6093, and NGC~6121}
 \label{fig:app:kinematics:cont}
\end{figure*}
\clearpage

\begin{figure*}
 \includegraphics[width=\textwidth]{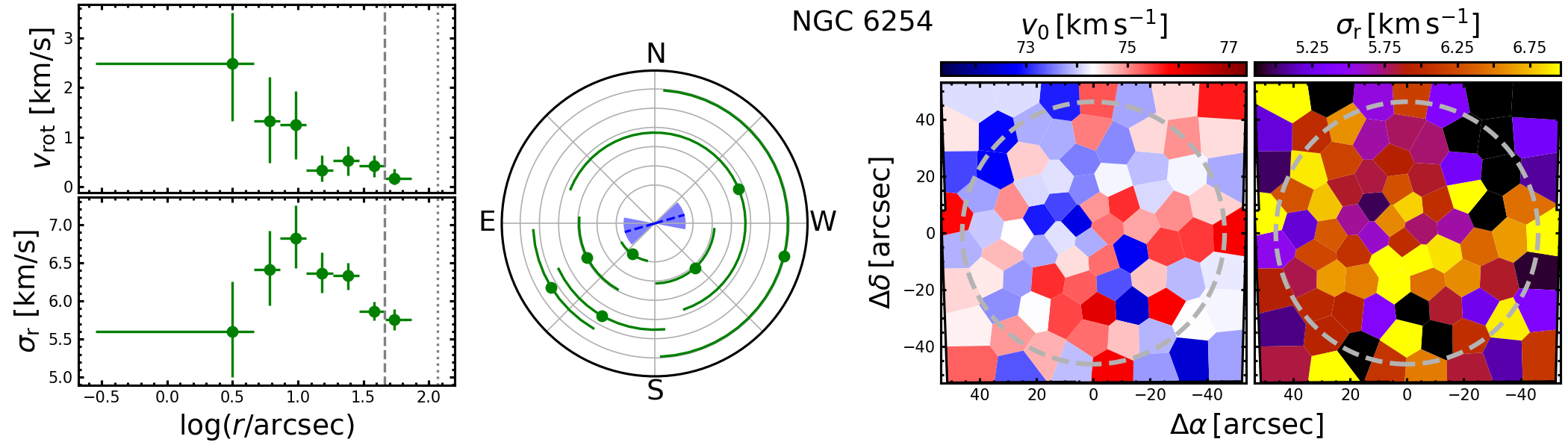}
 \includegraphics[width=\textwidth]{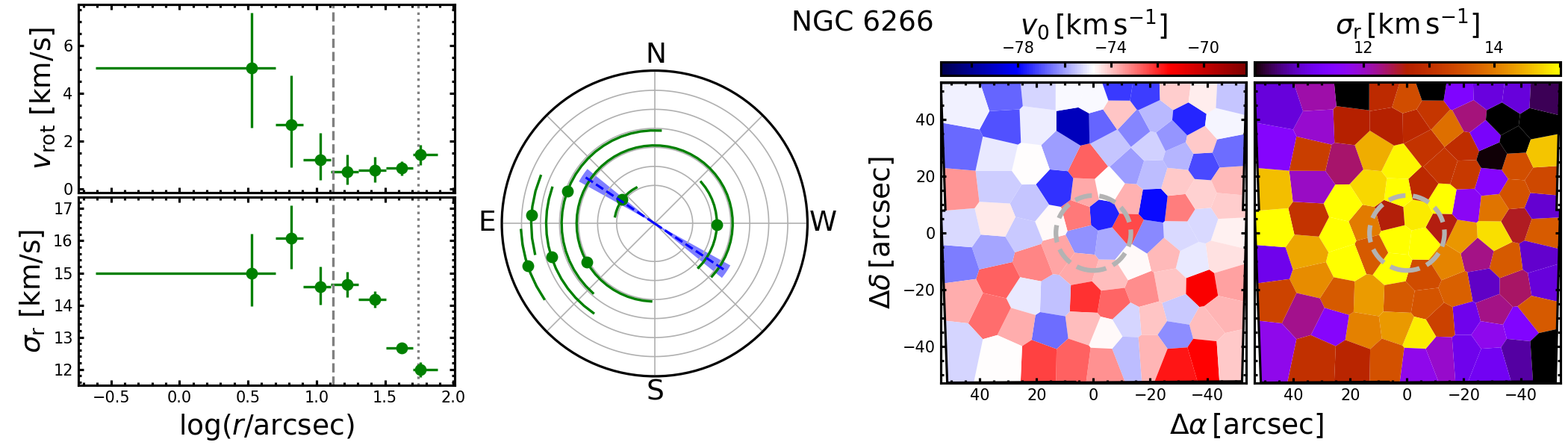}
 \includegraphics[width=\textwidth]{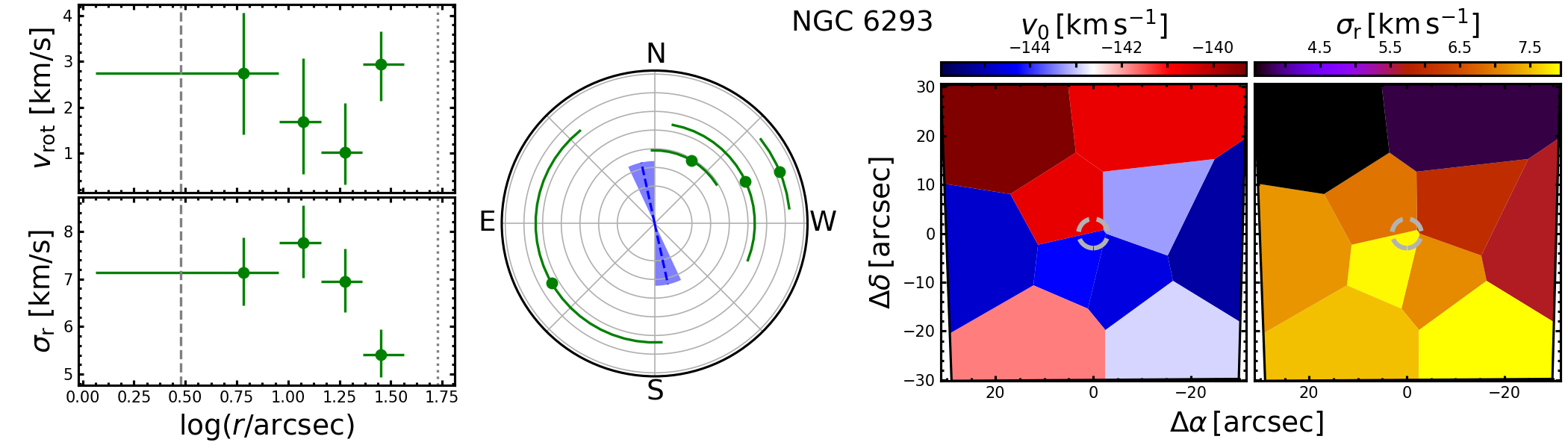}
 \includegraphics[width=\textwidth]{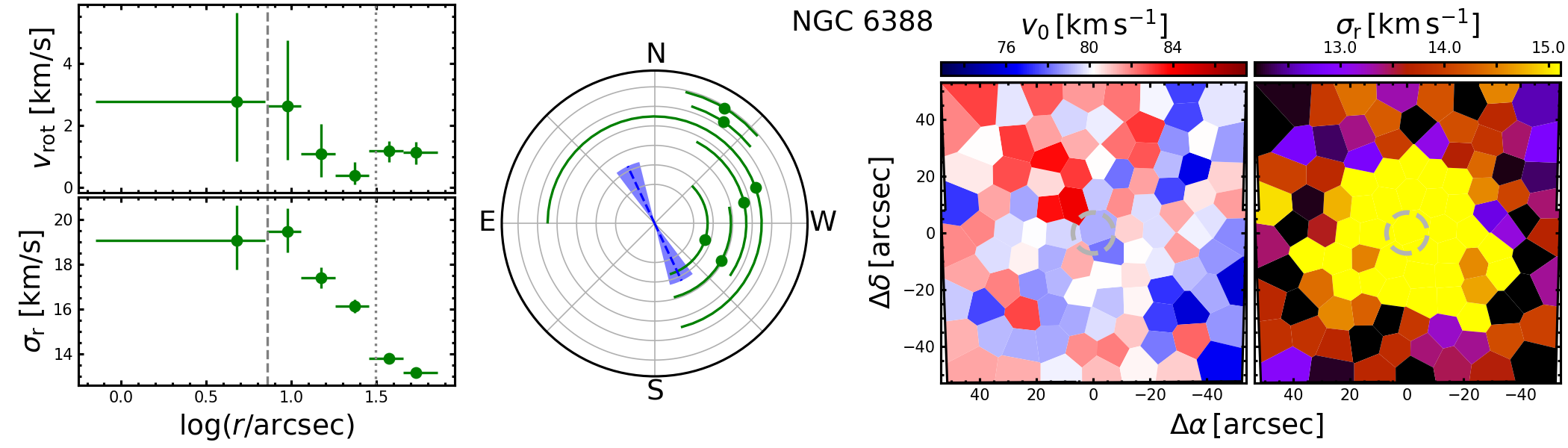}
 \contcaption{As Fig.~\ref{fig:example_kinematics} but for NGC~6254, NGC~6266, NGC~6293, and NGC~6388}
 \label{fig:app:kinematics:cont2}
\end{figure*}
\clearpage

\begin{figure*}
 \includegraphics[width=\textwidth]{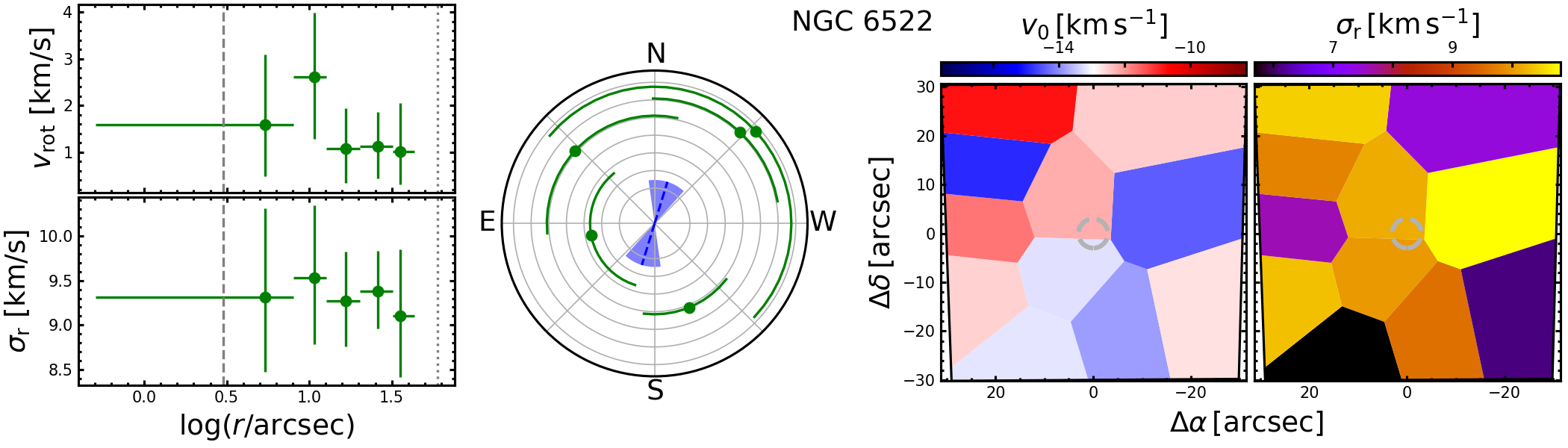}
 \includegraphics[width=\textwidth]{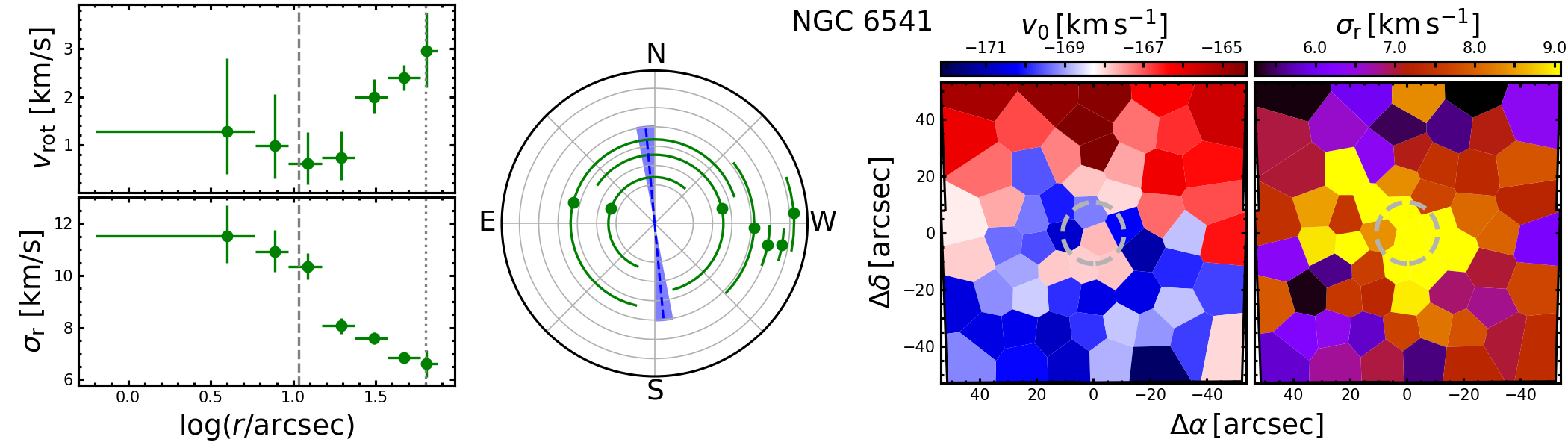}
 \includegraphics[width=\textwidth]{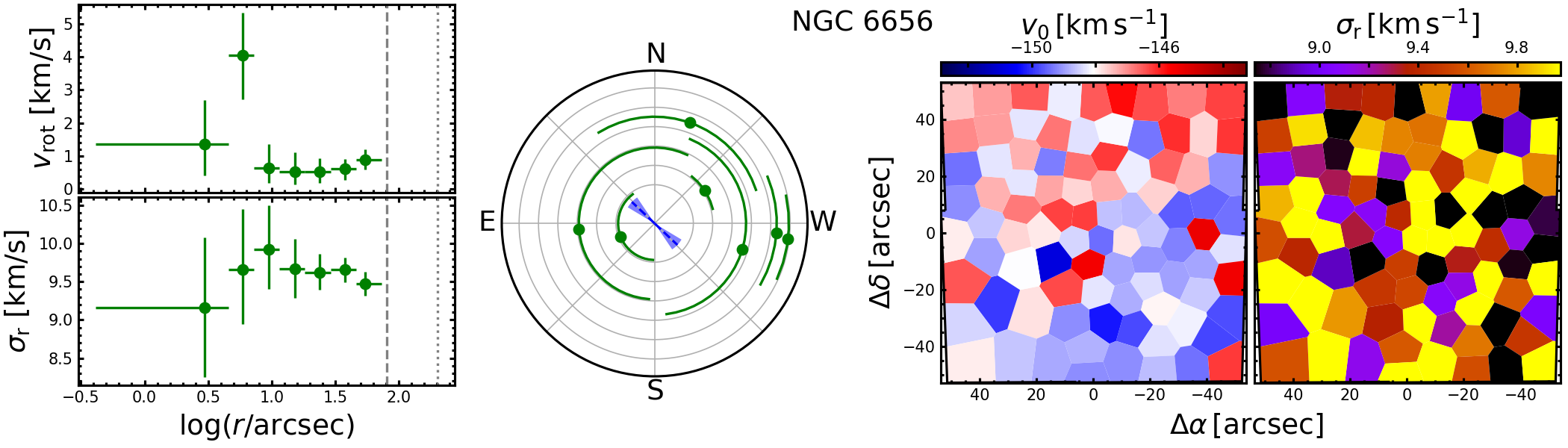}
 \includegraphics[width=\textwidth]{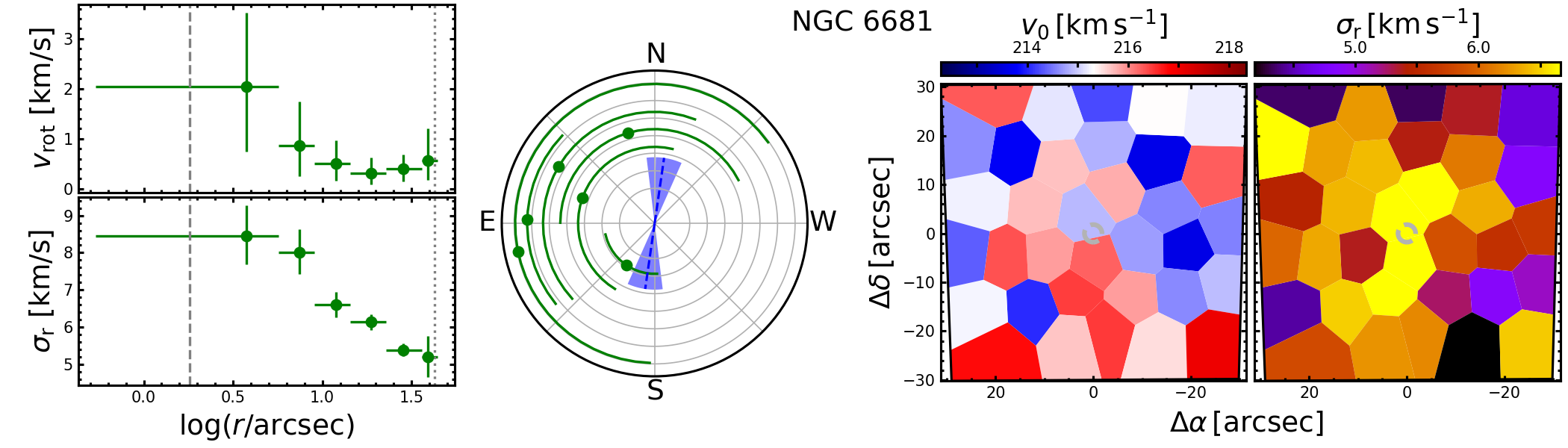}
 \contcaption{As Fig.~\ref{fig:example_kinematics} but for NGC~6522, NGC~6541, NGC~6656, and NGC~6681}
 \label{fig:app:kinematics:cont3}
\end{figure*}
\clearpage

\begin{figure*}
 \includegraphics[width=\textwidth]{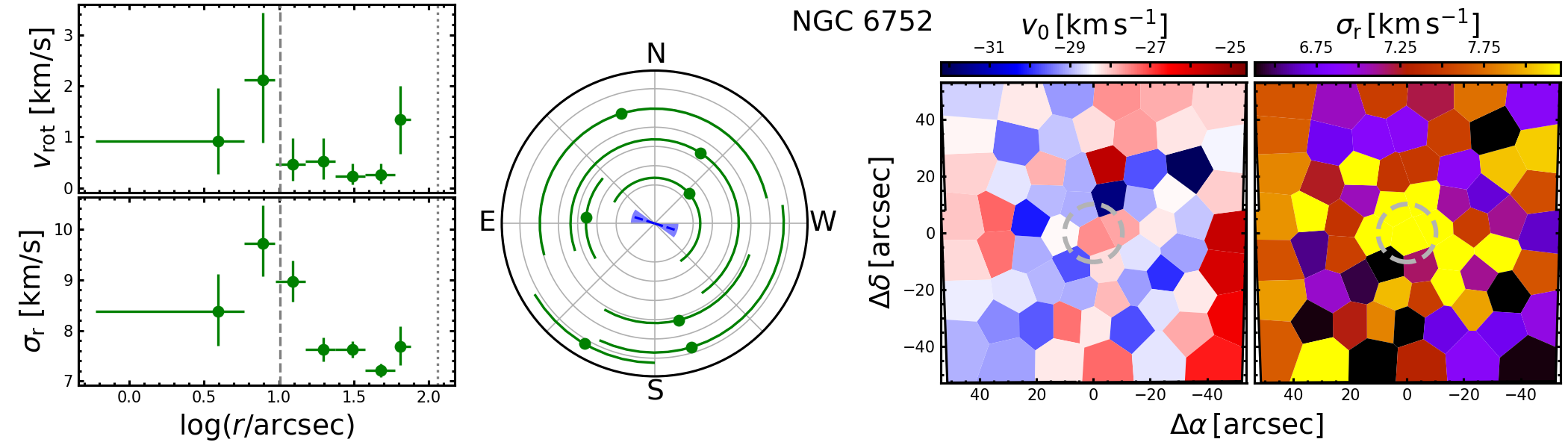}
 \includegraphics[width=\textwidth]{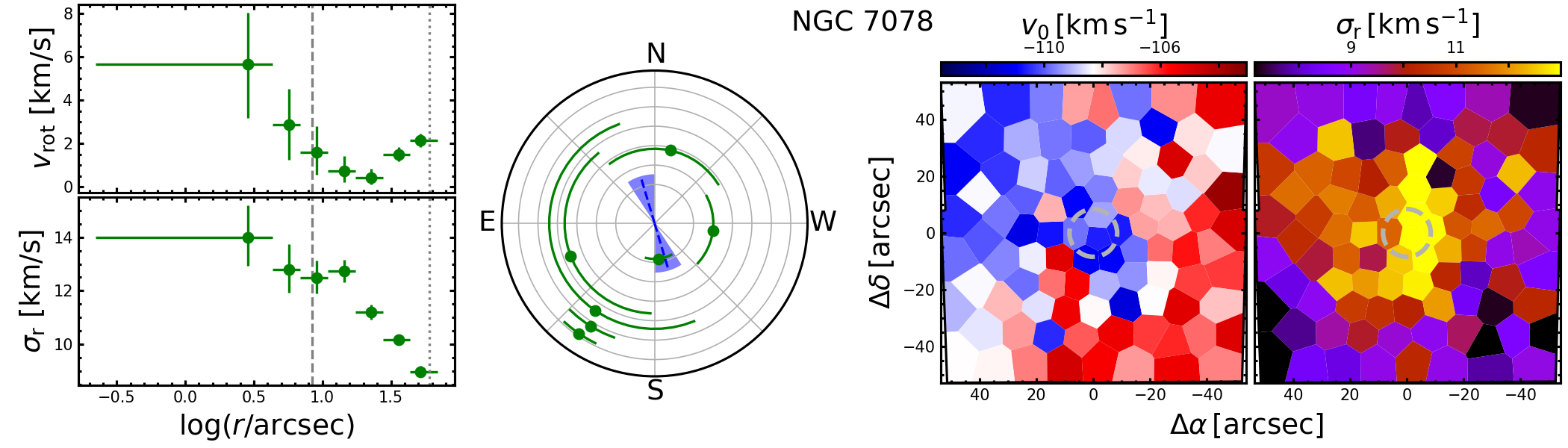}
 \includegraphics[width=\textwidth]{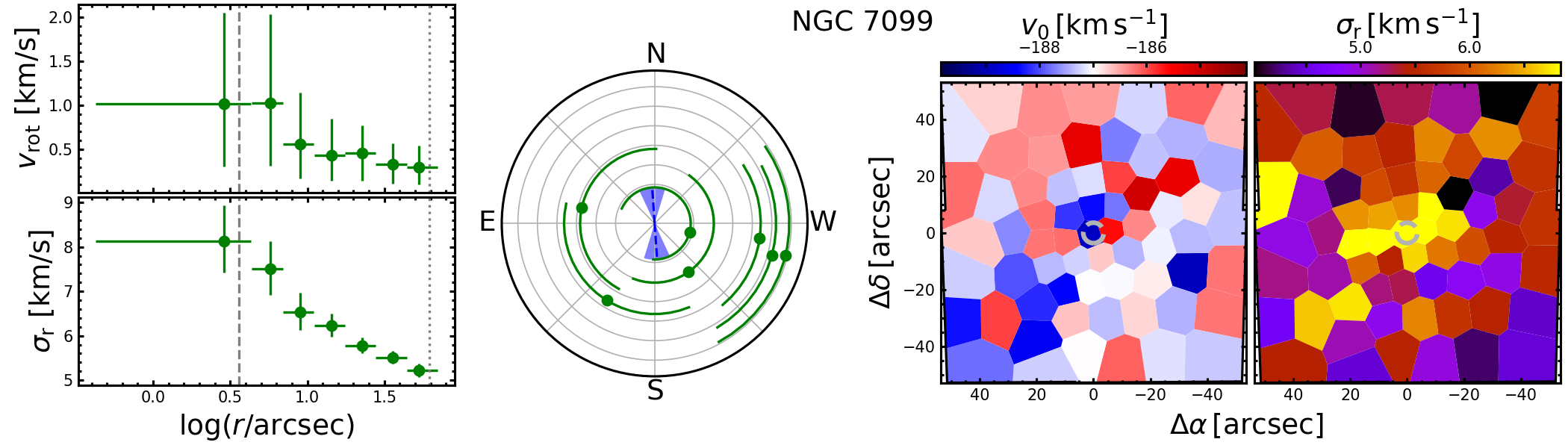}
 \contcaption{As Fig.~\ref{fig:example_kinematics} but for NGC~6752, NGC~7078, and NGC~7099}
 \label{fig:app:kinematics:cont4}
\end{figure*}
\clearpage

\section{Radial profiles data}
\label{sec:app:radial_profiles}

\begin{table}
 \caption{Radial profiles of the cluster dynamics from the MUSE sample. For each bin, $r$ indicates the mean distance of the stars to the cluster centre while the associated uncertainties give its full radial extent. Otherwise, we provide the median and the 16th and 84th percentiles of the parameter distributions obtained from the MCMC chains. The effective masses were determined as outlined in Sect.~\ref{sec:dispersions}.}
 \label{tab:app:radial_profiles}
 \begin{tabular}{ r | c | c | c | c | c }
\hline
 NGC & $r$ & $\theta_{\rm 0}$ & $v_{\rm rot}$ & $\sigma_{\rm r}$ & $M_{\rm eff.}$ \\
 & arcsec & rad. & ${\rm km\,s^{-1}}$ & ${\rm km\,s^{-1}}$ & $M_{\rm \odot}$ \\ \hline
 104 & $4.2^{+2.4}_{-3.8}$ & $2.63^{+0.85}_{-0.97}$ & $2.18^{+1.77}_{-1.43}$ & $12.37^{+1.05}_{-0.89}$ & $0.81$ \\
 104 & $8.9^{+1.6}_{-2.3}$ & $2.88^{+0.53}_{-0.48}$ & $3.50^{+1.69}_{-1.72}$ & $13.10^{+0.96}_{-0.81}$ & $0.81$ \\
 104 & $13.6^{+3.0}_{-3.2}$ & $2.07^{+1.08}_{-1.25}$ & $0.96^{+0.91}_{-0.68}$ & $12.29^{+0.53}_{-0.51}$ & $0.81$ \\
 104 & $21.9^{+4.4}_{-5.3}$ & $2.58^{+0.46}_{-0.51}$ & $1.29^{+0.60}_{-0.66}$ & $12.23^{+0.31}_{-0.31}$ & $0.81$ \\
 104 & $34.7^{+7.0}_{-8.4}$ & $2.26^{+0.15}_{-0.14}$ & $2.35^{+0.36}_{-0.35}$ & $12.02^{+0.18}_{-0.18}$ & $0.81$ \\
 104 & $53.6^{+12.6}_{-11.8}$ & $2.25^{+0.07}_{-0.08}$ & $3.22^{+0.24}_{-0.24}$ & $11.41^{+0.13}_{-0.13}$ & $0.80$ \\
 104 & $85.1^{+19.8}_{-18.9}$ & $2.49^{+0.09}_{-0.10}$ & $3.94^{+0.29}_{-0.26}$ & $11.30^{+0.12}_{-0.11}$ & $0.75$ \\
 104 & $114.4^{+24.4}_{-9.5}$ & $2.36^{+0.36}_{-0.44}$ & $4.03^{+0.70}_{-0.41}$ & $10.54^{+0.20}_{-0.20}$ & $0.72$ \\
 362 & $3.8^{+2.2}_{-3.2}$ & $1.90^{+0.59}_{-0.52}$ & $2.26^{+1.18}_{-1.21}$ & $7.66^{+0.69}_{-0.58}$ & $0.81$ \\
 362 & $7.8^{+1.8}_{-1.8}$ & $2.35^{+0.73}_{-0.74}$ & $1.56^{+1.16}_{-1.02}$ & $9.03^{+0.66}_{-0.64}$ & $0.81$ \\
 362 & $12.6^{+2.6}_{-3.0}$ & $2.26^{+0.73}_{-0.70}$ & $1.29^{+0.88}_{-0.81}$ & $9.34^{+0.48}_{-0.43}$ & $0.81$ \\
 362 & $20.0^{+4.2}_{-4.8}$ & $1.55^{+1.42}_{-2.22}$ & $0.39^{+0.44}_{-0.28}$ & $8.39^{+0.30}_{-0.29}$ & $0.81$ \\
 362 & $31.6^{+6.7}_{-7.5}$ & $-0.85^{+0.30}_{-0.30}$ & $1.16^{+0.35}_{-0.35}$ & $7.77^{+0.21}_{-0.21}$ & $0.80$ \\
 362 & $48.0^{+12.8}_{-9.6}$ & $-1.52^{+0.84}_{-0.80}$ & $0.36^{+0.30}_{-0.24}$ & $7.31^{+0.20}_{-0.19}$ & $0.78$ \\
 362 & $64.8^{+10.6}_{-4.1}$ & $-0.50^{+1.67}_{-1.54}$ & $0.65^{+0.68}_{-0.45}$ & $6.57^{+0.63}_{-0.59}$ & $0.79$ \\
 1851 & $4.0^{+1.5}_{-3.5}$ & $-1.35^{+0.56}_{-0.60}$ & $2.54^{+1.43}_{-1.40}$ & $9.09^{+0.81}_{-0.70}$ & $0.85$ \\
 1851 & $7.2^{+1.6}_{-1.7}$ & $-0.83^{+2.89}_{-1.86}$ & $0.87^{+0.96}_{-0.63}$ & $10.39^{+0.71}_{-0.65}$ & $0.85$ \\
 1851 & $11.8^{+2.3}_{-2.9}$ & $-0.54^{+1.04}_{-0.98}$ & $0.88^{+0.82}_{-0.60}$ & $9.54^{+0.46}_{-0.42}$ & $0.84$ \\
 1851 & $18.5^{+3.8}_{-4.4}$ & $0.12^{+0.48}_{-0.47}$ & $1.03^{+0.47}_{-0.53}$ & $8.60^{+0.26}_{-0.26}$ & $0.84$ \\
 1851 & $29.4^{+6.0}_{-7.1}$ & $0.23^{+0.15}_{-0.15}$ & $1.97^{+0.29}_{-0.30}$ & $8.24^{+0.17}_{-0.18}$ & $0.82$ \\
 1851 & $45.5^{+10.7}_{-10.1}$ & $-0.06^{+0.15}_{-0.15}$ & $1.40^{+0.21}_{-0.21}$ & $7.17^{+0.12}_{-0.12}$ & $0.80$ \\
 1851 & $61.7^{+14.9}_{-5.4}$ & $0.31^{+0.56}_{-0.58}$ & $0.91^{+0.49}_{-0.50}$ & $6.39^{+0.28}_{-0.26}$ & $0.80$ \\
 2808 & $4.8^{+2.4}_{-4.1}$ & $0.21^{+2.27}_{-2.50}$ & $1.40^{+1.57}_{-0.97}$ & $13.94^{+1.12}_{-1.00}$ & $0.83$ \\
 2808 & $9.6^{+1.9}_{-2.4}$ & $-0.03^{+1.47}_{-1.43}$ & $1.49^{+1.57}_{-1.03}$ & $13.60^{+0.92}_{-0.89}$ & $0.83$ \\
 2808 & $15.2^{+3.0}_{-3.7}$ & $-0.73^{+0.34}_{-0.35}$ & $3.06^{+1.07}_{-1.15}$ & $13.33^{+0.56}_{-0.53}$ & $0.83$ \\
 2808 & $23.8^{+5.0}_{-5.6}$ & $-0.67^{+0.29}_{-0.27}$ & $2.35^{+0.62}_{-0.63}$ & $13.43^{+0.35}_{-0.35}$ & $0.83$ \\
 2808 & $37.8^{+7.9}_{-9.0}$ & $-0.91^{+0.12}_{-0.12}$ & $3.50^{+0.41}_{-0.42}$ & $12.50^{+0.24}_{-0.22}$ & $0.83$ \\
 2808 & $54.0^{+18.4}_{-8.3}$ & $-0.79^{+0.11}_{-0.11}$ & $3.86^{+0.37}_{-0.38}$ & $11.33^{+0.22}_{-0.22}$ & $0.82$ \\
 3201 & $5.8^{+3.1}_{-5.3}$ & $2.44^{+1.08}_{-1.47}$ & $0.76^{+0.72}_{-0.54}$ & $5.36^{+0.50}_{-0.44}$ & $0.72$ \\
 3201 & $11.8^{+2.4}_{-2.8}$ & $0.01^{+1.54}_{-1.59}$ & $0.44^{+0.49}_{-0.31}$ & $4.60^{+0.33}_{-0.30}$ & $0.71$ \\
 3201 & $18.7^{+4.0}_{-4.4}$ & $-1.00^{+0.58}_{-0.54}$ & $0.70^{+0.39}_{-0.41}$ & $4.54^{+0.21}_{-0.21}$ & $0.70$ \\
 3201 & $29.7^{+6.3}_{-7.0}$ & $1.16^{+1.57}_{-2.16}$ & $0.18^{+0.19}_{-0.13}$ & $4.63^{+0.14}_{-0.13}$ & $0.70$ \\
 3201 & $46.5^{+10.6}_{-10.5}$ & $2.97^{+0.57}_{-0.63}$ & $0.32^{+0.18}_{-0.18}$ & $4.62^{+0.11}_{-0.10}$ & $0.70$ \\
 3201 & $63.0^{+14.9}_{-6.0}$ & $0.27^{+1.93}_{-2.20}$ & $0.26^{+0.29}_{-0.18}$ & $4.81^{+0.21}_{-0.21}$ & $0.71$ \\
 5139 & $3.9^{+1.6}_{-3.4}$ & $-2.58^{+1.15}_{-0.85}$ & $3.67^{+3.58}_{-2.56}$ & $22.82^{+1.97}_{-1.74}$ & $0.73$ \\
 5139 & $7.1^{+1.6}_{-1.6}$ & $-2.82^{+0.83}_{-0.73}$ & $3.02^{+2.24}_{-1.98}$ & $18.23^{+1.25}_{-1.15}$ & $0.74$ \\
 5139 & $11.4^{+2.4}_{-2.7}$ & $1.83^{+0.48}_{-0.46}$ & $3.20^{+1.38}_{-1.48}$ & $17.57^{+0.79}_{-0.74}$ & $0.74$ \\
 5139 & $18.2^{+3.7}_{-4.3}$ & $-0.20^{+0.25}_{-0.25}$ & $4.26^{+0.95}_{-0.98}$ & $18.22^{+0.53}_{-0.53}$ & $0.74$ \\
 5139 & $28.8^{+5.9}_{-6.9}$ & $0.17^{+0.45}_{-0.48}$ & $1.52^{+0.64}_{-0.74}$ & $18.41^{+0.33}_{-0.34}$ & $0.74$ \\
 5139 & $45.4^{+9.6}_{-10.7}$ & $0.34^{+0.12}_{-0.13}$ & $3.16^{+0.36}_{-0.36}$ & $17.98^{+0.22}_{-0.22}$ & $0.74$ \\
 5139 & $68.9^{+18.3}_{-13.9}$ & $0.15^{+0.22}_{-0.23}$ & $2.79^{+0.32}_{-0.30}$ & $18.09^{+0.21}_{-0.21}$ & $0.73$ \\
 5139 & $110.9^{+27.5}_{-23.6}$ & $-0.23^{+0.32}_{-0.23}$ & $4.27^{+0.95}_{-0.63}$ & $16.61^{+0.24}_{-0.22}$ & $0.72$ \\
 5139 & $177.1^{+42.3}_{-38.7}$ & $0.13^{+0.12}_{-0.15}$ & $4.33^{+0.38}_{-0.36}$ & $14.99^{+0.19}_{-0.18}$ & $0.69$ \\
 5139 & $267.5^{+57.8}_{-48.2}$ & $0.07^{+0.44}_{-0.11}$ & $7.21^{+2.90}_{-2.85}$ & $13.26^{+0.18}_{-0.17}$ & $0.67$ \\
 5904 & $3.4^{+1.3}_{-2.8}$ & $-2.87^{+0.40}_{-0.39}$ & $3.22^{+1.23}_{-1.22}$ & $7.33^{+0.81}_{-0.70}$ & $0.83$ \\
 5904 & $6.1^{+1.4}_{-1.4}$ & $-2.47^{+0.88}_{-0.90}$ & $1.46^{+1.13}_{-0.99}$ & $8.25^{+0.77}_{-0.70}$ & $0.83$ \\
 5904 & $9.9^{+2.0}_{-2.4}$ & $-1.05^{+0.38}_{-0.37}$ & $2.16^{+0.78}_{-0.83}$ & $7.91^{+0.49}_{-0.46}$ & $0.83$ \\
 5904 & $15.6^{+3.2}_{-3.7}$ & $-0.80^{+1.36}_{-1.21}$ & $0.48^{+0.50}_{-0.34}$ & $8.02^{+0.35}_{-0.32}$ & $0.82$ \\
 5904 & $24.6^{+5.2}_{-5.8}$ & $-0.51^{+0.21}_{-0.21}$ & $1.50^{+0.32}_{-0.31}$ & $6.95^{+0.21}_{-0.20}$ & $0.81$ \\
 5904 & $38.9^{+8.3}_{-9.2}$ & $-0.86^{+0.16}_{-0.15}$ & $1.51^{+0.22}_{-0.22}$ & $6.84^{+0.15}_{-0.14}$ & $0.79$ \\
 5904 & $55.1^{+20.2}_{-7.8}$ & $-1.06^{+0.15}_{-0.14}$ & $1.77^{+0.25}_{-0.26}$ & $6.79^{+0.16}_{-0.16}$ & $0.78$ \\
\hline
\end{tabular}

\end{table}

\begin{table}
 \contcaption{}
 \label{tab:app:radial_profiles_continued_1}
 \begin{tabular}{ r | c | c | c | c | c }
\hline
 NGC & $r$ & $\theta_{\rm 0}$ & $v_{\rm rot}$ & $\sigma_{\rm r}$ & $M_{\rm eff.}$ \\
 & arcsec & rad. & ${\rm km\,s^{-1}}$ & ${\rm km\,s^{-1}}$ & $M_{\rm \odot}$ \\ \hline
 6093 & $3.7^{+1.7}_{-3.0}$ & $0.22^{+0.75}_{-0.79}$ & $2.27^{+1.61}_{-1.49}$ & $10.47^{+0.86}_{-0.77}$ & $0.79$ \\
 6093 & $6.9^{+1.8}_{-1.5}$ & $2.24^{+1.14}_{-1.58}$ & $1.22^{+1.23}_{-0.86}$ & $10.42^{+0.84}_{-0.80}$ & $0.79$ \\
 6093 & $11.5^{+2.3}_{-2.8}$ & $0.41^{+2.13}_{-2.54}$ & $0.80^{+0.89}_{-0.56}$ & $13.16^{+0.66}_{-0.61}$ & $0.79$ \\
 6093 & $18.1^{+3.9}_{-4.2}$ & $-2.52^{+0.31}_{-0.31}$ & $2.04^{+0.72}_{-0.74}$ & $9.98^{+0.41}_{-0.37}$ & $0.79$ \\
 6093 & $28.4^{+6.4}_{-6.5}$ & $-2.63^{+0.28}_{-0.28}$ & $1.65^{+0.51}_{-0.52}$ & $9.22^{+0.27}_{-0.28}$ & $0.78$ \\
 6093 & $44.5^{+10.7}_{-9.7}$ & $-2.27^{+0.21}_{-0.21}$ & $1.96^{+0.41}_{-0.45}$ & $8.29^{+0.26}_{-0.25}$ & $0.78$ \\
 6093 & $60.5^{+12.9}_{-5.3}$ & $-1.73^{+0.34}_{-0.31}$ & $2.88^{+1.03}_{-1.02}$ & $8.21^{+0.62}_{-0.58}$ & $0.78$ \\
 6121 & $9.2^{+5.3}_{-8.2}$ & $-2.06^{+0.81}_{-0.66}$ & $1.71^{+1.17}_{-1.06}$ & $6.85^{+0.72}_{-0.64}$ & $0.74$ \\
 6121 & $19.2^{+3.8}_{-4.7}$ & $-1.86^{+3.23}_{-1.29}$ & $0.77^{+0.83}_{-0.52}$ & $6.57^{+0.69}_{-0.61}$ & $0.73$ \\
 6121 & $30.2^{+6.4}_{-7.1}$ & $-2.66^{+0.69}_{-0.69}$ & $1.07^{+0.75}_{-0.67}$ & $6.17^{+0.41}_{-0.40}$ & $0.73$ \\
 6121 & $46.8^{+11.4}_{-10.2}$ & $-2.98^{+0.86}_{-0.67}$ & $0.71^{+0.52}_{-0.43}$ & $5.99^{+0.31}_{-0.31}$ & $0.73$ \\
 6254 & $3.1^{+1.5}_{-2.8}$ & $2.51^{+0.42}_{-0.40}$ & $2.48^{+1.02}_{-1.16}$ & $5.60^{+0.66}_{-0.59}$ & $0.73$ \\
 6254 & $6.0^{+1.3}_{-1.4}$ & $-2.40^{+0.73}_{-0.71}$ & $1.32^{+0.89}_{-0.84}$ & $6.41^{+0.50}_{-0.48}$ & $0.75$ \\
 6254 & $9.6^{+2.0}_{-2.3}$ & $2.04^{+0.59}_{-0.54}$ & $1.25^{+0.68}_{-0.68}$ & $6.82^{+0.43}_{-0.38}$ & $0.74$ \\
 6254 & $15.2^{+3.2}_{-3.6}$ & $-1.18^{+2.36}_{-1.63}$ & $0.32^{+0.32}_{-0.23}$ & $6.36^{+0.27}_{-0.26}$ & $0.75$ \\
 6254 & $24.2^{+5.1}_{-5.8}$ & $2.62^{+0.61}_{-0.61}$ & $0.52^{+0.30}_{-0.30}$ & $6.33^{+0.18}_{-0.18}$ & $0.75$ \\
 6254 & $38.2^{+8.2}_{-8.9}$ & $2.12^{+0.49}_{-0.50}$ & $0.42^{+0.21}_{-0.22}$ & $5.87^{+0.12}_{-0.12}$ & $0.73$ \\
 6254 & $54.3^{+19.6}_{-7.9}$ & $-1.82^{+1.75}_{-1.27}$ & $0.17^{+0.19}_{-0.12}$ & $5.75^{+0.14}_{-0.13}$ & $0.74$ \\
 6266 & $3.4^{+1.6}_{-3.1}$ & $0.94^{+0.46}_{-0.46}$ & $5.07^{+2.27}_{-2.49}$ & $15.00^{+1.23}_{-1.03}$ & $0.83$ \\
 6266 & $6.6^{+1.4}_{-1.5}$ & $-1.60^{+0.75}_{-0.77}$ & $2.70^{+2.04}_{-1.76}$ & $16.08^{+1.01}_{-0.94}$ & $0.83$ \\
 6266 & $10.6^{+2.2}_{-2.6}$ & $2.09^{+1.01}_{-4.41}$ & $1.25^{+1.10}_{-0.86}$ & $14.58^{+0.61}_{-0.57}$ & $0.83$ \\
 6266 & $16.7^{+3.5}_{-3.9}$ & $1.22^{+1.19}_{-1.28}$ & $0.73^{+0.71}_{-0.51}$ & $14.65^{+0.39}_{-0.40}$ & $0.82$ \\
 6266 & $26.7^{+5.4}_{-6.4}$ & $1.89^{+0.66}_{-0.65}$ & $0.80^{+0.55}_{-0.49}$ & $14.17^{+0.26}_{-0.25}$ & $0.82$ \\
 6266 & $42.0^{+9.0}_{-9.8}$ & $1.51^{+0.31}_{-0.33}$ & $0.88^{+0.29}_{-0.29}$ & $12.66^{+0.16}_{-0.16}$ & $0.80$ \\
 6266 & $57.4^{+19.3}_{-6.5}$ & $1.89^{+0.28}_{-0.27}$ & $1.44^{+0.41}_{-0.42}$ & $12.00^{+0.23}_{-0.23}$ & $0.78$ \\
 6293 & $6.1^{+2.9}_{-4.9}$ & $-0.52^{+0.56}_{-0.48}$ & $2.75^{+1.31}_{-1.34}$ & $7.13^{+0.74}_{-0.69}$ & $0.77$ \\
 6293 & $11.8^{+2.7}_{-2.8}$ & $-1.14^{+0.95}_{-0.81}$ & $1.69^{+1.38}_{-1.15}$ & $7.77^{+0.78}_{-0.74}$ & $0.77$ \\
 6293 & $18.8^{+4.2}_{-4.3}$ & $2.09^{+1.10}_{-1.42}$ & $1.02^{+1.07}_{-0.70}$ & $6.94^{+0.70}_{-0.63}$ & $0.77$ \\
 6293 & $28.2^{+8.3}_{-5.2}$ & $-1.18^{+0.28}_{-0.28}$ & $2.93^{+0.73}_{-0.79}$ & $5.41^{+0.54}_{-0.46}$ & $0.77$ \\
 6388 & $4.8^{+2.2}_{-4.1}$ & $-1.89^{+1.08}_{-0.94}$ & $2.77^{+2.85}_{-1.92}$ & $19.06^{+1.55}_{-1.30}$ & $0.91$ \\
 6388 & $9.5^{+1.8}_{-2.4}$ & $-2.08^{+0.71}_{-0.79}$ & $2.63^{+2.12}_{-1.74}$ & $19.44^{+1.03}_{-0.93}$ & $0.91$ \\
 6388 & $15.0^{+3.1}_{-3.6}$ & $-1.33^{+0.85}_{-0.83}$ & $1.09^{+0.95}_{-0.75}$ & $17.38^{+0.46}_{-0.45}$ & $0.91$ \\
 6388 & $23.5^{+5.1}_{-5.5}$ & $-1.23^{+2.79}_{-1.65}$ & $0.39^{+0.42}_{-0.27}$ & $16.11^{+0.29}_{-0.29}$ & $0.91$ \\
 6388 & $37.3^{+8.0}_{-8.8}$ & $-0.60^{+0.28}_{-0.29}$ & $1.17^{+0.33}_{-0.35}$ & $13.78^{+0.18}_{-0.18}$ & $0.89$ \\
 6388 & $53.5^{+18.6}_{-8.2}$ & $-0.54^{+0.31}_{-0.31}$ & $1.13^{+0.35}_{-0.37}$ & $13.16^{+0.20}_{-0.19}$ & $0.88$ \\
 6441 & $3.1^{+1.8}_{-2.9}$ & $-0.73^{+1.90}_{-1.64}$ & $1.49^{+1.72}_{-1.02}$ & $14.30^{+1.12}_{-0.98}$ & $0.93$ \\
 6441 & $6.4^{+1.4}_{-1.5}$ & $-0.70^{+3.09}_{-2.03}$ & $1.85^{+1.75}_{-1.27}$ & $19.13^{+1.21}_{-1.11}$ & $0.93$ \\
 6441 & $10.3^{+2.2}_{-2.4}$ & $-0.53^{+0.34}_{-0.34}$ & $4.01^{+1.43}_{-1.52}$ & $18.23^{+0.72}_{-0.68}$ & $0.93$ \\
 6441 & $16.4^{+3.4}_{-3.9}$ & $0.80^{+1.80}_{-2.33}$ & $0.51^{+0.59}_{-0.36}$ & $17.13^{+0.40}_{-0.39}$ & $0.93$ \\
 6441 & $25.8^{+5.5}_{-6.0}$ & $2.38^{+1.03}_{-2.87}$ & $0.48^{+0.48}_{-0.34}$ & $16.33^{+0.29}_{-0.28}$ & $0.92$ \\
 6441 & $40.7^{+9.0}_{-9.3}$ & $2.51^{+0.99}_{-1.25}$ & $0.37^{+0.34}_{-0.25}$ & $14.96^{+0.19}_{-0.19}$ & $0.92$ \\
 6441 & $56.4^{+17.9}_{-6.7}$ & $-1.97^{+4.58}_{-1.07}$ & $0.65^{+0.50}_{-0.43}$ & $14.31^{+0.26}_{-0.26}$ & $0.91$ \\
 6522 & $5.4^{+2.6}_{-4.9}$ & $1.76^{+1.08}_{-1.07}$ & $1.58^{+1.52}_{-1.11}$ & $9.31^{+1.00}_{-0.84}$ & $0.79$ \\
 6522 & $10.7^{+2.0}_{-2.7}$ & $-2.75^{+0.52}_{-0.50}$ & $2.61^{+1.37}_{-1.33}$ & $9.53^{+0.81}_{-0.74}$ & $0.79$ \\
 6522 & $16.7^{+3.5}_{-3.9}$ & $0.83^{+0.83}_{-1.04}$ & $1.07^{+0.88}_{-0.73}$ & $9.27^{+0.56}_{-0.51}$ & $0.79$ \\
 6522 & $26.1^{+6.0}_{-5.8}$ & $-0.75^{+0.77}_{-0.64}$ & $1.13^{+0.73}_{-0.70}$ & $9.38^{+0.45}_{-0.42}$ & $0.78$ \\
 6522 & $35.5^{+7.8}_{-3.4}$ & $-0.83^{+1.70}_{-1.49}$ & $1.00^{+1.04}_{-0.69}$ & $9.10^{+0.75}_{-0.68}$ & $0.78$ \\
 6541 & $4.0^{+1.9}_{-3.3}$ & $1.24^{+1.52}_{-1.94}$ & $1.28^{+1.53}_{-0.89}$ & $11.53^{+1.15}_{-1.04}$ & $0.78$ \\
 6541 & $7.8^{+1.6}_{-1.9}$ & $-1.35^{+2.32}_{-1.51}$ & $0.98^{+1.07}_{-0.68}$ & $10.92^{+0.82}_{-0.79}$ & $0.78$ \\
 6541 & $12.2^{+2.6}_{-2.9}$ & $1.31^{+1.60}_{-2.55}$ & $0.61^{+0.65}_{-0.43}$ & $10.34^{+0.52}_{-0.47}$ & $0.77$ \\
 6541 & $19.5^{+4.1}_{-4.6}$ & $-1.61^{+0.69}_{-0.73}$ & $0.73^{+0.54}_{-0.46}$ & $8.07^{+0.30}_{-0.30}$ & $0.76$ \\
 6541 & $30.8^{+6.6}_{-7.2}$ & $-1.76^{+0.16}_{-0.18}$ & $2.00^{+0.36}_{-0.35}$ & $7.59^{+0.23}_{-0.23}$ & $0.75$ \\
 6541 & $47.0^{+12.2}_{-9.6}$ & $-1.74^{+0.12}_{-0.12}$ & $2.40^{+0.26}_{-0.27}$ & $6.84^{+0.18}_{-0.18}$ & $0.74$ \\
 6541 & $63.8^{+11.1}_{-4.6}$ & $-1.49^{+0.25}_{-0.27}$ & $2.95^{+0.79}_{-0.76}$ & $6.62^{+0.52}_{-0.50}$ & $0.74$ \\
\hline
\end{tabular}

\end{table}

\begin{table}
 \contcaption{}
 \label{tab:app:radial_profiles_continued_2}
 \begin{tabular}{ r | c | c | c | c | c }
\hline
 NGC & $r$ & $\theta_{\rm 0}$ & $v_{\rm rot}$ & $\sigma_{\rm r}$ & $M_{\rm eff.}$ \\
 & arcsec & rad. & ${\rm km\,s^{-1}}$ & ${\rm km\,s^{-1}}$ & $M_{\rm \odot}$ \\ \hline
 6656 & $3.0^{+1.6}_{-2.5}$ & $1.94^{+1.16}_{-1.29}$ & $1.37^{+1.34}_{-0.95}$ & $9.16^{+0.92}_{-0.91}$ & $0.68$ \\
 6656 & $5.9^{+1.3}_{-1.4}$ & $-0.99^{+0.32}_{-0.33}$ & $4.05^{+1.27}_{-1.34}$ & $9.66^{+0.79}_{-0.71}$ & $0.69$ \\
 6656 & $9.5^{+2.0}_{-2.3}$ & $1.65^{+1.42}_{-2.09}$ & $0.64^{+0.72}_{-0.45}$ & $9.92^{+0.58}_{-0.51}$ & $0.69$ \\
 6656 & $15.2^{+3.1}_{-3.7}$ & $-1.86^{+1.46}_{-1.15}$ & $0.53^{+0.59}_{-0.38}$ & $9.66^{+0.39}_{-0.38}$ & $0.70$ \\
 6656 & $23.9^{+5.0}_{-5.6}$ & $-0.33^{+0.89}_{-0.92}$ & $0.54^{+0.40}_{-0.35}$ & $9.62^{+0.25}_{-0.23}$ & $0.70$ \\
 6656 & $37.8^{+8.1}_{-8.9}$ & $-1.65^{+0.47}_{-0.45}$ & $0.61^{+0.31}_{-0.32}$ & $9.66^{+0.16}_{-0.17}$ & $0.69$ \\
 6656 & $54.2^{+18.6}_{-8.3}$ & $-1.69^{+0.32}_{-0.31}$ & $0.89^{+0.30}_{-0.30}$ & $9.48^{+0.16}_{-0.16}$ & $0.69$ \\
 6681 & $3.7^{+1.9}_{-3.2}$ & $2.54^{+0.65}_{-0.75}$ & $2.04^{+1.47}_{-1.30}$ & $8.45^{+0.81}_{-0.75}$ & $0.78$ \\
 6681 & $7.5^{+1.5}_{-1.8}$ & $1.23^{+1.36}_{-1.47}$ & $0.86^{+0.88}_{-0.62}$ & $8.01^{+0.62}_{-0.59}$ & $0.77$ \\
 6681 & $11.9^{+2.4}_{-2.8}$ & $0.29^{+1.27}_{-1.38}$ & $0.50^{+0.47}_{-0.35}$ & $6.60^{+0.35}_{-0.33}$ & $0.76$ \\
 6681 & $18.7^{+4.1}_{-4.3}$ & $1.04^{+1.27}_{-1.41}$ & $0.31^{+0.32}_{-0.22}$ & $6.14^{+0.21}_{-0.21}$ & $0.74$ \\
 6681 & $28.3^{+7.7}_{-5.6}$ & $1.54^{+0.72}_{-0.73}$ & $0.40^{+0.29}_{-0.26}$ & $5.39^{+0.17}_{-0.16}$ & $0.73$ \\
 6681 & $38.8^{+5.4}_{-2.8}$ & $1.77^{+1.33}_{-2.72}$ & $0.57^{+0.64}_{-0.40}$ & $5.20^{+0.57}_{-0.53}$ & $0.73$ \\
 6752 & $3.9^{+2.0}_{-3.3}$ & $-0.85^{+1.91}_{-1.67}$ & $0.92^{+1.03}_{-0.65}$ & $8.37^{+0.74}_{-0.67}$ & $0.80$ \\
 6752 & $7.8^{+1.5}_{-1.9}$ & $1.48^{+0.61}_{-0.61}$ & $2.11^{+1.31}_{-1.23}$ & $9.72^{+0.74}_{-0.64}$ & $0.79$ \\
 6752 & $12.4^{+2.6}_{-2.9}$ & $-0.57^{+2.46}_{-1.96}$ & $0.46^{+0.51}_{-0.33}$ & $8.97^{+0.42}_{-0.39}$ & $0.77$ \\
 6752 & $19.7^{+4.1}_{-4.7}$ & $-2.90^{+1.00}_{-0.76}$ & $0.53^{+0.44}_{-0.35}$ & $7.62^{+0.25}_{-0.23}$ & $0.75$ \\
 6752 & $31.0^{+6.7}_{-7.2}$ & $0.30^{+1.54}_{-1.64}$ & $0.23^{+0.24}_{-0.16}$ & $7.62^{+0.17}_{-0.16}$ & $0.74$ \\
 6752 & $47.6^{+12.2}_{-9.9}$ & $-2.85^{+1.42}_{-0.72}$ & $0.26^{+0.22}_{-0.18}$ & $7.22^{+0.13}_{-0.13}$ & $0.72$ \\
 6752 & $64.5^{+11.4}_{-4.7}$ & $2.61^{+0.52}_{-0.50}$ & $1.34^{+0.66}_{-0.67}$ & $7.68^{+0.41}_{-0.37}$ & $0.72$ \\
 7078 & $2.9^{+1.5}_{-2.6}$ & $-3.03^{+0.38}_{-0.35}$ & $5.66^{+2.35}_{-2.48}$ & $14.00^{+1.19}_{-1.06}$ & $0.76$ \\
 7078 & $5.7^{+1.2}_{-1.4}$ & $-1.69^{+0.61}_{-0.62}$ & $2.87^{+1.64}_{-1.63}$ & $12.78^{+0.96}_{-0.86}$ & $0.76$ \\
 7078 & $9.1^{+1.8}_{-2.2}$ & $-0.21^{+0.85}_{-0.81}$ & $1.60^{+1.21}_{-1.03}$ & $12.48^{+0.65}_{-0.60}$ & $0.76$ \\
 7078 & $14.4^{+3.0}_{-3.4}$ & $1.94^{+1.15}_{-1.27}$ & $0.72^{+0.70}_{-0.50}$ & $12.73^{+0.44}_{-0.43}$ & $0.76$ \\
 7078 & $22.7^{+4.9}_{-5.3}$ & $2.55^{+0.98}_{-2.20}$ & $0.42^{+0.42}_{-0.29}$ & $11.20^{+0.28}_{-0.27}$ & $0.75$ \\
 7078 & $36.0^{+7.7}_{-8.4}$ & $2.59^{+0.21}_{-0.22}$ & $1.51^{+0.34}_{-0.33}$ & $10.17^{+0.20}_{-0.19}$ & $0.75$ \\
 7078 & $52.2^{+17.1}_{-8.4}$ & $2.54^{+0.14}_{-0.14}$ & $2.16^{+0.31}_{-0.33}$ & $8.98^{+0.20}_{-0.20}$ & $0.74$ \\
 7089 & $3.5^{+1.9}_{-3.4}$ & $-0.41^{+1.03}_{-1.02}$ & $2.02^{+1.96}_{-1.42}$ & $12.35^{+1.02}_{-0.97}$ & $0.80$ \\
 7089 & $7.1^{+1.5}_{-1.7}$ & $1.62^{+0.78}_{-0.76}$ & $1.99^{+1.53}_{-1.33}$ & $10.83^{+0.81}_{-0.76}$ & $0.80$ \\
 7089 & $11.4^{+2.3}_{-2.7}$ & $1.33^{+0.80}_{-0.80}$ & $1.06^{+0.86}_{-0.71}$ & $10.25^{+0.43}_{-0.43}$ & $0.80$ \\
 7089 & $17.9^{+3.8}_{-4.2}$ & $0.76^{+0.31}_{-0.30}$ & $1.85^{+0.56}_{-0.56}$ & $10.64^{+0.32}_{-0.31}$ & $0.80$ \\
 7089 & $28.5^{+5.9}_{-6.8}$ & $0.75^{+0.11}_{-0.11}$ & $3.24^{+0.33}_{-0.34}$ & $9.75^{+0.20}_{-0.21}$ & $0.79$ \\
 7089 & $44.4^{+10.0}_{-10.0}$ & $0.69^{+0.07}_{-0.07}$ & $3.44^{+0.25}_{-0.24}$ & $8.96^{+0.14}_{-0.14}$ & $0.78$ \\
 7089 & $60.3^{+16.4}_{-5.9}$ & $0.80^{+0.13}_{-0.12}$ & $3.67^{+0.43}_{-0.45}$ & $8.18^{+0.27}_{-0.27}$ & $0.78$ \\
 7099 & $2.9^{+1.4}_{-2.5}$ & $-1.81^{+2.95}_{-1.37}$ & $1.02^{+1.03}_{-0.71}$ & $8.13^{+0.79}_{-0.71}$ & $0.76$ \\
 7099 & $5.7^{+1.2}_{-1.4}$ & $-2.53^{+1.90}_{-0.98}$ & $1.02^{+1.01}_{-0.71}$ & $7.51^{+0.62}_{-0.59}$ & $0.75$ \\
 7099 & $9.0^{+2.0}_{-2.0}$ & $1.36^{+1.28}_{-1.38}$ & $0.56^{+0.59}_{-0.39}$ & $6.54^{+0.43}_{-0.40}$ & $0.75$ \\
 7099 & $14.4^{+3.1}_{-3.4}$ & $2.59^{+0.94}_{-1.23}$ & $0.43^{+0.42}_{-0.29}$ & $6.23^{+0.26}_{-0.26}$ & $0.74$ \\
 7099 & $22.8^{+5.0}_{-5.2}$ & $-1.71^{+0.72}_{-0.73}$ & $0.46^{+0.31}_{-0.31}$ & $5.77^{+0.18}_{-0.17}$ & $0.73$ \\
 7099 & $35.8^{+8.3}_{-8.0}$ & $-1.84^{+0.75}_{-0.76}$ & $0.33^{+0.24}_{-0.21}$ & $5.50^{+0.15}_{-0.13}$ & $0.72$ \\
 7099 & $52.9^{+17.1}_{-8.8}$ & $-1.81^{+0.84}_{-0.84}$ & $0.30^{+0.24}_{-0.19}$ & $5.22^{+0.16}_{-0.14}$ & $0.71$ \\
\hline
\end{tabular}

\end{table}
\clearpage

\section{Bias estimates for the ($v/\sigma$) and $\lambda_{\rm R}$ profiles}
\label{app:vs_lambdar}

As mentioned above in Sect.~\ref{sec:vsigma_lambdar}, the values obtained for $(v/\sigma)$ and $\lambda_{\rm R}$ have to be corrected for biases introduced by the finite accuracy of the data. The reason for this is that any deviations from the mean velocity that enter eqs.~\ref{eq:vsigma} and \ref{eq:lambdar}, regardless of whether they are caused by cluster rotation or statistic scatter, contribute to the final result.

\begin{figure*}
 \includegraphics[width=\textwidth]{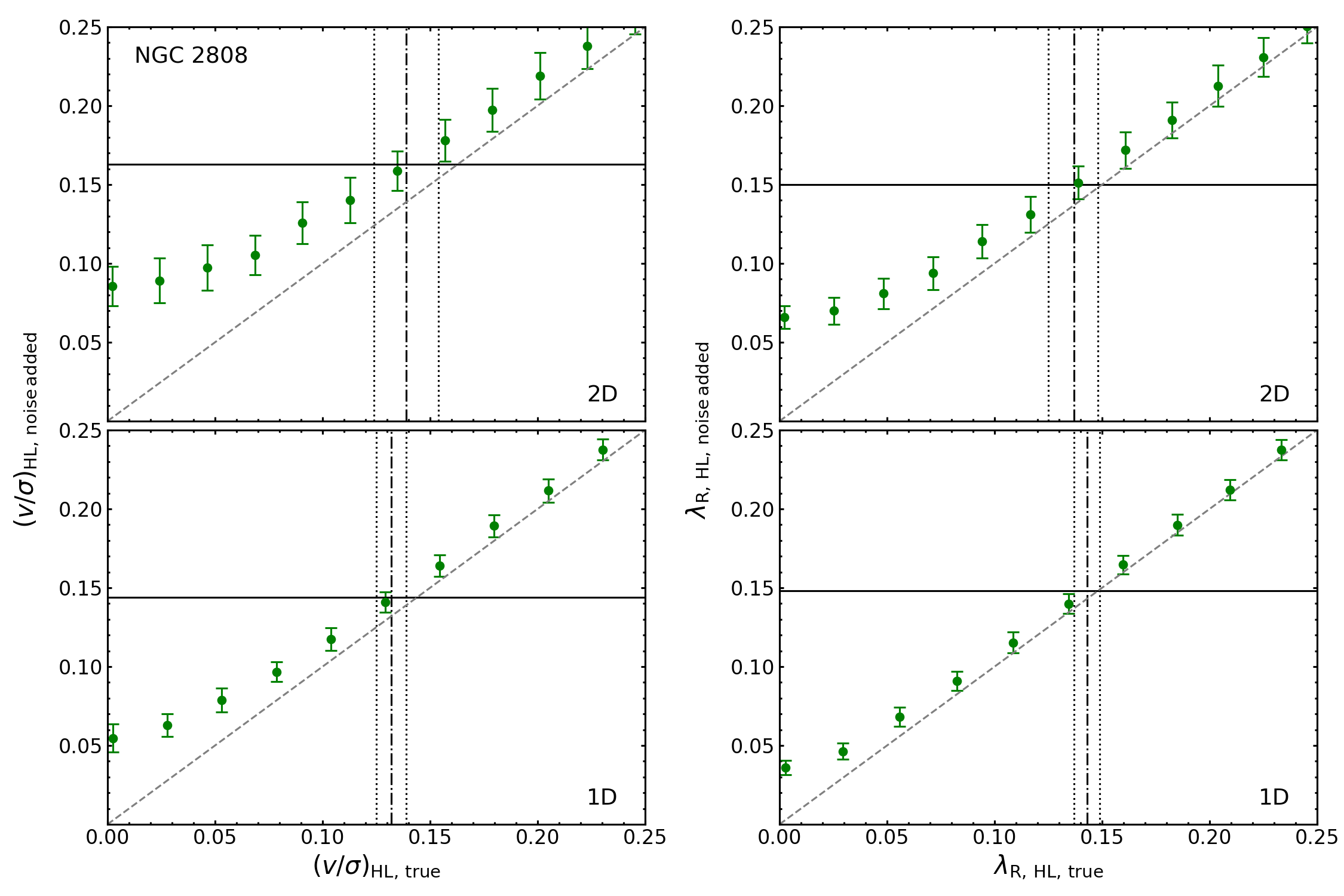}
 \caption{Bias correction applied when determining $(v/\sigma)_{\rm HL}$ (\textit{left}) or $\lambda_{\rm R,\,HL}$ (\textit{right}), using the example of NGC~2808. The \textit{top} panels show the results obtained on the original 2D grid, the \textit{bottom} panels the results obtained when creating a new grid from the 1D radial profiles. Each plot shows the recovered value (after adding noise) as a function of the intrinsic one (without noise). Green circles indicate the results from our simulations. The solid horizontal line is the value measured from the actual data, the dashed and dotted vertical lines provide the value and the $1\sigma$ uncertainty interval after correction for the noise bias.}
 \label{fig:app:vs_lambdar_bias}
\end{figure*}

To estimate the contribution of the statistic scatter to our values for $(v/\sigma)$ and $\lambda_{\rm R}$, we made the following approach. We used the same Voronoi bins depicted in Figs~\ref{fig:example_kinematics} and \ref{fig:app:kinematics} and assigned each bin a new mean velocity under the assumption that the rotation velocity scales linearly with radius, i.e. $v_{\rm rot}(r)=\alpha r$. From this data, we determined unbiased values for ($v/\sigma$) and $\lambda_{\rm R}$. Then we added scatter to the simulated rotation field, according to the noise estimates we obtained for each Voronoi bin from our data and recalculated $(v/\sigma)$ and $\lambda_{\rm R}$. The latter values are biased to high values because of the additional noise  distribution. We repeated this analysis for different velocity gradients $\alpha$, leading to a relation between the biased values and the intrinsic ones that is displayed in  Fig.~\ref{fig:app:vs_lambdar_bias} for the example of NGC~2808. Using this relation, we can correct the value measured on the real data (the horizontal line in Fig.~\ref{fig:app:vs_lambdar_bias}) to get an unbiased measurement (the vertical dashed line) and a corresponding uncertainty interval (the dotted vertical lines). The uncertainties were obtained by rerunning the simulation $N=100$ times for each value of $\alpha$.

In Fig.~\ref{fig:app:vs_lambdar_bias}, the bias correction is displayed for two different types of binning. In upper panels, we used the original Voronoi bins as displayed in Figs.~\ref{fig:example_kinematics} and \ref{fig:app:kinematics}. The results displayed in the bottom panels were obtained by creating new two-dimensional maps from the radial profiles as outlined in Sect.~\ref{sec:vsigma_lambdar}. It is obvious that for this approach, the biases are significantly reduced because of the smaller uncertainties per Voronoi bin. We observed a similar behaviour or all of our clusters, hence we decided to use the maps created from the radial profiles. It is further visible in Fig.~\ref{fig:app:vs_lambdar_bias} that the bias was stronger for the $(v/\sigma)$ measurements than it was for the $\lambda_{\rm R}$ measurements. The reason for this is likely the weighting with projected radius when determining $\lambda_{\rm R}$, see eq.~\ref{eq:lambdar}, which enhances the rotation signal compared to random scatter. For this reason, we decided to prefer the values obtained for $\lambda_{\rm R}$ over those obtained for $(v/\sigma)$ (keeping in mind that both values gave very similar results).


\bsp	
\label{lastpage}
\end{document}